\documentclass[fleqn,usenatbib]{mnras}

\usepackage{newtxtext,newtxmath}
\usepackage[T1]{fontenc}
\usepackage{ae,aecompl}

\usepackage{graphicx}	% Including figure files
\usepackage{amsmath}	% Advanced maths commands

\usepackage{amssymb}	% Extra maths symbols
\usepackage{xspace}
\usepackage[dvipsnames]{xcolor}
\usepackage{lineno}
\newcommand{\Alslr}{$^{26}$Al\xspace}

\newcommand{\Clslr}{$^{36}$Cl\xspace}
\newcommand{\Caslr}{$^{41}$Ca\xspace}

\newcommand{\Feslr}{$^{60}$Fe\xspace}
\newcommand{\Mnslr}{$^{53}$Mn\xspace}

\newcommand{\Nbslr}{$^{92}$Nb\xspace}
\newcommand{\Tcslrseven}{$^{97}$Tc\xspace}
\newcommand{\Tcslreight}{$^{98}$Tc\xspace}
\newcommand{\Pdslr}{$^{107}$Pd\xspace}
\newcommand{\Snslr}{$^{126}$Sn\xspace}
\newcommand{\Islr}{$^{129}$I\xspace}
\newcommand{\Csslr}{$^{135}$Cs\xspace}
\newcommand{\Smslr}{$^{146}$Sm\xspace}
\newcommand{\Hfslr}{$^{182}$Hf\xspace}
\newcommand{\Pbslr}{$^{205}$Pb\xspace}
\newcommand{\NiC}{$^{56}$Ni\xspace}

\newcommand{\SiC}{$^{28}$Si\xspace}
\newcommand{\NeC}{$^{20}$Ne\xspace}
\newcommand{\OC}{$^{16}$O\xspace}
\newcommand{\CC}{$^{12}$C\xspace}
\newcommand{\HeC}{$^{4}$He\xspace}
\newcommand{\msun}{\stmass}

\newcommand{\MgpgAl}{$^{25}$Mg(p,$\gamma$)$^{26}$Al\xspace}
\newcommand{\Fedoublen}{$^{58}$Fe(n,$\gamma$)$^{59}$Fe(n,$\gamma$)\Feslr}
\newcommand{\NeAlphaN}{$^{22}$Ne($\alpha$,n)$^{25}$Mg\xspace}

\newcommand{\stmass}{ M$_{\odot}$\xspace}
\newcommand{\FoE}{$\times$10$^{51}$ ergs\xspace}

\newcommand{\gammaproc}{$\gamma$-process\xspace} 

\newcommand{\alp}{$\alpha$\xspace}

\title[Radioactive nuclei in the early Solar System]{Radioactive nuclei in the early Solar System: 
analysis of the 15 isotopes produced by core-collapse supernovae
}

\author[Lawson et al.]{Thomas V. Lawson$^{1,2,3,4,5}$,
Marco Pignatari$^{1,2,3,4,5}$,
Richard J. Stancliffe$^{1,3,5,6}$,
\newauthor
Jacqueline den Hartogh$^{2,3}$,
Sam Jones$^{3,7}$,
Chris L. Fryer$^{3,7}$,
Brad K. Gibson$^{1,4,5}$,
\newauthor
Maria Lugaro$^{2,8,9}$
\\
$^{1}$E.~A.~Milne Centre for Astrophysics, Department of Physics and
               Mathematics, University of Hull, HU6 7RX, United Kingdom\\
$^{2}$Konkoly Observatory, Research Centre for Astronomy and Earth Sciences, E\"otv\"os Lor\'and Research Network (ELKH),\\ Konkoly Thege Mikl\'{o}s \'{u}t 15-17, H-1121 Budapest, Hungary\\
$^{3}$NuGrid Collaboration, \url{http://nugridstars.org}\\
$^{4}$Joint Institute for Nuclear Astrophysics - Center for the Evolution of the Elements\\
$^{5}$BridGCE: Bridging Disciplines of Galactic Chemical Evolution\\
$^{6}$H. H. Wills Physics Laboratory, Tyndall Avenue, Bristol BS8 1TL, UK\\
$^{7}$X Computational Physics (XCP) Division, Los Alamos National Laboratory, Los Alamos, NM 87545, USA\\
$^{8}$School of Physics and Astronomy, Monash University, VIC 3800, Australia\\
$^{9}$ELTE E\"{o}tv\"{o}s Lor\'and University, Institute of Physics, Budapest 1117, P\'azm\'any P\'eter s\'et\'any 1/A, Hungary\\
}

\date{Accepted 2021 December 13. Received 2021 December 13; in original form 2021 September 10}

\pubyear{2021}
 
\begin{document}
\label{firstpage}
\pagerange{\pageref{firstpage}--\pageref{lastpage}}
\maketitle

\begin{abstract}
Short-lived radioactive isotopes (SLRs) with half-lives between 0.1 to 100 Myr can be used to probe the origin of the Solar System. In this work, we examine the core-collapse supernovae production of the 15 SLRs produced: \Alslr, \Clslr, \Caslr, \Mnslr, \Feslr, \Nbslr, \Tcslrseven, \Tcslreight, \Pdslr, \Snslr, \Islr, \Csslr, \Smslr, \Hfslr, and \Pbslr. We probe the impact of the uncertainties of the core-collapse explosion mechanism by examining a collection of 62 core-collapse models with initial masses of 15, 20, and 25\stmass, explosion energies between 3.4$\times$10$^{50}$ and 1.8$\times$10$^{52}$ ergs and compact remnant masses between 1.5\stmass and 4.89\stmass. We identify the impact of both explosion energy and remnant mass on the final yields of the SLRs. Isotopes produced within the innermost regions of the star, such as \Nbslr and \Tcslrseven, are the most affected by the remnant mass, \Nbslr varying by five orders of magnitude. Isotopes synthesised primarily in explosive C-burning and explosive He-burning, such as \Feslr, are most affected by explosion energies. \Feslr increases by two orders of magnitude from the lowest to the highest explosion energy in the 15\stmass model. The final yield of each examined SLR is used to compare to literature models.
% from the lowest energy 15\stmass model to the highest.
% from mass fraction 4$\times$10$^{-5}$ in the lowest energy 15\stmass model (3.4$\times$10$^{50}$ ergs) to 1.4$\times$10$^{-3}$ in the highest energy 15\stmass model (1.1$\times$10$^{52}$ ergs) .
% Explosion energy determines the neutron and proton budget in 
% We use these models to identify which models can best reproduce the early Solar System abundances.

\end{abstract}

\begin{keywords}
Stellar types: Massive stars -- Supernovae -- Nucleosynthesis
\end{keywords}

%%%%%%%%%%%%%%%%%%%%%%%%%%%%%%%%%%%%%%%%%%%%%%%%%%
%%%%%%%%%%%%%%%%% BODY OF PAPER %%%%%%%%%%%%%%%%%%
%%%%%%%%%%%%%%%%%%%%%%%%%%%%%%%%%%%%%%%%%%%%%%%%%%

\section{Introduction} 
\label{ss:Intro}

Radioactive isotopes are useful tools to determine the age of objects. On Earth, $^{14}$C (with a half-life, t$_{\frac{1}{2}}$, of 5730 yr) is used to measure timescales relative to human history. 
In the Universe, long lived isotopes (t$_{\frac{1}{2}}$ $\approx$ Gyr) such as $^{232}$Th or $^{238}$U are used to measure cosmological timescales like the age of the Galaxy \citep{Dauphas2005}. 
The process of the formation of the Sun, starting from the formation of its molecular cloud, lasts between 15 years \citep[][]{Hartmann2001,Murray2011}, therefore, in order to probe this time period we examine isotopes that have comparable decay times.
We use radioactive isotopes with half-lives between 0.1 and 100 Myr (short-lived radionuclides: SLRs, henceforth) as chronometers for understanding the birth of the Sun. 
The abundances of SLRs in the early Solar System (ESS) are derived from meteoritic analysis.
% Based on Table 2 in \cite{Lugaro2018} a list of SLRs are chosen, these are detailed in Table \ref{tab:ess_ref_table}. 
% We do not consider $^{7}$Be and $^{10}$Be in this work as they are not produced significantly in supernovae, rather by cosmic ray spallation \citep{Desch2004}. 
% Notice, however, that \cite{Banerjee2016} proposed that low-mass CCSNe could produce $^{10}$Be by neutrino interaction with ejecta.
% % finds that low-mass supernova could produce $^{10}$Be by neutrino interactions.
% We do not consider $^{244}$Pu or $^{147}$Cm as CCSN do not produce r-process isotopes.
% To probe the formation of the Solar System we examine isotopes that have a decay time that is comparable to the timescale of the early Solar System (ESS).

\begin{table}
\centering
\begin{tabular}{c|ccc}
\hline
SLR & Daughter & Reference & T$_{1/2}$ (Myr) \\ 
\hline
$^{26}$Al & $^{26}$Mg & $^{27}$Al & 0.72  \\ 
$^{36}$Cl & $^{36}$S & $^{35}$Cl & 0.30  \\ 
$^{41}$Ca & $^{41}$K & $^{40}$Ca & 0.099  \\ 
$^{53}$Mn & $^{53}$Cr & $^{55}$Mn & 3.7  \\ 
$^{60}$Fe & $^{60}$Ni & $^{56}$Fe & 2.6  \\ 
$^{92}$Nb & $^{92}$Zr & $^{92}$Mo & 34  \\ 
$^{97}$Tc & $^{97}$Mo & $^{98}$Ru & 4.2  \\ 
$^{98}$Tc & $^{98}$Ru & $^{98}$Ru & 4.2  \\ 
$^{107}$Pd & $^{107}$Ag & $^{108}$Pd & 6.5  \\ 
$^{126}$Sn & $^{126}$Te & $^{124}$Sn & 0.23  \\ 
$^{129}$I & $^{129}$Xe & $^{127}$I & 15  \\ 
$^{135}$Cs & $^{135}$Ba & $^{133}$Cs & 2.3  \\ 
$^{146}$Sm & $^{142}$Nd & $^{144}$Sm & 68  \\ 
$^{182}$Hf & $^{182}$W & $^{180}$Hf & 8.9  \\ 
$^{205}$Pb & $^{205}$Tl & $^{204}$Pb & 17  \\ 
\end{tabular}
\label{tab:ess_ref_table}
\caption{SLRs considered in this work. SLRs are listed with daughter isotopes, reference isotopes and Half-lives (T$_{1/2}$). Data gathered from \protect\cite{Lugaro2018}.}
% If no error is given this value is an upper limit. 
% We do not consider $^{10}$Be in this work as it is not produced significantly in supernovae.
\end{table}

% Short-lived radioactive isotopes (SLRs) with half-lives between 0.1 to 100 Myr can be used to probe the origins of the Solar System. 
SLR signatures can be used alongside expected contributions from galactic chemical evolution (GCE) to identify the
time elapsed from production to being incorporated into the ESS; this is referred to as the so-called `isolation time' \citep{Cote2019,Cote2019a,YagueLopez2021}. 
SLRs with the shortest half-lives completely decay over long timescales ($\sim$100 Myr).
% if the isolation times.
% over long isolation times.% , and their signature would not be observable in meteorites formed in the ESS
Therefore, to explain the abundance of the shortest lived isotopes, such as \Alslr, \Clslr and \Caslr (half lives of these isotopes can be found in Table \ref{tab:ess_ref_table}), we require additional, local sources that are close in both time and space \citep{Huss2009}. These local sources may have polluted either the molecular cloud or the proto-solar disk \citep[see the review by][]{Lugaro2018}.

\cite{Huss2009} presented a comparison of the potential sources of SLRs into the ESS, concluding that intermediate-mass asymptotic giant branch (AGB) stars and massive stars between 20--60\stmass provide the most feasible sources. The AGB scenario has been considered \citep[see][]{Wasserburg2017} and compared to massive stars \citep{Vescovi2018}. The latter find that the contamination of a solar nebula by a single core-collapse supernova (CCSN) event should also pollute stable isotopes to excess.
\cite{Meyer2000} examines the different SLR yields in the Solar System and the values expected from GCE approaches, including a comparison to the massive star models of \cite{Meyer1995}.

A popular ESS pollution scenario is a single CCSN event, which polluted the material that formed the Solar System. A supernova could also explain the formation of the Solar System itself, by shocking the gas cloud that would go on to form the Solar System (as first proposed by \cite{Cameron1977}). During this shock we can expect the ejecta of this supernova event to enrich the cloud \citep{Gritschneder2012} or the disk \citep[][]{PortegiesZwart2018}. 
\cite{Boss2014} and \cite{Boss2017} show that a collapse caused by a supernova can explain both \Feslr and \Alslr, but a problem with this is that it can lead to an over-enhancement of \Mnslr up to three orders of magnitude. The \Mnslr overproduction can be solved by allowing a larger fallback during the CCSN \citep[see][]{Takigawa2008}, where fallback is defined as the inner region of the star that collapses onto the neutron star or black hole remnant.
A chemo hydro-dynamical simulation of the entire Milky Way Galaxy by \cite{Fujimoto2018} examined the impact of individual CCSNe to SLR abundances, finding that pollution by CCSNe provides significant amounts of \Alslr and \Feslr. However, this work considered only two SLRs.
% with plans for an extended analysis of \Caslr and \Mnslr.

% Another potential source of SLRs are AGB winds.
% \cite{Wasserburg2006} reviews the potential for AGB stellar sources. This work is expanded on by \cite{Wasserburg2017} who compare enrichemnt of \Pdslr and \Hfslr to the enrichment of \Alslr and \Feslr, finding that in order to best balance the production of these isotopes a 4--5\stmass \citep{Vescovi2018}.
% Massive AGB stars also produce large amounts of \Alslr via proton capture at the base of the convective envelope.
% We excluded this scenario as the timescale of AGB winds yields would be too long for a single instantaneous pollution event.

Previous works that examine CCSN scenario yields have only examined a selection of SLRs.
% \cite{Woosley1995} present yields for supernovae of different masses, between 11--40\stmass and $Z$ = 0--1 $Z_{\odot}$. These models had an upper isotopic limit of $^{71}$Ge, which allows a comparison to the lower mass isotopes we examine (\Alslr, \Clslr, \Caslr, \Mnslr, \Feslr).
\cite{Timmes1995} examined the GCE component of \Alslr and \Feslr in the Milky Way using the CCSN yields from \cite{Woosley1995}. 
\cite{Meyer1995} examined the yield for the ejecta of a 25\stmass star, including yields for \Alslr, \Clslr, \Caslr, \Mnslr and \Feslr.
In this work we continue the work done by \cite{Jones2019} and \cite{Andrews2020} to study the production of radioactive isotopes in CCSNe. Our nucleosynthesis simulations cover the full range of explosion parameters as described in \cite{Fryer2018}. We aim to study the production of all SLRs presented in Table \ref{tab:ess_ref_table} in CCSNe \citep[][]{Lugaro2018}.
%In this work we will expanding on the work done by expanding on work done by \cite{Jones2019} and \cite{Andrews2020}, by including for the first time the complete suite of SLRs as described in \cite{Lugaro2018} (Table \ref{tab:ess_ref_table}). 
%Our nucleosynthesis simulations cover the full range of explosion parameters described in \cite{Fryer2018}.
In particular, we discuss the impact of the explosion parameters affecting the production of the SLRs in CCSN ejecta. 

The paper is organized as follows. In Section~\ref{ss:Methods} we describe the stellar models and the nucleosynthesis code. In Section~\ref{sec:results}, we present the results for the production and abundances of the SLRs.
The discussion about the impact of model parameters on the SLR abundances and the comparison with other models are given in Section~\ref{sec:disc}. In Section~\ref{sec:Conc}, we summarize our results.

%%%%%%%%%%%%%%%%%%%%%%%%%%%%%%
%%%%%%%%%%%%%%%%%%%%%%%%%%%%%%
%%%%%%%%%%%%%%%%%%%%%%%%%%%%%%
%%%%%%%%%%%%%%%%%%%%%%%%%%%%%%
%%%%%%%%%%%%%%%%%%%%%%%%%%%%%%
%%%%%%%%%%%%%%%%%%%%%%%%%%%%%%
%%%%%%%%%%%%%%%%%%%%%%%%%%%%%%
%%%%%%%%%%%%%%%%%%%%%%%%%%%%%%
%%%%%%%%%%%%%%%%%%%%%%%%%%%%%%

\section{Simulations}
\label{ss:Methods}

We first describe the CCSN models used in this work, in particular which parameters are altered, as well as the method of pre-explosive and explosive nucleosynthesis. We then give a brief explanation of the nucleosynthesis that has taken place, calling into focus the key explosive nucleosynthesis sites.

\subsection{Methods and Models}

We use the CCSN models calculated by \cite{Fryer2018} on the basis of three progenitor models with initial masses of 15, 20 and 25 M$_{\odot}$ computed by \cite{Heger2010} with the KEPLER code \citep{Weaver1978,Woosley1995} and with  initial metallicity (Z) of 0.02 \citep[][]{Grevasse1993}. The explosions of each progenitor were calculated using a 1-dimensional Lagrangian hydrodynamic code, to mimic a 3-dimensional convective engine \citep{Fryer1999,Herant1994} by injecting energy above the proto-neutron star with three free parameters: the mass at which the energy is injected, the length of time of the energy injection, and the total energy injected.
\cite{Fryer2018} use these three parameters to study the uncertainty of the explosive engine, and to determine the uncertainty on the remnant mass and isotopic yield of each explosion model. 
The compact remnant left after core-collapse and fallback will form either a neutron star or black hole, depending on the final mass of the compact object.
The heaviest neutron star known is MSP J0740+6620, with a mass of 2.14$^{+0.18}_{-0.20}$M$_{\odot}$ \citep[][]{cromartie:20}. In general, we may assume that remnants with masses lower than MSP J0740+6620 form neutron stars, while heavier remnants will form black holes. Theoretical simulations of the critical mass boundary dividing neutron star and black hole formation are affected from several uncertainties, among others from the Equation of State to use for the neutron star interior \citep[e.g.,][]{ozel:16}. Therefore, at present the observations of the most massive neutron stars provides the best available constraint of such a mass boundary.
% \textbf{The remnant mass describes the mass of the compact remnant, either a neutron star of black hole, that remains after the shock.}

From this parameter study we examine 17, 23 and 22 models of progenitor mass 15, 20, and 25\stmass, respectively.
% The details of the parameter study are described in 
The ranges of the values of the explosion energy and remnant mass are listed in
Table \ref{tab:parameters}, with each model's specific resultant explosion energy and remnant mass of the compact object listed in Tables \ref{yield_table:1}--\ref{yield_table:5} \citep[][]{Fryer2018}.
As a result of the parameter space explored by \cite{Fryer2018}, a large range of explosion energies and remnant masses are given for each stellar progenitor.
We note that when describing a model we use the term "mass cut" to define a point in mass below which material falls back back onto the compact remnant.
% at which material is not ejected, rather falling back into the compact remnant. 
Mass cut is commonly determined by placing it where the ejected yield of \NiC is in agreement with CCSN light curves,
however in the models examined here mass cut is based on the physical properties of the explosion \citep[See Section 2 of][for a description]{Fryer2018}.

\begin{table}
 \centering
 \begin{tabular}{ccccc}
  \hline
  M$_{\rm prog}$ & Model & E$_{\rm exp}$ & M$_{\rm rem}$ \\
  (\stmass) & count & (\FoE) & (\stmass) \\
  \hline
  15& 17 & 0.34 -- 10.7 & 1.50 -- 1.94  \\
  20& 23 & 0.53 -- 8.86 & 1.74 -- 3.40  \\
  25& 22 & 0.89 -- 9.73 & 1.83 -- 4.89  \\
 \end{tabular}
 \caption{Parameters of CCSN explosion models: M$_{\rm prog}$ is the zero age main sequence mass, E$_{\rm exp}$ is the explosion energy, and M$_{\rm rem}$ is the %compact remnant mass.
 final mass of the compact object left from the SN explosion (a Neutron Star or a Black Hole). }
% \marco{[MP: Please correct everywhere font in the brackets, e.g., M$_{rem}$ should be M$_{\rm rem}$.}}
 \label{tab:parameters}
\end{table}

We perform the nucleosynthesis calculations using NuGrid post-processing tools, described in detail in \cite{Jones2019} and \cite{Andrews2020}. For the massive star progenitors we use the Multi-zone Post-Processing Network -- Parallel (MPPNP) code \citep{Pignatari2016a,Ritter2018}. This code calculates the nuclear reactions at each time step then performs mixing across the stellar model, separate to calculating the impact of nuclear reactions. The reaction network consists of roughly 1100 isotopes and 14000 reactions. For the core-collapse simulations we used the Tracer particle Post-Processing Network -- Parallel (TPPNP) \citep{Jones2019}. 
TPPNP functions by considering single tracer particles anchored to a mass co-ordinate, post-processing with no mixing between particles. 
We use a network of up to 5200 isotopes and 67000 reactions for CCSN nucleosynthesis \citep[for a full description of post-processing techniques see][]{Jones2019}. This network is larger than that in the MPPNP calculations because explosive nucleosynthesis experiences higher temperatures, as well as significantly higher neutron and proton densities. Therefore, more isotopes and reactions are needed.
We note that our \Alslr yields are higher than those reported in \cite{Jones2019} and \cite{Andrews2020}. In those calculations, the pre-explosive abundance of only the \Alslr isomeric state was inadvertently used as initial abundance of \Alslr during the supernova nucleosynthesis calculations. In this work %as here 
this issue with the treatment of the initial abundance of the isomeric and ground states of \Alslr is corrected. As a result, higher final \Alslr yields are obtained %initial abundance of \Alslr within the progenitor, which results in higher \Alslr yields, 
since also the pre-explosive contribution is consistently taken into account when the final integrated yields are calculated from the SN ejecta.

\subsection{General nucleosynthesis results} \label{ss:nucleosyn}

\begin{figure}
 \centering
 \includegraphics[width=\columnwidth]{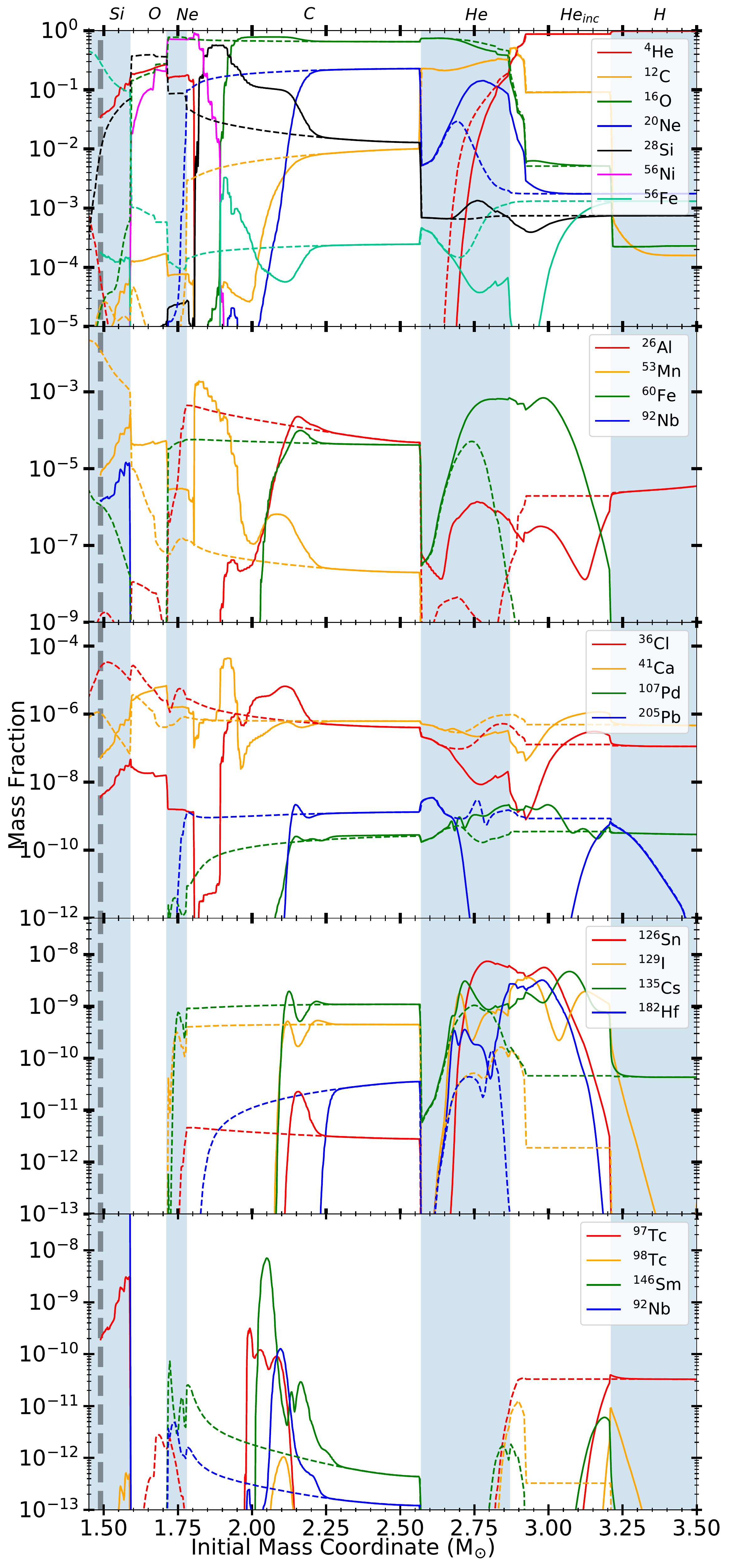}
 \caption{Abundance profile (in mass fraction) of selected isotopes for a selected 15\stmass model with
  % in massive star structure and 
 mass cut of 1.50\stmass (the lowest of the 15\stmass models) shown as a thick dashed vertical line and explosion energy of 4.79~\FoE. 
 % The mass-coordinates shown here 
 The top panel shows isotopes that represent the pre-CCSN structure of the model, with alternating shaded and non-shaded sections showing the different burning ashes. Each burning phase is labelled at the top of the plot. The other panels show the abundance profiles of the SLRs considered in Table \ref{tab:ess_ref_table}, all radioactive isotopes including the radiogenic contribution from isotopes with half-lives of less than 10$^{5}$ years.
 The abundance profiles are plotted at both pre-CCSN (dashed line) and post-CCSN profile (solid line). A thick gray, dashed vertical line shows the mass cut, below which we do not plot the explosive profiles.
 \Nbslr is plotted in both the second from top and the bottom panels, to show its large peak in the innermost region of the ejecta.
 % \marco{[MP: careful with the selection of colors. Are these color blind-friendly?]}
 }
 \label{fig:long_abu_plot_15}
\end{figure}

\begin{figure}
 \centering
 \includegraphics[width=\columnwidth]{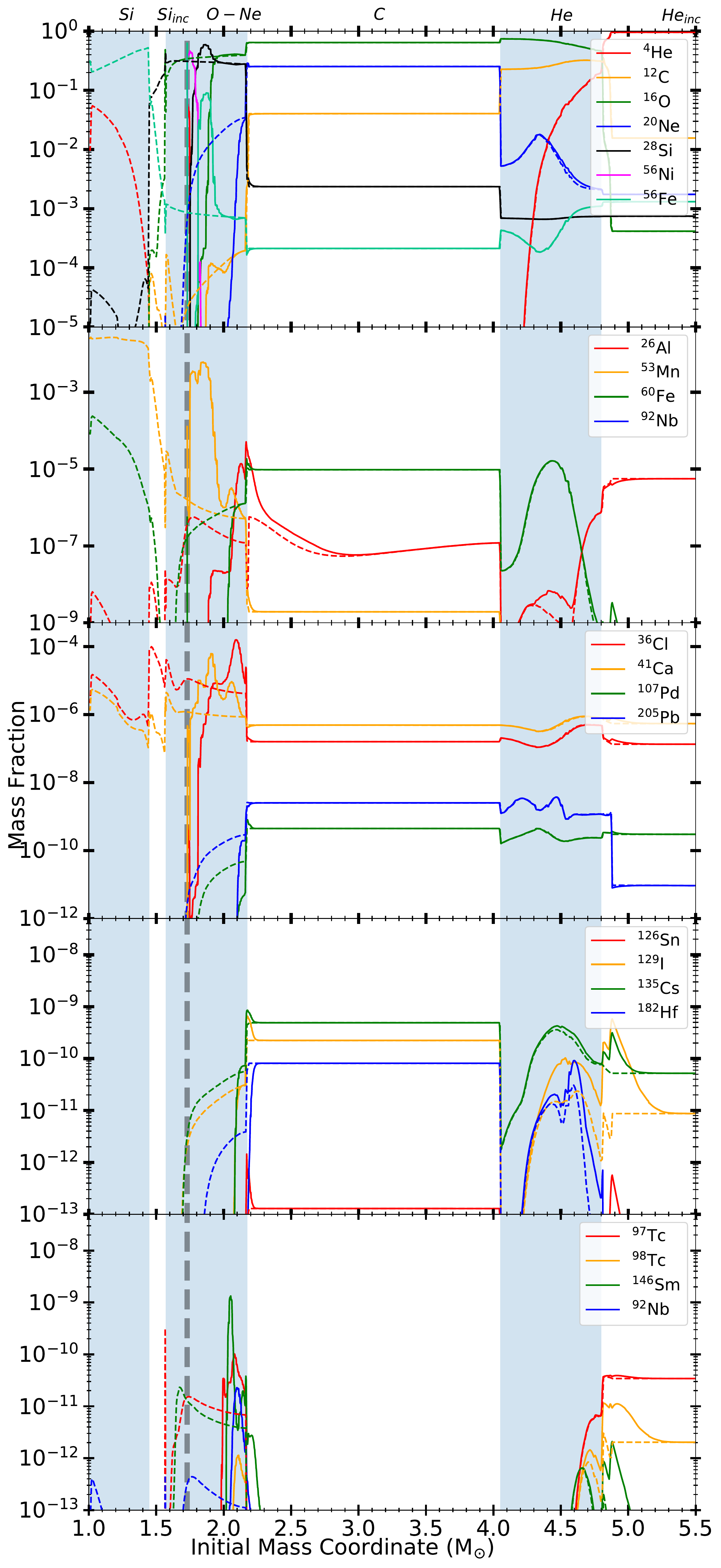}
 \caption{
 Same as Fig.~\ref{fig:long_abu_plot_15} for 20\stmass progenitor, with mass cut 1.74\stmass and explosion energy 2.85\FoE.
 }
 \label{fig:long_abu_plot_20}
\end{figure}

\begin{figure}
 \centering
 \includegraphics[width=\columnwidth]{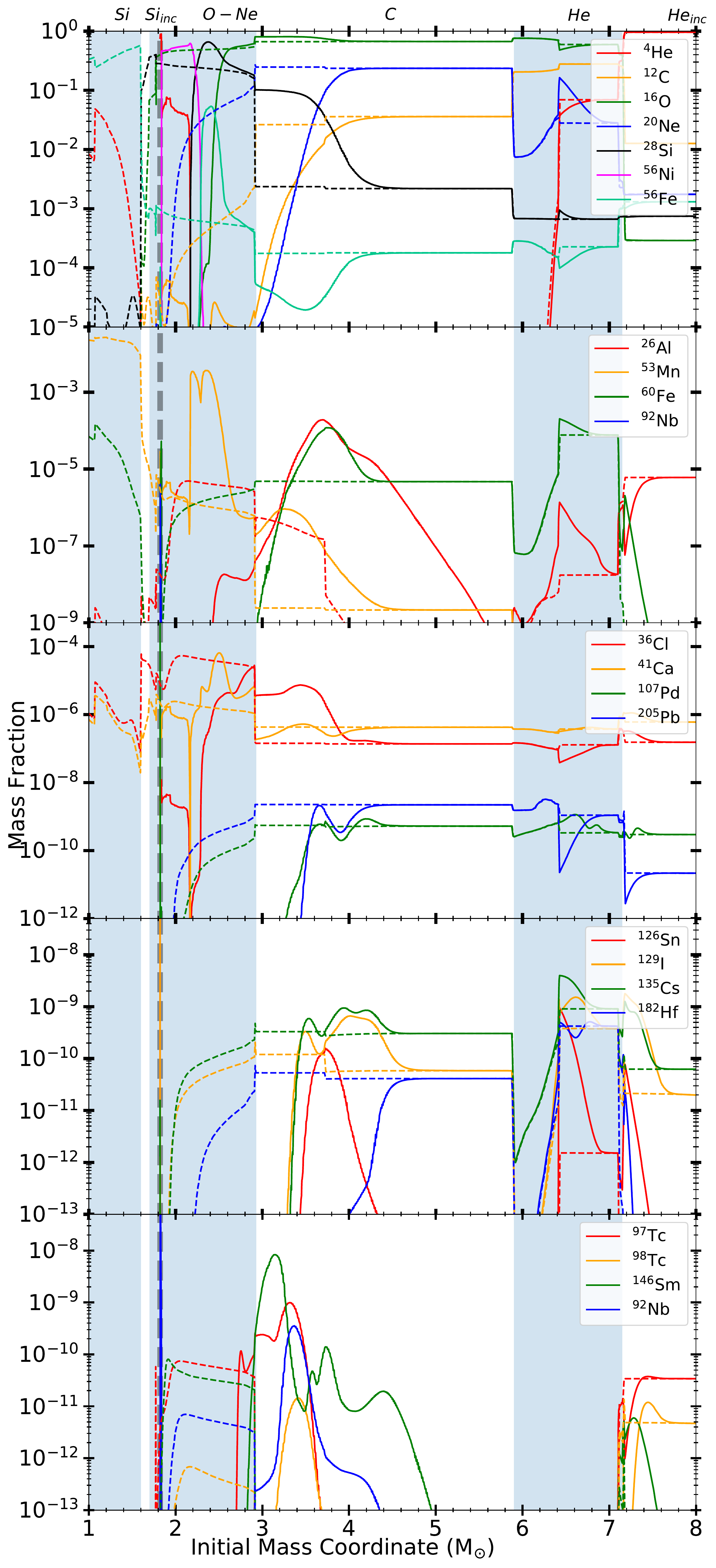}
 \caption{
 Same as Fig.~\ref{fig:long_abu_plot_15} for 25\stmass progenitor, with mass cut 1.83\stmass and explosion energy 3.07\FoE.}
 \label{fig:long_abu_plot_25}
\end{figure}

% We study the 
We examine the
% production of the SLRs listed in Table \ref{tab:ess_ref_table} 
general nucleosynthesis
in the 62 CCSN models described in the previous subsection\footnote{Full data of these models can be found at:
\hyperlink{https://ccsweb.lanl.gov/astro/nucleosynthesis/nucleosynthesis_astro.html}{https://ccsweb.lanl.gov/astro/nucleosynthesis/nucleosynthesis\_astro.html}}
% of progenitor ZAMS mass 15, 20, and 25\stmass models 17, 23, 22, respectively
\citep[][]{Fryer2018,Andrews2020}. 
The final yields of SLRs and their reference isotopes, explosive energy and remnant masses are reported in Table \ref{yield_table:1}.
Here we focus on three selected models to discuss the nucleosynthesis both pre-explosion and post-explosion. 
Figures \ref{fig:long_abu_plot_15}, \ref{fig:long_abu_plot_20}, and \ref{fig:long_abu_plot_25} show selected abundance profiles for pre- and post-CCSN of selected models. These models are selected due to their low mass cut values, allowing the largest range of explosive nucleosynthesis to be observed.
% The 15\stmass model (Fig.~\ref{fig:long_abu_plot_15}) has a remnant mass of 1.50\stmass and explosion energy of 4.79\FoE, the 20\stmass model (Fig.~\ref{fig:long_abu_plot_20}) has a remnant mass of 1.74\stmass and explosion energy of 2.85\FoE, and the 25\stmass model (Fig.~\ref{fig:long_abu_plot_25}) has a remnant mass of 1.83\stmass and explosion energy of 3.07\FoE.

The shaded areas of Figs. \ref{fig:long_abu_plot_15}, \ref{fig:long_abu_plot_20} and \ref{fig:long_abu_plot_25} show the location of the pre-CCSN burning ashes, as labeled at the top of the figure. 
% The next burning phase has a lower upper mass coordinate limit, so the ashes of the previous burning phase remain.
These regions can be identified by the significant reduction of the relevant isotope, due to the previous burning phases.
Using the top panel of Fig.~\ref{fig:long_abu_plot_15} as an illustrative example, at mass coordinate 3.21\stmass a sharp increase in both \CC and \OC represent the interface between the He-burning ashes and the H-burning ashes.
The He-burning ashes extend to 2.57\stmass, and is split at 2.87\stmass into He and He$_{\rm inc}$, which are the ashes of complete and incomplete He-burning, respectively.
During He-burning the triple-$\alpha$ reaction has produced $^{12}$C, with a further $\alpha$ capture generating $^{16}$O.
% The outermost He$_{\rm inc}$-burning ashes are the result of partial He fusion,
% Using the top panel of Fig.~\ref{fig:long_abu_plot_15} as an illustrative example at mass coordinate 2.92\stmass a sharp increase in \CC and shallow \HeC decrease, this is the interface between the H-burning ashes and the He-burning ashes. 
% At 2.57\stmass we see a sharp drop in \CC and a sharp increase of \NeC and \SiC, defining the interface between He-burning ashes and the C-burning ashes. 
The ashes of C-burning are identifiable by the significant reduction in the abundance of $^{12}$C between 1.78\stmass and 2.57\stmass. During C-burning, $^{12}$C is consumed via the C-fusion main channels %reaction
$^{12}$C($^{12}$C,$\alpha$)$^{20}$Ne and $^{12}$C($^{12}$C,p)$^{23}$Na. 

At 1.77\stmass the sharp drop in \NeC 
and the increase of \OC (created via the $^{20}$Ne($\gamma$,$\alpha$)$^{16}$O reaction) represent the interface between the C-burning and Ne-burning ashes.
% During primarily creating \OC (via the $^{20}$Ne($\gamma$,$\alpha$)$^{16}$O reaction). 
%\marco{[MP: This is fine. Really hard to distinguish sometime from Ne-burning ashes and mild O-burning, since sometime they overlap and from Si28 rise and C decrease would not be enough to tell. You do not need to specify, but can see that O16 is increasing in this zone compared to the zone above... this is the Ne20ga making O16 only partly balanced from the o16ag. o16 fusion clearly is not active. No need to add this, just comment but keep in mind in case of the referee is asking.]} OK, you say this here below.
At 1.71\stmass the sharp drop in \OC and the sharp increase of \SiC define the beginning of the O-burning ashes, whose burning is dominated by the $^{16}$O($^{16}$O,$\alpha$)$^{28}$Si reaction. 
% This region has a significant depletion of \OC and excess of \SiC due to the 
% At 1.59\stmass we see a sharp drop in \SiC and a sharp increase of \FeC, everything below this point is Si-burning ashes. 
% We use the top panel of Fig.~\ref{fig:long_abu_plot_15} as an example to illustrate the location of the different pre-CCSN burning ashes.
% Within this figure, the dashed vertical line is the position of the mass cut, the alternating shaded areas are the different burning ashes, the dashed coloured lines are the isotopic abundance profiles pre-CCSN and the solid lines are post-CCSN abundance profiles. 
The Si-burning ashes are located below 1.59\stmass, where there is a sharp drop in $^{28}$Si. Silicon burning is characterised by hundreds of reactions, where a significant fraction of %activation of Quasi-Statistical Equilibrium (QSE), in which the 
forward and backward nuclear reactions %rates 
balance \citep{Woosley1973,Chieffi1998}. 
The main product of Si-burning is $^{56}$Fe.

The impact of the shock wave on the final yields of a CCSN model can be determined by examining the temperature profiles, as explosive nucleosynthesis processes have a strong temperature dependence. The mass of the remnant, and thus the location of the mass cut, is also key to determining the yields of a CCSN model, as everything above the remnant mass is ejected. In other words, if the mass cut is above a specific explosive nucleosynthesis region, then the products of that region are not ejected, therefore they are not included in the final yields.

\begin{figure}
 \centering
 \includegraphics[width=\columnwidth]{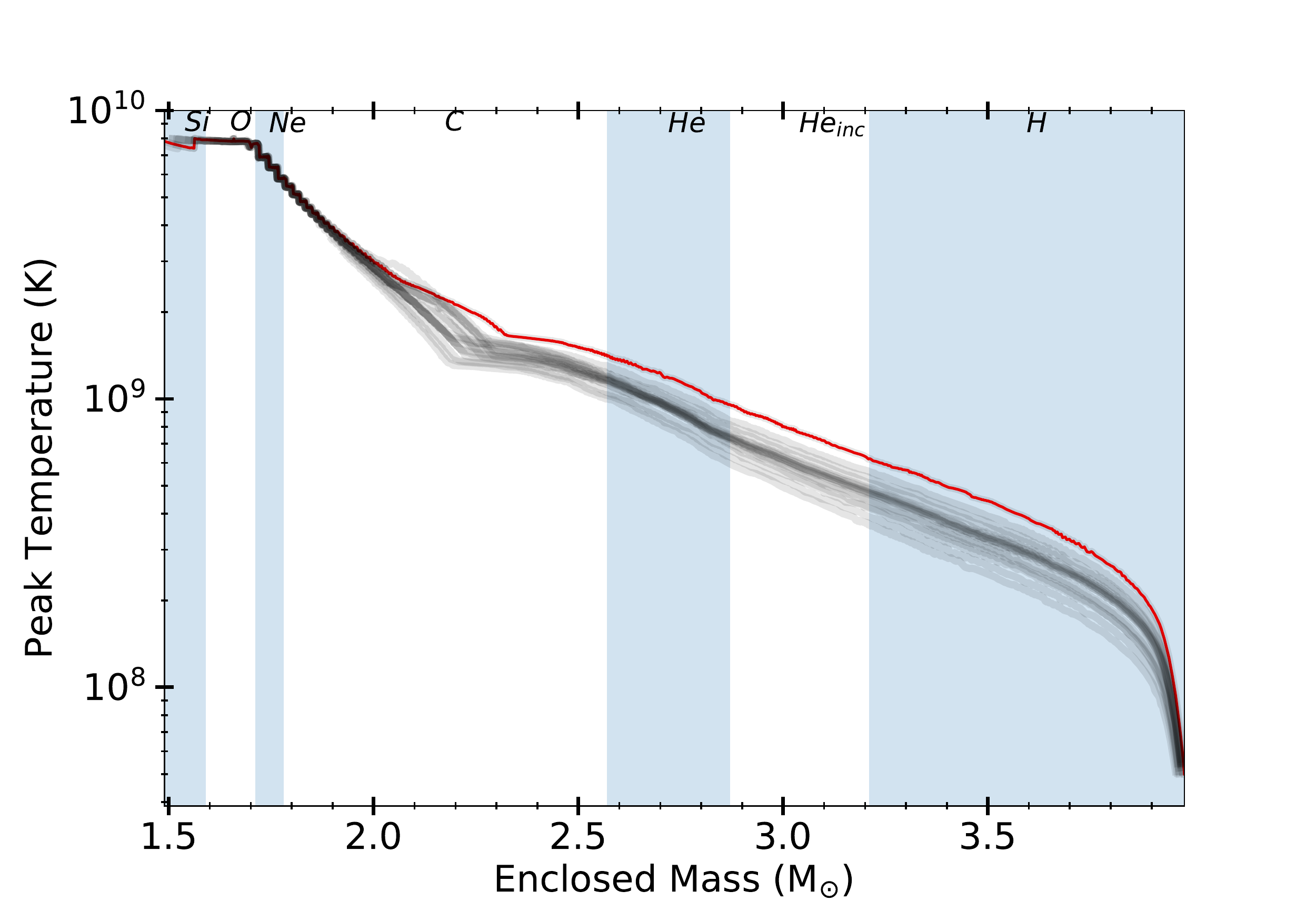}
 \includegraphics[width=\columnwidth]{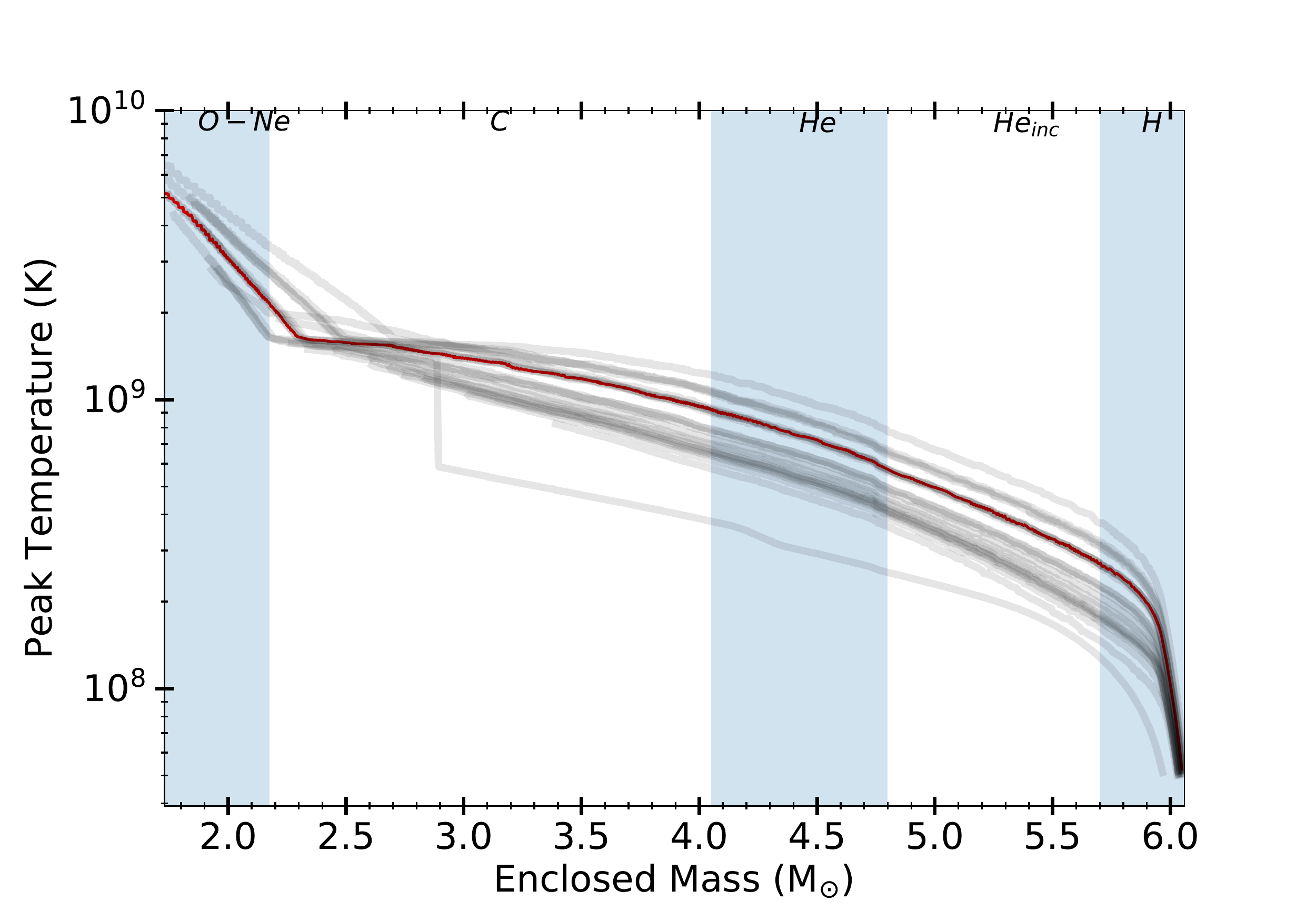}
 \includegraphics[width=\columnwidth]{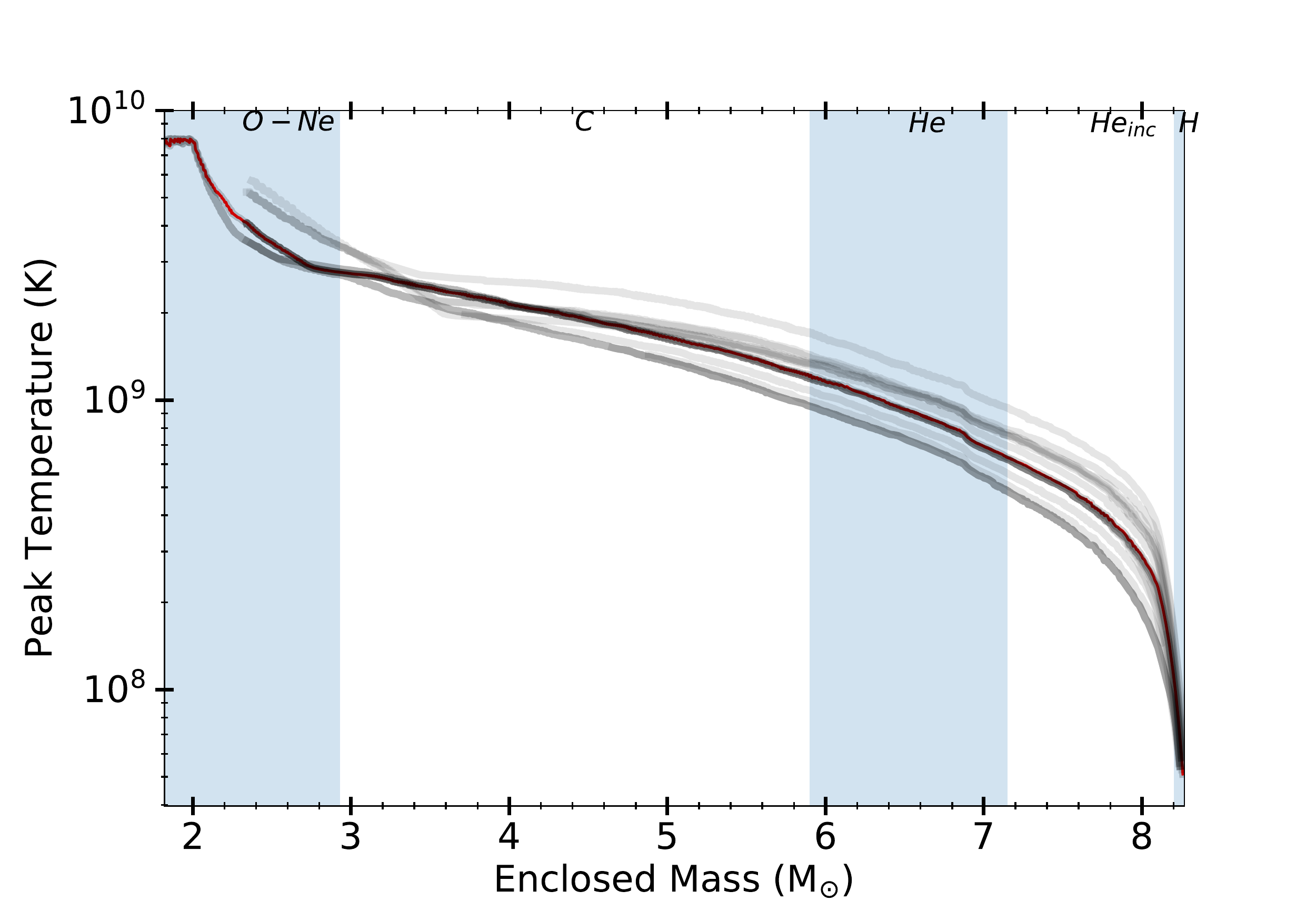}
 \caption{Peak temperature profiles of all the 62 models, constructed by selecting the maximum temperature reached at each mass coordinate as the shock passes through it. 
 The 15\stmass, 20\stmass and 25\stmass models are shown in the upper, middle and lower panels, respectively. 
 Alternating shaded and non-shaded areas highlight pre-CCSN burning ashes as labelled on the top.
 The profiles marked in red are those of the models shown in Fig.\ref{fig:long_abu_plot_15} -- Fig.\ref{fig:long_abu_plot_25}.
%  , the models with the lowest mass cut value, showing the most explosive nucleosynthesis processes.
 }
 \label{fig:t_MASS}
\end{figure}

% These injection parameters provide models with different explosion temperature profiles, shown in Fig.~\ref{fig:t_MASS}.
As the explosion shock-wave passes through the pre-CCSN material, the temperature and densities rise dramatically (to the peak values shown in Fig.~\ref{fig:t_MASS}) and subsequently cool.
These profiles determine the outcome of the explosive nucleosynthesis.
% Explosive nucleosynthesis is determined by the specifics of these shock-wave characteristics.
During explosive nucleosynthesis (solid lines in Figures \ref{fig:long_abu_plot_15}--\ref{fig:long_abu_plot_25}) 
%light isotopes 
in the inner %portion 
stellar ejecta %of the star, 
isotopes made in the pre-CCSN phase are mostly destroyed to make new species,
%via photodisintegration
due to extreme temperatures and densities (see Fig.~\ref{fig:t_MASS}). Explosive Si-burning mostly feeds the production of $^{56}$Ni and iron-group elements, with some unconsumed \HeC. 
This can be seen in Fig.~\ref{fig:long_abu_plot_15}, below 1.8\stmass.
The $\alpha$-rich freeze-out can also trigger the nucleosynthesis of elements heavier than iron by charged particle reactions, up to the mass region of Sr and Zr \citep[e.g.,][]{Woosley1992,Pignatari2016a}.
%This inner most portion of the star experiences the $\alpha$-rich freeze-out, where high temperatures and densities break down material into protons and neutrons, which subsequently form $\alpha$-particles as the explosion shock cools and expands.

As the CCSN shock expands
, moving outwards in space,
the peak temperature decreases (Fig.~\ref{fig:t_MASS}). 
Note that this temperature peak is not reached in all the structure at the same time, rather only the peak temperature is plotted per mass coordinate. 
% Simulating the temperature of the CCSN shock as it moves outward.
Between $\sim$1.7\stmass and $\sim$1.85\stmass the temperatures decreases to $\sim$5~GK and explosive Si-burning occurs (top panel of Fig.~\ref{fig:long_abu_plot_15}) \citep[][]{Woosley1995}. This consumes \SiC~and produce mostly \NiC~and iron-group elements under nuclear statistical equilibrium, possibly leaving some \HeC~unconsumed if the rate of expansion is sufficiently fast \citep[][]{Bodanskyi1968,Woosley1973}. %as well as a large amount of \HeC. 

Where the peak temperature of the shock falls to $\sim$3.6GK (between $\sim$1.85\stmass and $\sim$1.95\stmass in Fig.~\ref{fig:long_abu_plot_15}), explosive O-burning products are mostly obtained \citep[][]{Truran1970,Woosley1973}. 
In the figure, this consumes \OC and produces \SiC. 
%The shock wave triggers nuclear statistical equilibrium in both Si- and O-burning, during which the abundance of isotopes with a ratio of neutrons and protons (i.e. the electron fraction, \ye) of approximately 0.5 is favoured \marco{[TO CHECK WITH TOM]}.
% , reaching toward
% an electron fraction (\ye) 

As the peak temperature of the shock falls between $\sim$3.2GK and $\sim$2.5GK \citep[][]{rapp:06} also the signature of explosive Ne-burning can be obtained (between $\sim$1.95\stmass and $\sim$2.1\stmass in Fig.~\ref{fig:long_abu_plot_15}). As in pre-explosive nucleosynthesis \NeC is consumed by photodisintegration and $\alpha$-captures \citep[e.g.,][]{Thielemann1985}. Heavy isotopes provide successive ($\gamma$,n), ($\gamma$,p), and ($\gamma$,$\alpha$) photodisintegration reactions, allowing for lighter seed nuclei to produce proton rich isotopes such as \Nbslr and \Smslr. This process is the \gammaproc \citep[e.g.][]{rauscher:13,Pignatari2016b}.

As the peak temperature drops to $\sim$2~GK the explosive C-burning signature becomes dominant (between $\sim$2.1\stmass and $\sim$2.55\stmass in Fig.~\ref{fig:long_abu_plot_15}) \citep[][]{Hansen1971,Truran1978}. This consumes \CC left in the C-shell ashes, %and \NeC, 
modifying local pre-CCSN abundances of C-burning ashes like %\OC and some \SiC. 
Ne, Na and Mg. 
The neutron density reaches $\sim$10$^{18}$ cm$^{-3}$ and the proton density reaches above 10$^{20}$ cm$^{-3}$ \citep[see Fig.~8 of][]{Jones2019}. 
This allows for the nucleosynthesis of both neutron-rich and proton-rich isotopes. Above $\sim$2.2\stmass, as temperatures drop to $\sim$2GK,
% the timescale of the $^{22}$Ne($\alpha$,n)$^{25}$Mg neutron source reaction increases
the rate of the $^{22}$Ne($\alpha$,n)$^{25}$Mg neutron source reaction decreases, as such we see little to no nucleosynthesis between 2.2\stmass and 2.55\stmass.

As the shock expands and peak temperatures drop to $\sim$1~GK, explosive He-burning occurs between $\sim$2.55\stmass and $\sim$3.1\stmass in the 15\stmass model shown in Fig.~\ref{fig:long_abu_plot_15}. Here \HeC is consumed to produce isotopes in the $\alpha$-capture chain. At mass coordinate 2.77\stmass (the peak of $^{20}$Ne production), neutrons are significantly made by $^{22}$Ne($\alpha$,n)$^{25}$Mg. The neutrons made by $^{17}$O($\alpha$,n)$^{20}$Ne and $^{21}$Ne($\alpha$,n)$^{24}$Mg, are recycled from those previously captured by $^{16}$O and $^{20}$Ne, respectively \citep[e.g.,][]{raiteri:91,pignatari:10}.
%the reactions that provide neutrons to the system are $^{22}$Ne($\alpha$,n)$^{25}$Mg, $^{21}$Ne($\alpha$,n)$^{24}$Mg and $^{17}$O($\alpha$,n)$^{20}$Ne. 
The peak neutron density here is $\sim$10$^{18}$ cm$^{-3}$ and proton density is $\sim$5$\times$10$^{15}$ cm$^{-3}$ \citep[see Fig.~8 of][]{Jones2019}. While the neutron density is similar within explosive helium burning and explosive oxygen burning, the production of  neutron-rich  isotopes is considerably higher in explosive helium burning, as at these lower temperatures 
%as there is less competition with
photodisintegration reactions are not active for the heavy isotopes beyond iron.
% (p,$\gamma$) reactions due to comparably lower proton densities.

The 20\stmass and 25\stmass models (Fig.~\ref{fig:long_abu_plot_20} and Fig.~\ref{fig:long_abu_plot_25} respectively) show %different nucleosynthesis regimes, 
some important differences in the CCSN ejecta compared to the 15\stmass model of Fig.~\ref{fig:long_abu_plot_15}. 
The O and Ne-burning ashes within the pre-CCSN structure of the 20\stmass and 25\stmass models cannot be easily distinguished, unlike these ashes in the 15\stmass models. %are combined. %to form a O-Ne-burning ashes. 
They are identified only by a reduction in both \OC and \NeC, while \SiC increases considerably more than in Ne-burning \citep[e.g.,][]{Chieffi1998}. Both the 20\stmass and 25\stmass models show the ashes of incomplete Si-burning 
(labelled as Si$_{\rm inc}$), characterised by the partial destruction of \SiC without significant production of \HeC. 
However, these regions will not be ejected by the CCSN explosion.% which results from QSE.

The steep temperature profile of the 20\stmass model (see Fig.~\ref{fig:t_MASS})
causes the ignition of explosive Si-, O-, and C-burning to be located in the inner portion of the star. Due to a higher mass cut the contribution of explosive Si-burning is less present in the final yields, with a sharp peak of \NiC reaching 0.46 mass fraction at mass coordinate 1.75\stmass (0.01\stmass above the mass cut). Explosive O-burning occurs at $\sim$1.87\stmass, and explosive C-burning is located at $\sim$2.15\stmass. As explosive C-burning is within the Ne-burning ashes, there is less \CC to be used as fuel, reducing the impact of this explosive nucleosynthesis site on the final yields of isotopes created by these conditions. As the shock reaches $\sim$1~GK, explosive He-burning would normally begin, but as this happens in the upper C-burning ashes there is no \HeC to be consumed, and therefore %there is no explosive He-burning. 
the nucleosynthesis signature of explosive He-burning is marginal.

The 25\stmass model shown in Fig.~\ref{fig:long_abu_plot_25} has a structure that resembles the 15\stmass model, however, as the mass cut is higher, there is no ejected $\alpha$-rich freeze-out material outside of the mass cut. Figure \ref{fig:t_MASS} shows that for all 25\stmass models, the CCSN peak temperatures is above 1~GK up %until He-burning ashes are reached, 
at the bottom of the former convective He shell, ensuring a strong activation of explosive He-burning. % is reached, unlike in the 20\stmass model.

\begin{figure}
 \centering
 \includegraphics[width=\columnwidth]{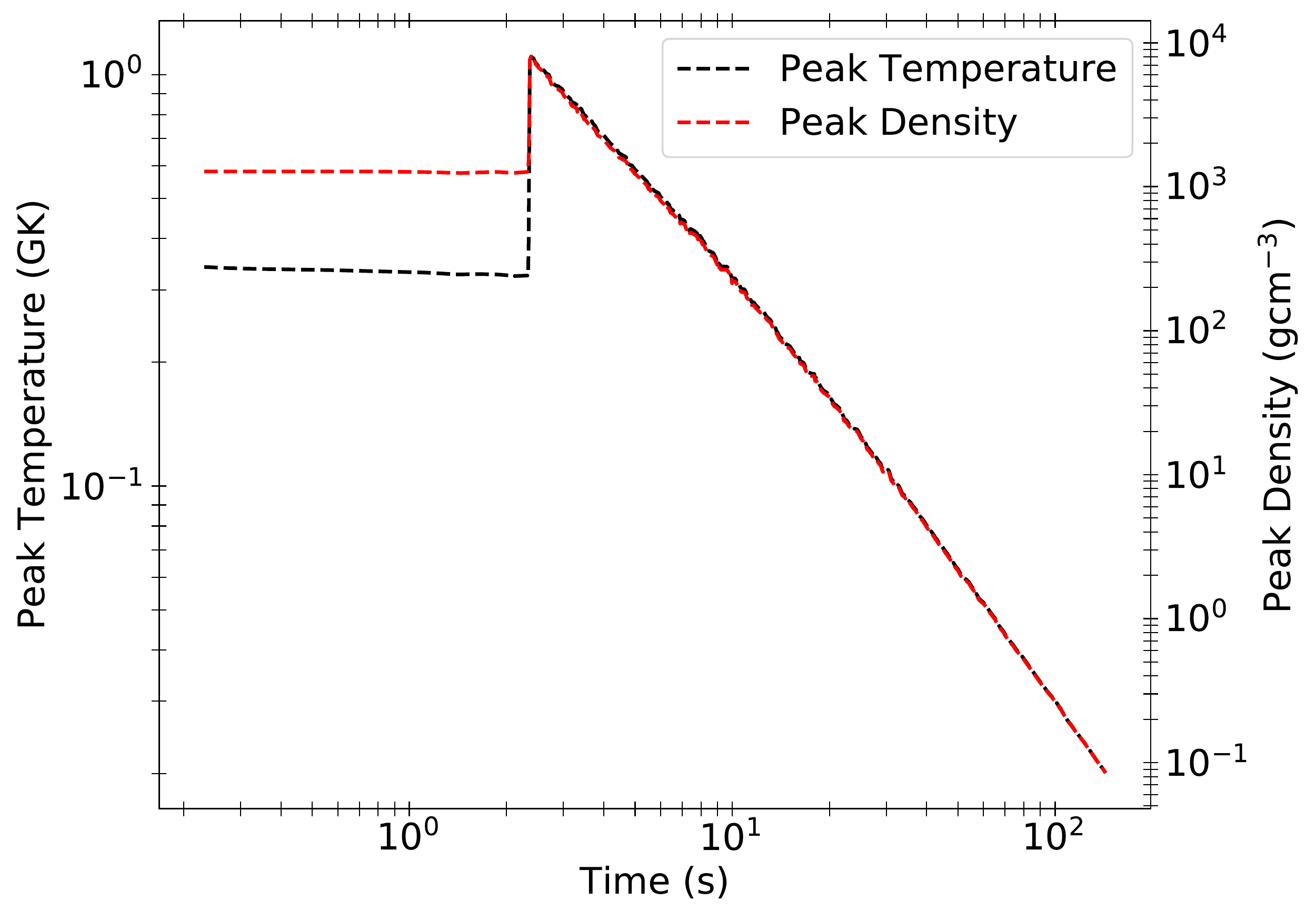}
 \caption{An example trajectory of temperature and density. 
 Trajectory generated from mass coordinate 2.77\stmass in model shown in Fig.~\ref{fig:long_abu_plot_15}.}
 \label{fig:traj_example}
\end{figure}

To examine in detail the specific nucleosynthesis sites of SLRs in the CCSN ejecta we extracted some trajectories from the models, where we took the temperature and density profiles over the course of the CCSN event for a given mass coordinate (an example is given in Fig.~\ref{fig:traj_example}). We then use these trajectories to examine the flux associated to each reaction and determine which are the most important for production and destruction.

%%%%%%%%%%%%%%%%%%%%%%%%%%%%%%
%%%%%%%%%%%%%%%%%%%%%%%%%%%%%%
%%%%%%%%%%%%%%%%%%%%%%%%%%%%%%
%%%%%%%%%%%%%%%%%%%%%%%%%%%%%%
%%%%%%%%%%%%%%%%%%%%%%%%%%%%%%
%%%%%%%%%%%%%%%%%%%%%%%%%%%%%%
%%%%%%%%%%%%%%%%%%%%%%%%%%%%%%
%%%%%%%%%%%%%%%%%%%%%%%%%%%%%%
%%%%%%%%%%%%%%%%%%%%%%%%%%%%%%

\section{Results} 
\label{sec:results}

For all SLRs we check the impact of radiogenic contributions from the decay of unstable isotopes
and we include where significant. In the next section we address the nucleosynthesis of each isotope separately.

We do not consider $^{7}$Be and $^{10}$Be in this work as they are not produced significantly in supernovae, rather by cosmic ray spallation \citep{Desch2004}. 
Notice, however, that \cite{Banerjee2016} proposed that low-mass CCSNe could produce $^{10}$Be by neutrino interaction with ejecta.
% finds that low-mass supernova could produce $^{10}$Be by neutrino interactions.
We do not consider $^{244}$Pu or $^{147}$Cm as CCSN do not produce r-process isotopes.

\subsection{\protect\Alslr} \label{ss:al}

In stars the \MgpgAl reaction is the main nucleosynthesis channel to produce \Alslr. The proton capture on $^{25}$Mg is first activated in the Mg-Al chain during hydrogen burning in the pre-CCSN phase. In more advanced burning stages, \Alslr can be made during C-burning because of the combined presence of abundant $^{25}$Mg and protons, where protons are made directly by C-fusion
(via the $^{12}$C($^{12}$C,p)$^{23}$Na reaction). During O-fusion $^{25}$Mg is destroyed to make heavier elements, and therefore these conditions are not suited for the production of \Alslr. Similar considerations can be made for explosive nucleosynthesis triggered by the CCSN shock. The dominant depletion channels for \Alslr under explosive conditions are the (n,p) and (n,$\alpha$) neutron capture reactions. These reactions deplete \Alslr in He-burning conditions and mitigate its production during C-burning, where neutrons are released by the $^{22}$Ne($\alpha$,n)$^{26}$Mg reaction. 
Above approximately 3.2\stmass the CCSN ejecta are carrying the signature of pre-CCSN H-burning ashes (see Fig.~\ref{fig:long_abu_plot_15}). \Alslr abundance is of the order of a few 10$^{-6}$ by mass fraction. The He-shell ashes above 2.92\stmass only show a minor depletion with respect to these quantities.
% [MP: CAN WE CHECK IN THE PROGENITOR IF THIS AL26 IS THERE SINCE THE ONSET OF THE SHELL AND JUST THERE WAS NO TIME TO BURN IT ALL, OR INSTEAD THERE IS SOME LEAKING FROM THE LAYERS ABOVE?]. 
This is due to a marginal activation of the $^{22}$Ne($\alpha$,n)$^{26}$Mg reaction, as also indicated by the negligible depletion of the s-process seed, $^{56}$Fe. In the 15$_{\odot}$ model the explosive H-burning contribution is not relevant for the nucleosynthesis of \Alslr, rather the bulk of production lies with C-burning (see Fig.~\ref{fig:long_abu_plot_15}). Instead, the neutron burst activated in the explosive shell He-burning by the CCSN shock 
%due to the strong activation of the $^{22}$Ne($\alpha$,n)$^{26}$Mg 
\citep[n-process, e.g.,][]{blake:76,meyer:00,pignatari:18} depletes \Alslr via the neutron capture channels mentioned above. Interestingly, some production of \Alslr also occurs during explosive He-burning. In this case, the source of protons to activate the production channel \MgpgAl is a combination of ($\alpha$,p) and (n,p) reactions. 

%During pre-CCSN He-burning \Alslr is destroyed via the (n,p) and (n,$\alpha$) neutron capture reactions, with the neutrons released via $^{22}$Ne($\alpha$,n)$^{26}$Mg reaction. 
In the 15\stmass model the pre-CCSN largest production of \Alslr occurs during carbon burning, with protons provided by the $^{12}$C($^{12}$C,p)$^{23}$Na reaction.
Within a region of about 0.8\msun in thickness, \Alslr reaches values larger than 10$^{-4}$ in mass fraction. The CCSN shock will deplete most of the C-burning ashes, but the region between about 2.2 and 2.55M$_{\odot}$ is ejected mostly unchanged, including the \Alslr reservoir made before the explosion. An explosive C-burning contribution is also present at about 2.15\msun, shown by a small increase in ejected \Alslr compared to the pre-CCSN abundance.
No significant production of \Alslr is obtained in CCSN ejecta deeper than the mass coordinates where explosive C-burning is taking place. 

\begin{figure}
 \centering
 \includegraphics[width=\columnwidth]{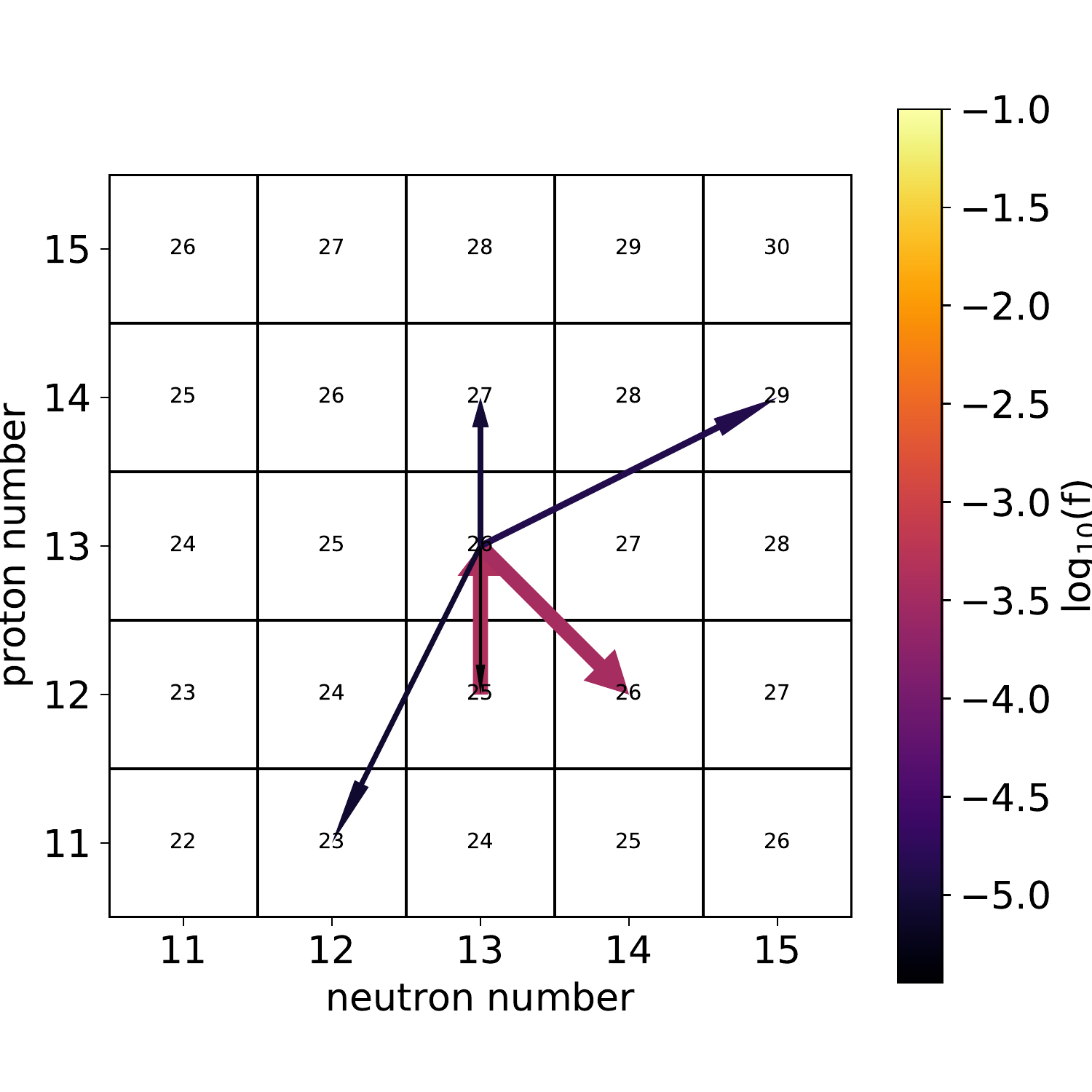}
  \caption{The distribution of fluxes ($\delta$X/$\delta$t, where X is mass fraction) for each reaction during explosive carbon burning at mass coordinate 2.16\stmass in the 15\msun star integrated over the shock time for~\Alslr.
% Reactions in equilibrium are represented by flux lines with the same thickness and colour in both directions. 
 Stellar conditions are extracted at mass coordinate 2.16\stmass in Fig.~\ref{fig:long_abu_plot_15}. The $^{26}$Al(n,p)$^{26}$Mg and $^{26}$Al($\beta^{-}$)$^{26}$Mg fluxes are both present as two arrows, the smaller arrow denotes the $\beta^{-}$ decay.
 % \textcolor{purple}{Re-position for Al, MN, Mn and Fe? need to change the connection in text if so?}
 }
 \label{fig:Flux_matrix_1_al26}
\end{figure}

The reaction fluxes for these conditions are shown in %the bottom left of 
Fig.~\ref{fig:Flux_matrix_1_al26}.
%[WHY THE BOTTOM LEFT, AND NOT THE TOP? THIS IS THE FIRST FLUXES DESCRIBED..]
As expected, the strongest fluxes are those related to production via \MgpgAl and destruction by $^{26}$Al(n,p)$^{26}$Mg driven by the high neutron densities. The relevance of the other nuclear reaction channels are at least an order of magnitude smaller. As no radioactive isotopes decaying into \Alslr are produced to a significant level, there is no radiogenic contribution 
% to take into account 
for \Alslr.

%The \Alslr production described for the 15M$_{\odot}$ star is [LATER] 
%is destroyed in the layers located below explosive C-burning, due to photodisintegration. 
%In the explosive C-burning there is a high proton density ($\approx$10$^{21}$ cm$^{-3}$) and therefore \Alslr is produced. 
%The reaction fluxes are shown in the bottom left of Fig.~\ref{fig:Flux_matrix_1}. The strongest fluxes are those related to production via \MgpgAl and destruction by $^{26}$Al(n,p)$^{26}$Mg driven by the high neutron densities. This competing fluxes cause the overall abundance of \Alslr to remain below 10$^{-3}$ in mass fraction. During explosive He-burning there is also \Alslr some production due to high proton density. However, the proton density here is lower compared to explosive C-burning ($\approx$10$^{15}$ cm$^{-3}$) and the \Alslr abundance is reduced as its production cannot compete with destruction via the $^{26}$Al(n,p)$^{26}$Mg reaction during the neutron burst. 

Comparing the 20\stmass (Fig.~\ref{fig:long_abu_plot_20}) and 25\stmass (Fig.~\ref{fig:long_abu_plot_25}) models with the 15\stmass model discussed above, the most remarkable difference is the smaller abundance of pre-CCSN \Alslr in the C-burning ashes (ejecta between 2.2 and 4\stmass, and between 2.9 and 5.9\stmass, respectively for the 20 and 25\stmass models). 
This discrepancy is explained by a less efficient or inactive C-burning in the last days of stellar evolution: the \Alslr previously made decays to $^{26}$Mg without being produced. 
Interestingly, in Fig.~\ref{fig:long_abu_plot_25} we can still find \Alslr abundance of the order of a few 10$^{-7}$ in the upper part of the O-Ne ashes, where final pre-explosive O-burning activation is too weak to destroy the isotope. 
%However, the abundance is lower in carbon burning for these models compared to the because of a less efficient $^{12}$C($^{12}$C,p)$^{23}$Na reaction, which results in a higher \CC abundance in the C-ashes, and less \Alslr.
The 25\stmass star shows a much stronger C-burning explosive nucleosynthesis compared to the 20\stmass star, leading to a higher production of \Alslr. 

In the 20\stmass model shown in Fig.~\ref{fig:long_abu_plot_20}, the peak temperature for explosive C-burning (about 2 GK) is reached only at the interface between the O-Ne and C ashes. 
%As the mass-fraction of $^{25}$Mg in the Ne-burning ashes is reduced from 9$\times10^{-3}$ to 4$\times10^{-3}$ as there are less seeds for proton captures, resulting in reduced overall \Alslr production.
The same conditions are achieved in the C-ashes for the 
%due to the temperature gradient of the 
25\stmass model, showing a broad 
%the structure of the 
production peak of \Alslr.
%within the C-burning ashes is broader than in the 15\stmass model, while the same peak value is reached. 

In summary most of \Alslr in the 20\stmass model (Fig.~\ref{fig:long_abu_plot_20}) is made by the pre-CCSN stage, in the 25\stmass model (Fig.~\ref{fig:long_abu_plot_25}) the yields are instead mostly due to explosive nucleosynthesis. In Fig.~\ref{fig:long_abu_plot_15} \Alslr is produced by both. From this we may already argue that the relative contribution to the total ejected \Alslr from different parts of the ejecta and the fraction of \Alslr made by the explosion do not depend solely on the initial mass of the progenitor or on the explosion energy \citep[e.g.,][]{Limongi2006}. 
%  Details of the late shell structure evolution of the stellar progenitor may indeed generate significant star-to-star variations.

%occurs in the same explosive burning conditions for all the stellar masses, i.e., during explosive C-burning and explosive He-burning. 
% In the 20\stmass model the peak shock temperature reaches 2~GK at the threshold between the C- and Ne-burning ashes.
% Explosive C-burning takes place with a set of different initial abundances than in the 15\stmass model where the peak shock temperature reaches 2~GK at the

\begin{figure}
 \centering
 \includegraphics[width=\columnwidth]{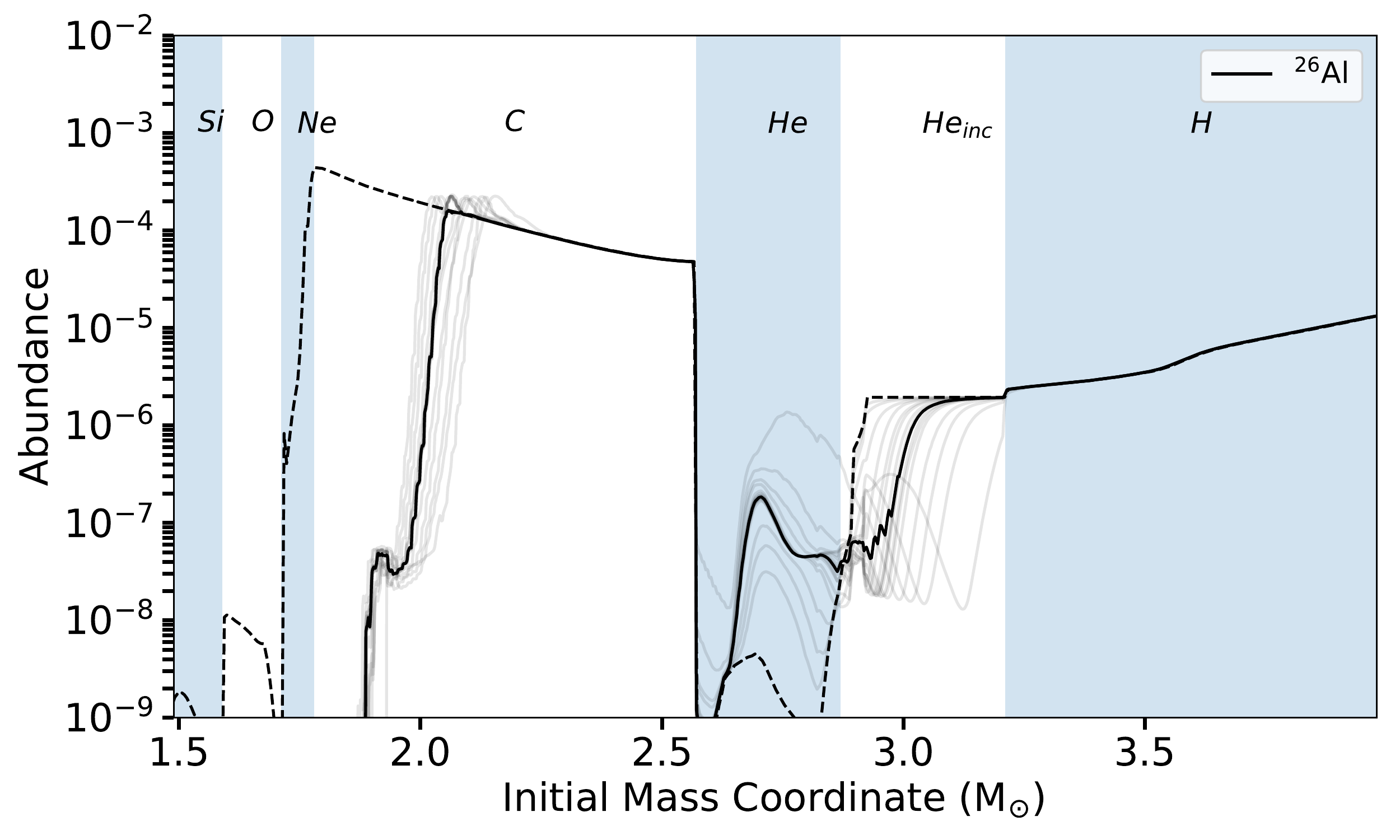}
 \includegraphics[width=\columnwidth]{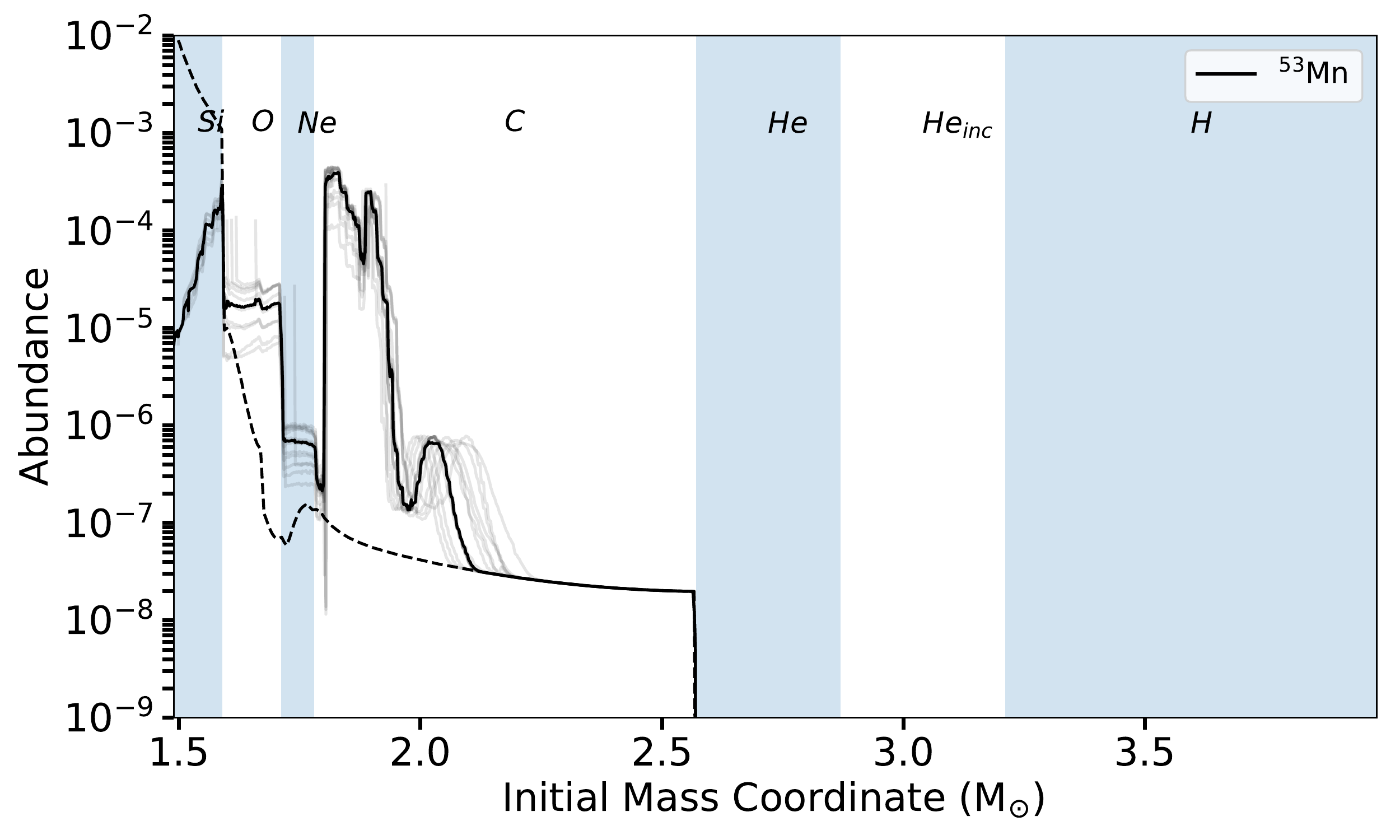}
 \includegraphics[width=\columnwidth]{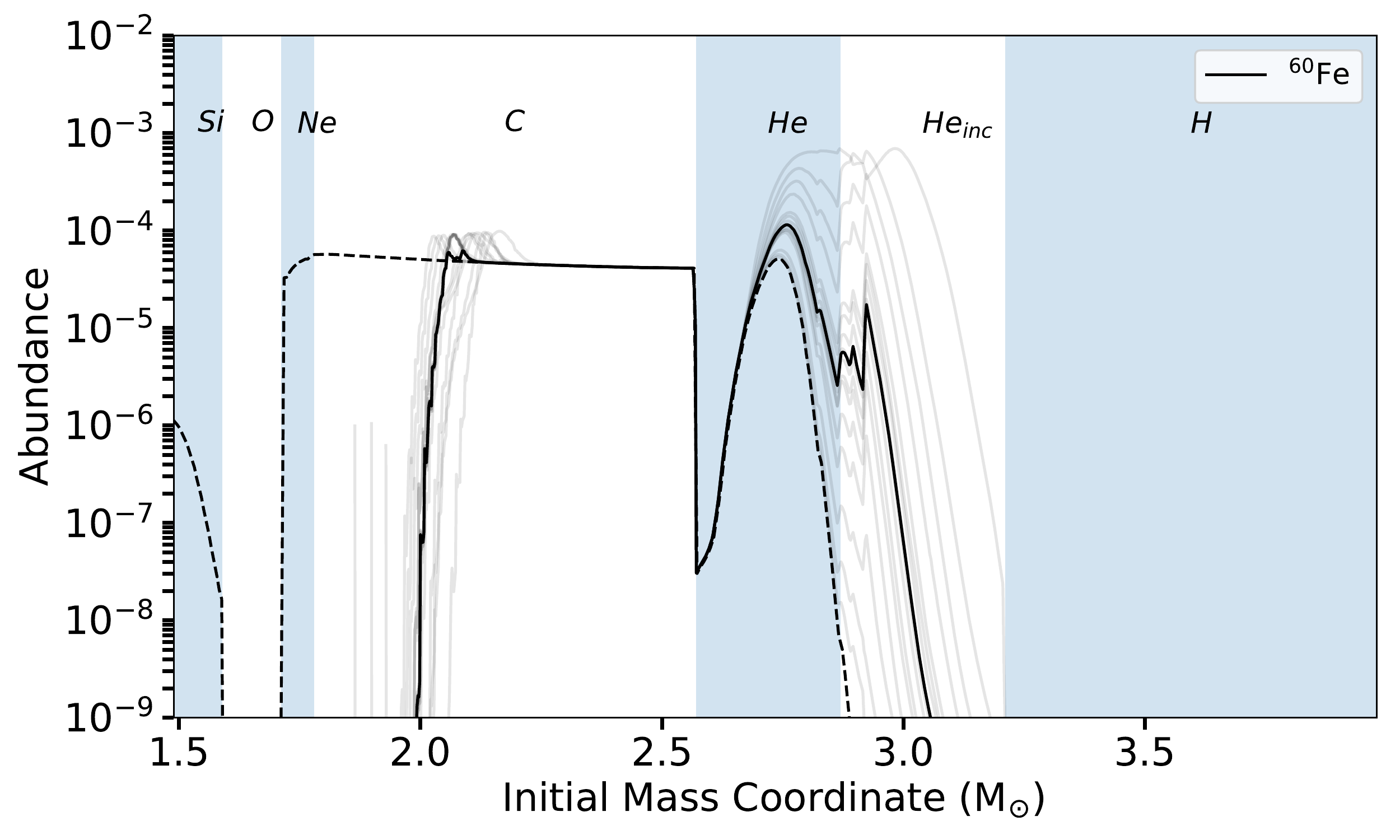}
 \caption{The isotopic abundances of \Alslr, \Mnslr and \Feslr (upper panel, central panel and lower panel, respectively) are shown with respect to mass coordinate of all 15\stmass models: pre-CCSN abundances are shown by the dashed line, the median of all models is shown by a black solid line, and thin lines show the abundance profiles for single models. %and maximum and minimum vales are shown by a red solid line. 
 % {\textbf Add mention to the mass cut within the plots}
 Alternating shaded and non-shaded areas highlight pre-CCSN burning ashes, from left to right: Si ashes, O ashes, Ne ashes, C ashes, He ashes and He$_{\rm inc}$ ashes.
 % \marco{[In the reference 15Msun model there is no double peak for Mn53 in the O-Si region, seen in this figure. Why? Can you check about the source of the difference? I am wondering if one is decayed profile, and the other is just non-decayed? ]}
 %Isotopes shown are \Alslr in the upper plot, \Mnslr in the centre plot and \Feslr in the lower plot. 
 % \textcolor{purple}{M+M: I can add each models mass cut, but it causes the inner most portion to be covered up, is this a worthwhile addition?}
 }
 \label{fig:agg_plot_al26_fe60_mn53_15}
\end{figure}

We examine the abundance profiles of all 15\stmass models in Fig.~\ref{fig:agg_plot_al26_fe60_mn53_15}.
The top panel of Fig.~\ref{fig:agg_plot_al26_fe60_mn53_15} shows the abundance profiles of \Alslr, and shows several similarities, with the largest variations obtained in the explosive He-burning.
%it is clear that, \Alslr is always affected in similar ways as described above: destruction in nuclear statistical equilibrium (NSE), a peak of production in explosive C-burning, with somewhat less significant production during explosive He-burning. These production and destruction regimes depend on the peak shock temperature 
% and their location in mass are shifted in the different models in accordance with the temperature profiles (see Fig.~\ref{fig:t_MASS})
%Overall, t
The yields of \Alslr for the 15\stmass and 20\stmass models show variations by a factor of about 1.5. However, due to broad production peaks %and large differences in the production
during explosive He-burning the 25\stmass models show a significantly larger variation in yield, of about an order of magnitude
%differ only slightly (by a mass fraction of $\times10^{-5}$ at its most 1.5) from each other 
(see Table \ref{yield_table:1}). 
% T
% he same applies to the 20\stmass models, while there is a larger variation within the 25\stmass models \marco{[MP: what variation? by a factor of?]}, due to the broad production peaks %and large differences in the production
% during explosive He-burning. 
% % (see Fig.~\ref{fig:agg_appendix_1})
% %Broadly speaking 
% In general, the highest \Alslr~yields are made in the 25\stmass models, followed by the 15\stmass and the 20\stmass model yields, respectively.

%15\stmass model yields are much higher than the 20\stmass model yields and the 25\stmass model are higher than both the 15\stmass and 20\stmass model yields.

\subsection{\Clslr and \Caslr}

\begin{figure}
 \centering
 \includegraphics[width=\columnwidth]{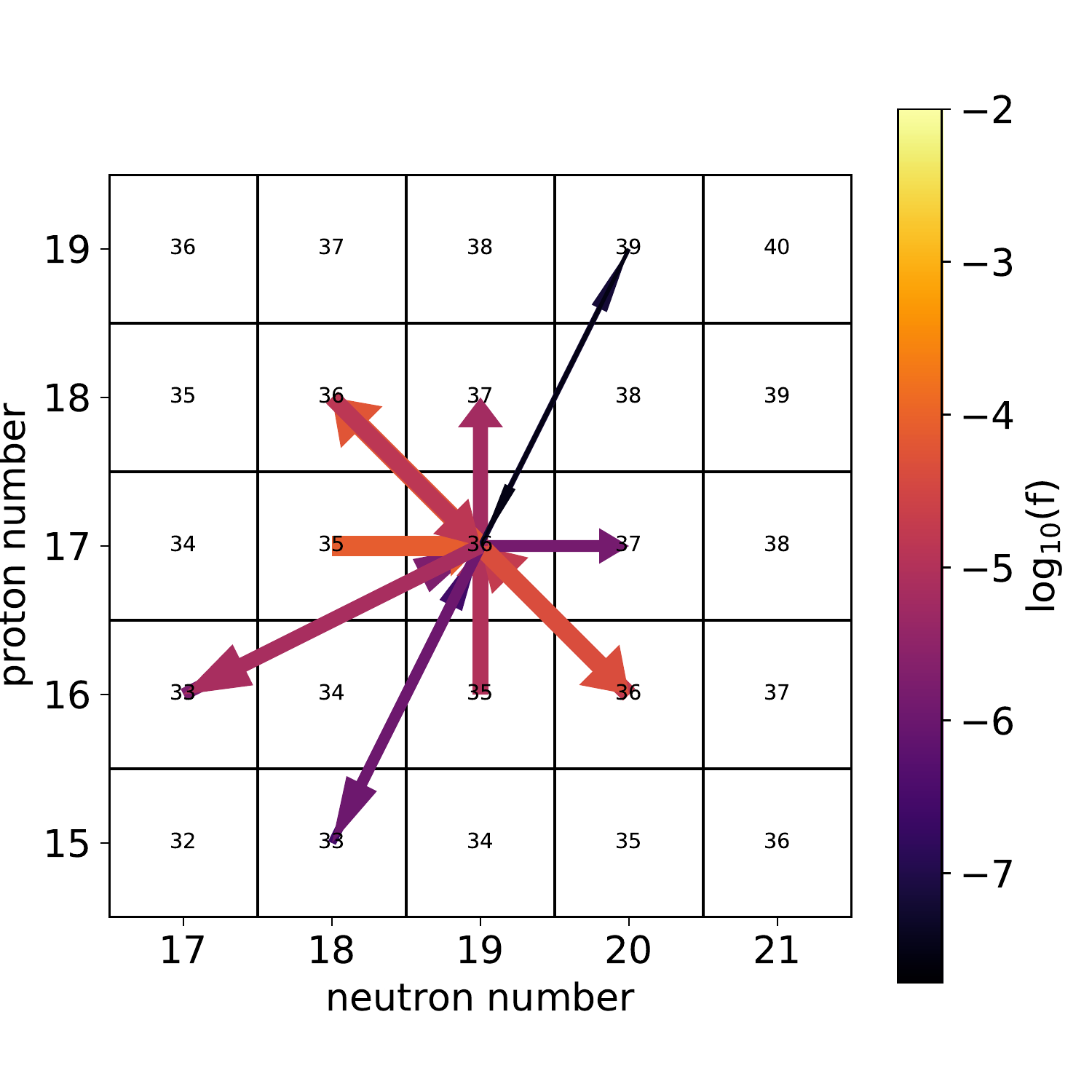}
 \caption{Same as in Fig.~\ref{fig:Flux_matrix_1_al26}, but for \Clslr in explosive carbon burning (at mass coordinate 2.1\stmass for the 15\msun model in Fig.~\ref{fig:long_abu_plot_15}).}
 \label{fig:Flux_matrix_2_cl36}
\end{figure}

\begin{figure}
 \centering
 \includegraphics[width=\columnwidth]{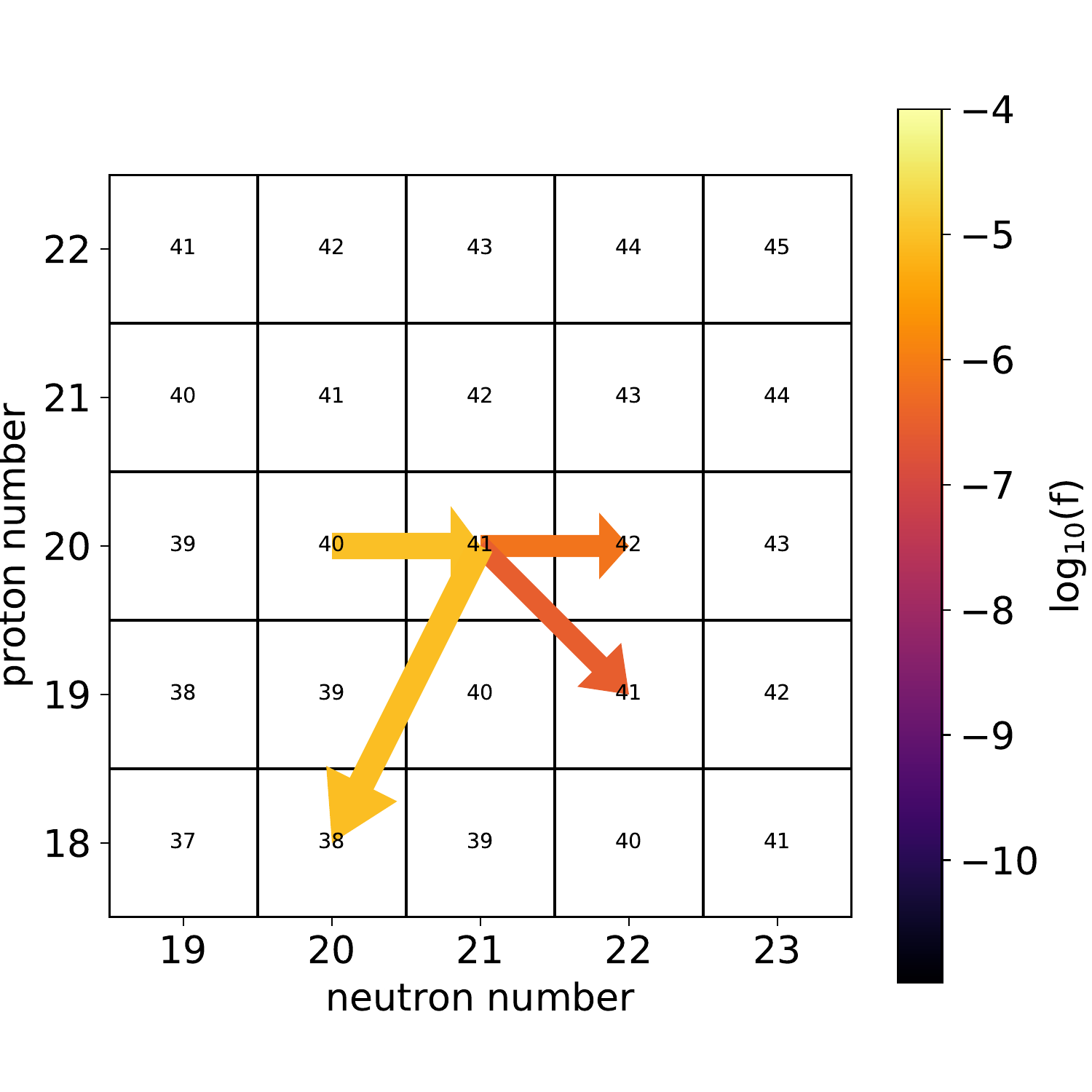}
 \caption{Same as in Fig.~\ref{fig:Flux_matrix_1_al26}, but for \Caslr in explosive helium burning (at mass coordinate 2.9\stmass for the 15\msun model in Fig.~\ref{fig:long_abu_plot_15}).
}
 \label{fig:Flux_matrix_2_ca41}
\end{figure}

% \marco{[MP: Their nucleosynthesis is not similar, excepting from the fact that their production and destruction up to C burning is dominated by neutron captures. ]}
The radioactive isotopes \Clslr and \Caslr are both located next to the neutron shell closure at N=20. In the 15\stmass model shown in Fig.~\ref{fig:long_abu_plot_15}, the two isotopes are already efficiently produced in the pre-CCSN He-burning and C-burning regions by a neutron capture on $^{35}$Cl and $^{40}$Ca, respectively. Significant variations due to the CCSN explosion are then obtained in the calculations.
%We discuss \Clslr and \Caslr together as they are have very similar nucleosynthesis.
%Both can be formed by one neutron capture on $^{35}$Cl and $^{40}$Ca, respectively. Pre-CCSN they are produced via neutron capture in He-burning and O-burning. 
In explosive nucleosynthesis \Clslr is efficiently made by explosive C-burning, with a peak of production up to one order of magnitude larger than in the pre-CCSN stages. \Clslr is largely destroyed from explosive O-burning and in general for all mass coordinates below 1.9\msun. %at temperatures above 4~GK. 
This is not the case for \Caslr, which can be made even in \alp-rich freeze-out conditions in the deepest ejecta. The highest production peak is obtained in explosive O-burning conditions (at about 1.9\msun), with abundances at least an order of magnitude higher than in the other ejected stellar layers. The dominant production channel of \Clslr is $^{36}$Ar(n,p)$^{36}$Cl, while it is mostly depleted by the reverse of this reaction, $^{36}$Cl(p,n)$^{36}$Ar.
% \marco{[MP: I cannot find in the literature a detailed description. You need to look in a trajectory. You do not have the add the flux plot to the paper, that will be crowded. There will be many reactions in balance. But you can see the sharp peak. This looks to me like a reaction making Ca41 until it goes to balance. Maybe take 1.91\stmass or 1.92\stmass to not miss it... careful that if you take it even only a little too deep it will be gone.]}
%\Caslr is instead produced, as $^{40}$Ca is made by \alp-chain reactions in the alpha freeze-out, requiring a neutron capture to form \Caslr. 
During explosive C-burning, \Clslr is produced by neutron capture on $^{35}$Cl and is mostly destroyed by $^{36}$Cl(n,p)$^{36}$S and $^{36}$Cl(p,n)$^{36}$Ar (see Fig.~\ref{fig:Flux_matrix_2_cl36}). \Caslr is largely unaffected by explosive C-burning as the neutron capture reaction that forms \Caslr is balanced by $^{41}$Ca(n,$\alpha$)$^{38}$Ar.
In explosive He-burning both \Clslr and \Caslr are typically destroyed by n-process neutron captures (see Fig.~\ref{fig:Flux_matrix_2_ca41} for \Caslr). %, and the lower proton density reduces production via proton capture reactions. 
%There is no major contribution to the final yields of \Clslr and \Caslr from radioactive decay.

\begin{figure}
 \centering
 \includegraphics[width=\columnwidth]{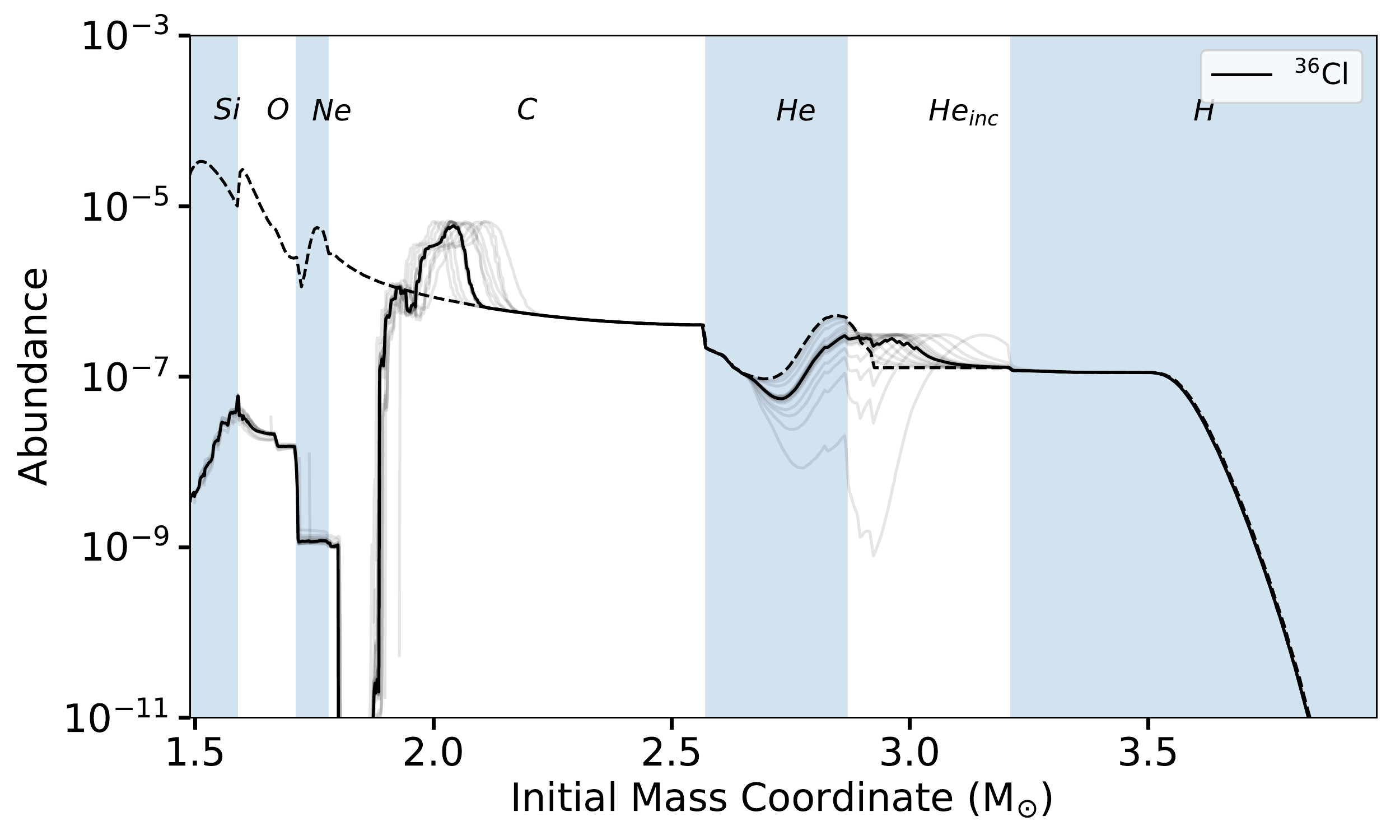}
 \includegraphics[width=\columnwidth]{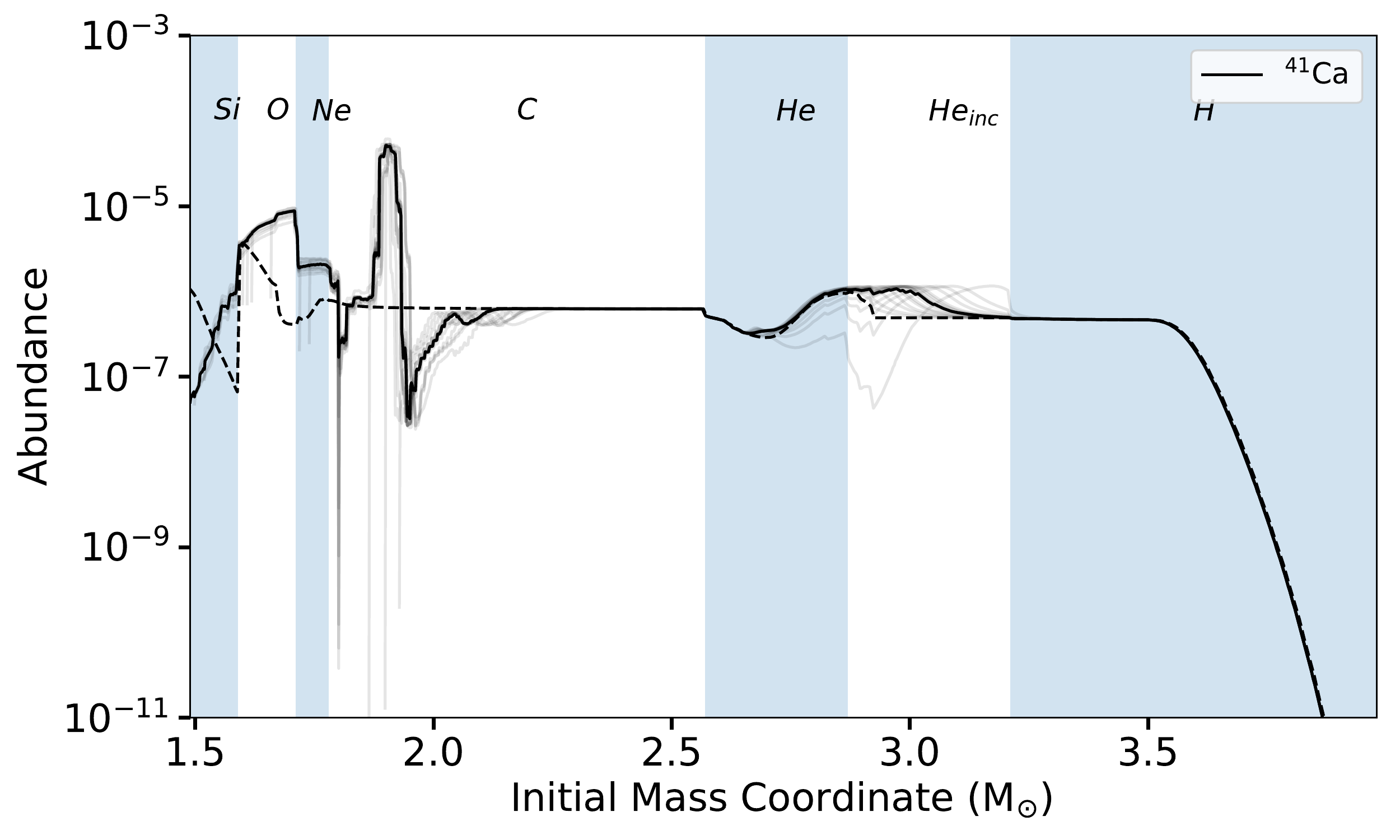}
 \caption{Same as Fig.~\ref{fig:agg_plot_al26_fe60_mn53_15}. Isotopes shown are \Clslr and \Caslr.}
 \label{fig:agg_plot_cl36_ca41_15}
\end{figure}

Comparing all the 15\stmass models (see Fig.~\ref{fig:agg_plot_cl36_ca41_15}) we see that they are all qualitatively consistent for both \Clslr and \Caslr. %, with the impact of explosive He-burning contributing to the largest differences in the yields between models. 
The largest differences in the yields are due to the amount of depletion in explosive He-burning.
The location of the mass cut %for each model 
is another key factor in the final yields of %both \Clslr and 
\Caslr, as a high mass cut 
prevents the innermost portions of the star from being ejected.
% causes all production in the innermost portions of the star to not being ejected. %fall back.
Examining the ejected yields from each 15\stmass model (Table \ref{yield_table:1}), %the largest impact on the final yields of \Clslr and \Caslr is the location of the mass cut, as there is a considerable component of production in the innermost portion of the star (see bottom panel of Fig.~\ref{fig:agg_plot_cl36_ca41_15}, for example). As \Clslr is largely produced also in explosive C-burning while \Caslr is produced in explosive O-burning,a high mass cut impacts the yield of \Caslr more than that of \Clslr.
\Caslr yield varies by a factor of four (9.1$\times$10$^{-7}$ -- 4.2$\times$10$^{-6}$\stmass), while \Clslr which is not made in the deepest ejecta, varies only by 20\% (8.3$\times$10$^{-7}$ -- 1.1$\times$10$^{-6}$\stmass) 
% [MP: here you said the same things twice, but without telling the numbers. Can you see if Ca41 is changing more? What are the variations factors in place? Put the numbers here. ]}

The pre-CCSN production of \Caslr is similar between the 15, 20 and 25\stmass models. However, the production of \Clslr in pre-CCSN carbon burning is higher in the 15\stmass model (Fig.\ref{fig:long_abu_plot_15}), than in the 20 and 25\stmass models (Figs. \ref{fig:long_abu_plot_20} and \ref{fig:long_abu_plot_25}, respectively). 
Post-CCSN production of \Clslr in the 20\stmass and 25\stmass models is close to the position on the mass cut, which causes variance in the ejected yield.
For \Clslr the ejected yields vary by two and one order of magnitude, for 20\stmass and 25\stmass models respectively.
For \Caslr the ejected yields vary by a factor of two and by one order of magnitude, for 20\stmass and 25\stmass models respectively (see Table \ref{yield_table:1}).

\subsection{\Mnslr} \label{ss:mn}

% \textcolor{red}{MOVE TO 4.2: During pre-CCSN nucleosynthesis \Mnslr is produced primarily in Si-burning, with a smaller contribution from O-burning. The peak in \Mnslr in the core is due to QSE during Si-burning (see innermost region of second from top panel in Figs. \ref{fig:long_abu_plot_15}, \ref{fig:long_abu_plot_20} and \ref{fig:long_abu_plot_25}), in agreement with \cite{Chieffi1998}.}

\Mnslr is mostly made by explosive nucleosynthesis. 
% For instance, let us consider as an illustrative case %We start to describe the production of \Mnslr using 
% the 15\stmass model in Fig.~\ref{fig:long_abu_plot_15}. 
In our fiducial 15\stmass model (presented in Fig.~\ref{fig:long_abu_plot_15}),
% In the CCSN ejecta, 
a small pre-CCSN component is present from the C-shell ashes, between about 1.78\stmass and 2.54\stmass.
During the CCSN explosion, additional nucleosynthesis channels contribute to most of the ejected \Mnslr abundance. At about 2.1\stmass, a production peak is obtained due to explosive C-burning. The largest contribution is obtained between 1.8\stmass and 1.9\stmass, triggered by explosive O-burning and partial Si-burning. Below this point the \Mnslr abundance drops by three orders of magnitude in the products of complete Si-burning (between the mass cut and 1.6\stmass). 
In the deepest CCSN layers ($\lesssim$1.7 \stmass) the abundance rises again due to production in \alp-rich freeze-out conditions \citep[e.g.,][]{Hoffman1996}.
% However, while this outer region is comparable in mass to the inner region, the abundacne is two orders of magnitude lower than that found in explosive 
% the relative contribution will be marginal compared to the main peak outwards. %explosive nucleosynthesis 
%\Mnslr is produced during explosive C, O and Si-burning as well as during the \alp-rich freeze-out. Within the Si-burning ashes (below 1.6\stmass) the shock-wave destroys the majority of the \Mnslr via photodisintegration, reducing its abundance by three orders of magnitude. The explosive production seen in the O and Ne-burning ashes is due to \alp-rich freeze-out during which isotopes up the \alp-chain are produced, as well as Fe-group isotopes.
The production peak at M$\approx$1.8\stmass is due to a near equilibrium state between the many forward and reverse reactions that are reached in explosive partial Si-burning. This produces isotopes up to the Fe-group with the largest contribution to \Mnslr from the decay of $^{53}$Fe, and its main destruction reaction being $^{53}$Mn(p,$\gamma$)$^{54}$Fe. During explosive O-burning a near equilibrium state is again reached at M$\approx$1.9\stmass. The main destruction reactions are $^{53}$Mn(p,$\gamma$)$^{54}$Fe and $^{53}$Mn($\gamma$,p)$^{52}$Cr, and the main production reactions are the reverse of these reactions. 
% \marco{[MP: the diagram is quite busy. How did you check that the reactions indicated are the most relevant? Did you check numbers?]}\TL{[TL: I mainly use this plot to show how messy explosive O can be, realistically this can be removed as it shows little required data]}

\begin{figure}
 \centering
 \includegraphics[width=\columnwidth]{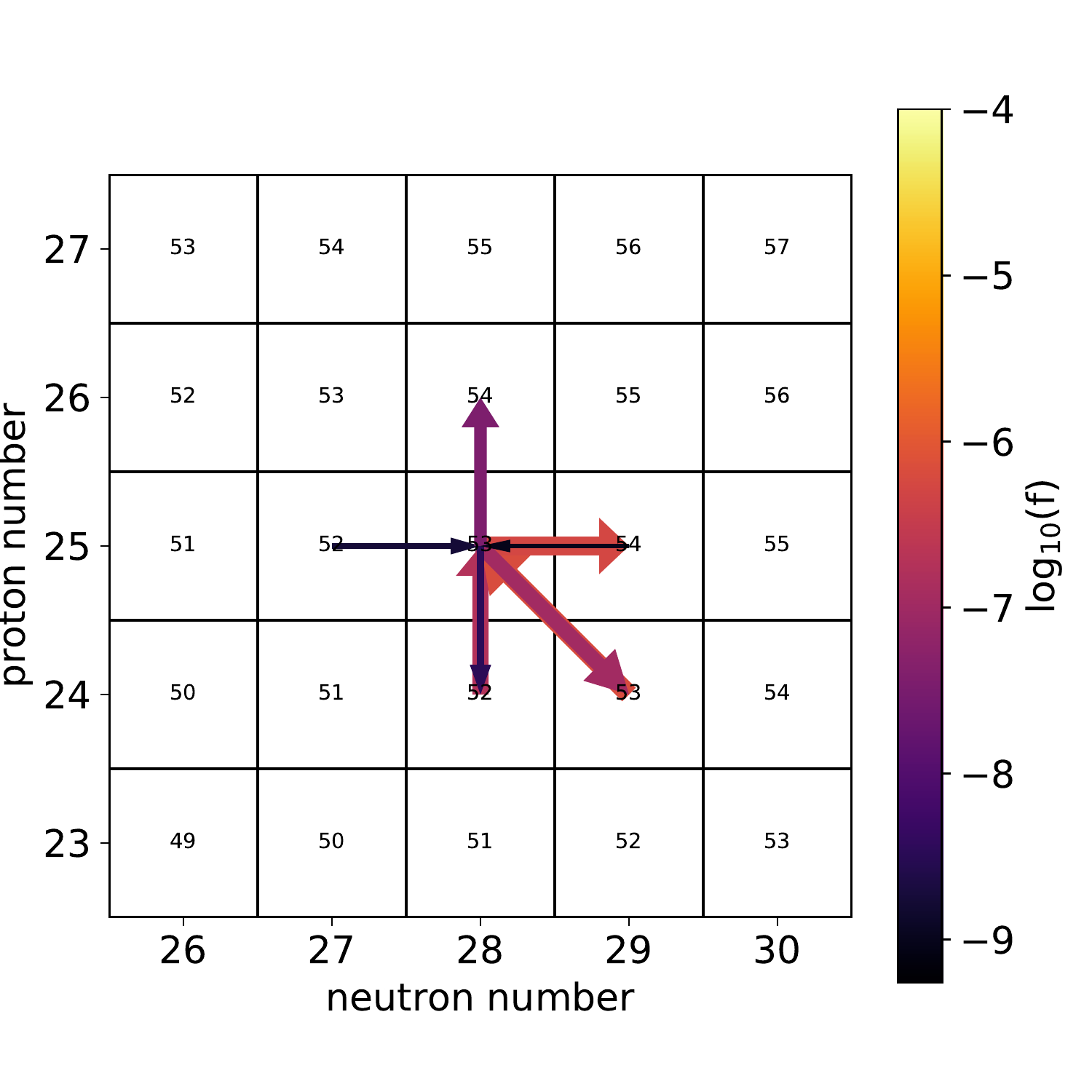}
 \caption{The same as in Fig.~\ref{fig:Flux_matrix_1_al26}, but for \Mnslr from mass coordinate 2.1\stmass in Fig.~\ref{fig:long_abu_plot_15}. 
 }
 \label{fig:Flux_matrix_1_mn53}
\end{figure}

The other smaller peak of production is due to explosive C-burning, taking place at M$\approx$2.1\stmass (Fig.~\ref{fig:long_abu_plot_15}). Here the \alp-capture reaction $^{24}$Mg($\alpha$,p)$^{27}$Al produces a large amount of protons,
% \marco{[MP: from where did you take this information about the mg24ap? During C-burning, the main source of protons I would expect is directly via c12c12p, no? Did you check integrated fluxes or just the fluxes for 1 dt?]}\TL{[TL: TL: During my flux plot script I print out the top 10 reactions for each trajectory, of which C12C12p is not even in the top 10 reactions (2 of mag orders less than the highest), the $^{24}$Mg($\alpha$,p)$^{27}$Al reaction is the highest, however, its reverse is the second highest flux. Also all fluxes checked were integrated - Explosive Ne Burning]}
leading to %(upper right panel in Fig.~\ref{fig:Flux_matrix_1}) 
the major production channels of \Mnslr: the proton-capture reactions $^{53}$Cr(p,n)$^{53}$Mn and $^{52}$Cr(p,$\gamma$)$^{53}$Mn. The major destruction channels are the neutron-capture reactions $^{53}$Mn(n,p)$^{53}$Cr and $^{53}$Mn(n,$\gamma$)$^{54}$Mn (see Fig. \ref{fig:Flux_matrix_1_mn53}).

The middle panel in Fig.~\ref{fig:agg_plot_al26_fe60_mn53_15} shows the \Mnslr abundance profiles of all the 15\stmass models, and demonstrate that the nucleosynthesis behaves consistently across them. The yields of \Mnslr for the model with the highest mass cut (1.9\stmass) is 5.26$\times$10$^{-7}$\stmass while it is 5.89$\times$10$^{-5}$\stmass for the model with the lowest mass cut (1.5\stmass), a variation of the order of two orders of magnitude (see Table \ref{yield_table:2}).
This makes \Mnslr a useful diagnostic of the mass cut, as we discuss further in Section \ref{ss:diag}.

Figs.~\ref{fig:long_abu_plot_20} and \ref{fig:long_abu_plot_25} for the 20 and 25 \stmass models show a similar %the same %mechanisms of 
production pattern as described above for the 15\stmass models,
% , due to explosive %Si, O and C-burning 
% nucleosynthesis is obtained %are also present 
% in the 20 and 25\stmass models. %However %, due to the different evolution of the peak temperature, 
although production peaks are closer to the mass cut. This is due to production occurring in the Ne-burning ashes, with no significant production in the C-burning ashes. 
Examining the yields of \Mnslr for the 20 and 25\stmass models, the abundance of \Mnslr varies by four orders of magnitude, from 6.4$\times$10$^{-9}$\stmass for the model with the highest mass cut (4.89\stmass) to 5.0$\times$10$^{-5}$\stmass with the lowest mass cut (1.7\stmass) (Table \ref{yield_table:2}).
The mass cut impact is even more significant here than for the 15\stmass model, because the %high mass cut 
models with the highest mass cut do not even %do not 
eject any explosive \Mnslr contribution. Therefore, within the possible range of progenitor mass and explosion energy, it is indeed possible to have CCNSNe with no substantial \Mnslr ejected. 
% \marco{[MP: Is this statement correct?] }

\subsection{\Feslr} \label{ss:fe60}

\begin{figure}
 \centering
 \includegraphics[width=\columnwidth]{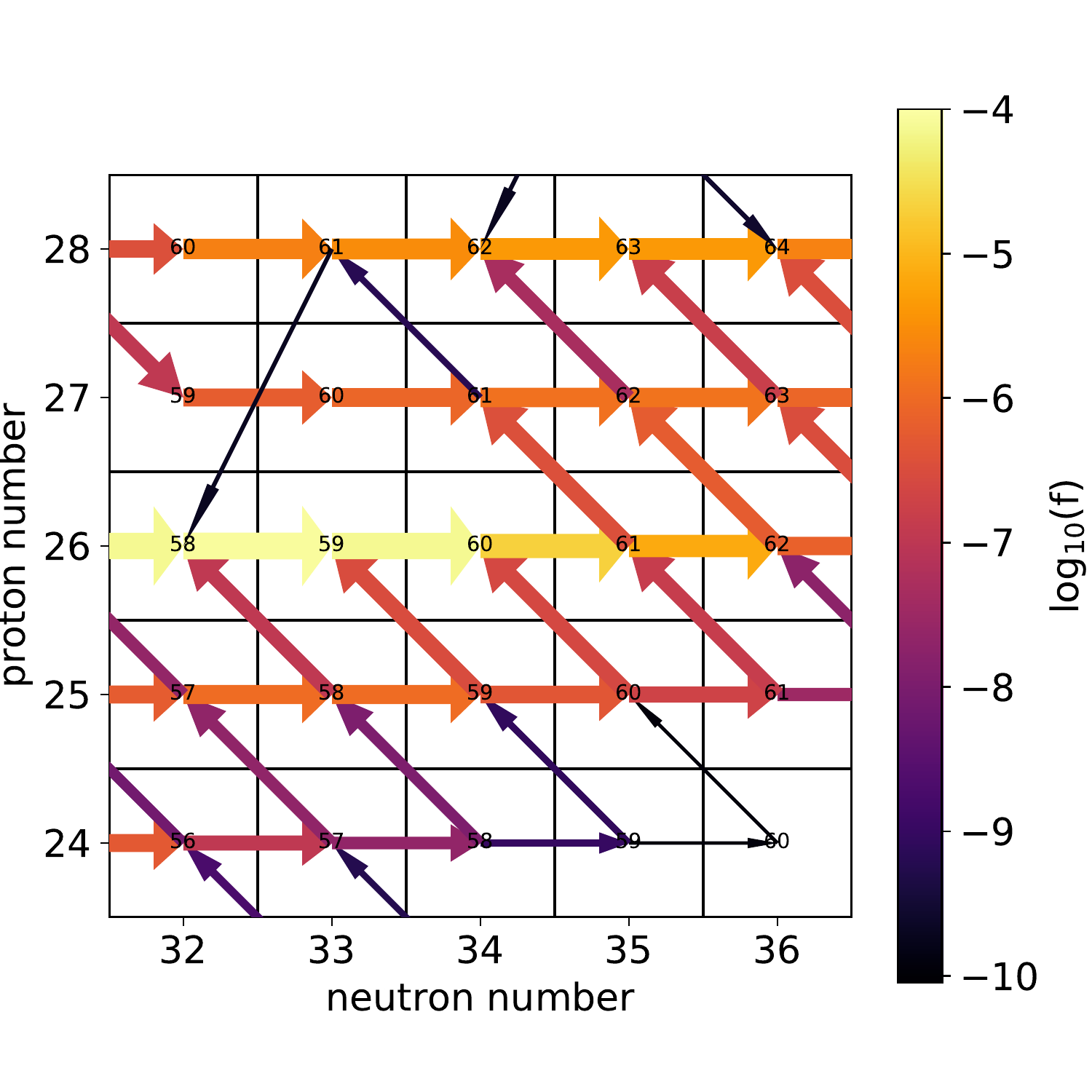}
 \caption{The same as in Fig.~\ref{fig:Flux_matrix_1_al26}, but for \Feslr from mass coordinate 3.0\stmass in Fig.~\ref{fig:long_abu_plot_15}, respectively. 
 }
 \label{fig:Flux_matrix_1_fe60}
\end{figure}

%Again we consider first the 15\stmass model to illustrate both the pre-CCSN and CCSN nucleosynthesis of \Feslr. 
Prior to core-collapse, \Feslr is produced in C-burning and also marginally in He-burning (see the second panel from top in Fig.~\ref{fig:long_abu_plot_15}, for the reference 15\stmass model). \Feslr is destroyed in O-burning when the temperature rises above about $\sim$2GK. During explosive nucleosynthesis all the pre-CCSN \Feslr exposed to conditions more extreme than %below 
explosive carbon burning is depleted. %by photodisintegration. 
Instead, \Feslr is significantly produced during explosive He-burning, as neutron densities reach $\geq$10$^{18}$ cm$^{-3}$ and neutron captures are very efficient during the n-process activation \citep[e.g.,][]{pignatari:18}. Here, \Feslr is mostly created via the double neutron-capture chain \Fedoublen, and marginally
% (one to two orders of magnitude less) 
via the $\beta^{-}$ decay of $^{60}$Mn (see Fig.~\ref{fig:Flux_matrix_1_fe60}), made by a sequence of neutron captures starting from stable $^{55}$Mn. The main destruction channel of \Feslr is the (n,$\gamma$) neutron capture forming $^{61}$Fe. %, $^{62}$Fe etc.
% While neutron densities are similar in explosive C-burning and explosive He-burning there is less 
%Note that during explosive C-burning competition from proton captures, and therefore neutron captures are less efficient than proton captures (specifically \Feslr(p,n)).
%This reduces the overall production of \Feslr in C-burning.
The \Feslr production peak due to C-burning at about 2.15\msun in Fig.~\ref{fig:long_abu_plot_15} is around one order of magnitude lower than the peak in explosive He-burning.
% /is less relevant for this specific model compared to explosive He-burning. 
This result may potentially be affected by parameters like the progenitor mass and the explosion energy, which we will discuss below. However, another important parameter to consider is the amount of $^{22}$Ne left to produce neutrons in the C shell during the CCSN shock. Indeed, a weaker s-process activation in the progenitor would cause a larger remaining amount of $^{22}$Ne to produce neutrons during the explosion, a lower pre-CCSN abundance of \Feslr, but also a higher amount of iron seeds to feed the explosive \Feslr production. 
Therefore, the relative relevance of explosive He burning and explosive C burning in producing \Feslr could potentially vary between different sets of stellar models.
% , and this cannot be directly linked to main physical parameters of the progenitor like the initial mass of the star.

Comparing the \Feslr abundance profile of the 15\stmass\ model (Fig.~\ref{fig:long_abu_plot_15}) to that of the 20\stmass\ (Fig.~\ref{fig:long_abu_plot_20}) and 25\stmass\ (Fig.~\ref{fig:long_abu_plot_25}) models,  the production sites for the radioactive isotope do not change. %the abundance behaviour is similar. % in relation to the destruction of \Feslr that takes place below the region of explosive carbon burning. 
However, in the 25\stmass\ model the %production in
relevance of explosive carbon burning is higher than for the other two masses (see peak at 3.8\stmass in Fig.~\ref{fig:long_abu_plot_25}). %, due to the lower proton densities resulting in lower destruction of \Feslr via (p,n) captures.
% \marco{[MP: I removed this part of the (p,n). I really doubt that the Fe60(p,n) will have any relevance in destroying Fe60... did you check the flux plots? Did you read it somewhere?]}\TL{[TL: I believe I read it in Sam's paper, but ill double check it]}
In the same model, a significant pre-CCSN \Feslr production is obtained in the most internal convective He shell, between about 6.4\msun and 7.1\msun. The explosion will boost further the isotope at the bottom of the shell.
In the 20\stmass model, there is no significant explosive production peak in the He-burning ashes compared to the other stellar masses considered. 
The reason is that in this model the temperature peak associated the CCSN shock is too low at the bottom of the former He shell to trigger the n-process within the short explosive timescales. %due to the steeper temperature gradient 
% (Fig.~\ref{fig:t_MASS}). %This steep temperature gradient causes the 1~GK threshold for explosive He-burning to be located below the location where He is available to burn (the upper He-burning ashes). 
%This results in no contribution from explosive He-burning.
%In the 25\stmass model there is a small contribution during explosive He-burning, where He can be found in the He-burning ashes.
% in the 20\stmass\ model and in the 20\stmass\ model is is in the middle of the He-burning ashes. 

The bottom panel of Fig.~\ref{fig:agg_plot_al26_fe60_mn53_15} shows the abundance profile of \Feslr in all the 15\stmass models. The qualitative behaviour is the same for all the models, with almost complete depletion %destruction in NSE 
for mass coordinates below explosive carbon burning and a large production in explosive He-burning. However, the amount of \Feslr made by explosive He-burning changes greatly by varying the CCSN explosion parameters used in this work, with the final yield varying by two orders of magnitude for the 15\stmass models (Table \ref{yield_table:2}). %large differences are present in the amount and extent of production in explosive He-burning. 
This is due to the temperature variation in the He-shell material during the CCSN shock, which falls below 10$^9$ K in the %lower energy 
models with lower explosion energy. %falling below the critical temperature of 1~GK in regions with He to burn. 
%This causes the final yield of \Feslr to vary by two orders of magnitude in the 15\stmass models ( Table \ref{yield_table:2}). 
Due to the lack of contribution from explosive He-burning in the 20\stmass models, the \Feslr yield is lower by up to one order of magnitude relative to the 15\stmass model. The 25\stmass models show an increase in yield of \Feslr when compared to the 15\stmass model, mainly due to higher %initial abundances of
contribution from the pre-CCSN stage \citep[see also, e.g.,][]{Timmes1995, Limongi2006}. % \Feslr pre-CCSN.

We note that there is little contribution from radiogenic decay in the final yields of \Feslr. For a more extended description of the production of \Feslr in massive stars, we refer to \citet{Jones2019} for the same models that we use here.

\subsection{\Pdslr and \Pbslr }

Beyond iron, \Pdslr and \Pbslr are SLRs located in-between stable isotopes ($^{106}$Pd and $^{108}$Pd, and $^{204}$Pb and $^{206}$Pb, respectively).
%Due to their similar nucleosynthesis we discuss \Pdslr and \Pbslr together.
Both SLRs are§ produced by a neutron capture on $^{106}$Pd and $^{204}$Pb, respectively, and they are also destroyed by neutron captures. Charged particle reactions are not relevant for the nucleosynthesis of such a heavy nuclei. the \gammaproc does not contribute to the production of these isotopes either: there are no seeds for making \Pbslr via photodisintegration, and \Pdslr is shielded from any $\gamma$-process contribution by its stable isobar $^{107}$Ag.

In Fig.~\ref{fig:long_abu_plot_15} (third panel from top),
%We begin by considering their profiles in the 15\stmass model shown in the third panel from top of Fig.~\ref{fig:long_abu_plot_15}. 
\Pdslr and \Pbslr show a relevant production by neutron captures
%primarily produced 
in the pre-CCSN phase during He- and C-burning, while they are destroyed in more advanced evolutionary stages. In stellar conditions the weak decay rates of both the two isotopes depend on temperature and density. However, electron captures on \Pbslr starts to affect its half-life at much lower temperatures than \Pdslr. For instance, at typical conditions at the onset of He-burning, the half-life of \Pbslr is reduced to a value of the order of a few decades, while the \Pdslr $\beta$-decay half-life is still of the order of a Myr. Such a difference may also introduce significant variations in the abundance profiles of the two isotopes, as can be seen for example in the drop in abundance above 3.2\stmass in Fig.~\ref{fig:long_abu_plot_15}. %in the deeper Ne-burning shell. 
%The \Pbslr abundance is further affected by its electron capture rate.
%As electron capture rate is both temperature and density dependant the half-life can decrease strongly in stellar conditions, for example the drop in abundance above 3.2\stmass. 

During the CCSN explosion both \Pdslr and \Pbslr are destroyed below the explosive C-burning peak, at about 2.15 \msun. %by photodisintegration into lighter species in the deep layers of the C-burning ashes. 
In explosive He-burning, the n-process neutron burst produces \Pdslr by radiogenic contribution from $\beta^{-}$ decay of the neutron-rich isotopes $^{107}$Rh and $^{107}$Ru. The %existing 
pre-CCSN \Pbslr is destroyed by neutron captures and cannot be formed by the $\beta^{-}$ decay of unstable isotopes %after the neutron burst is over 
because it is shielded by the stable isobar $^{205}$Tl.

In the 20 \msun model shown in Fig.~\ref{fig:long_abu_plot_20}, there is no effect of the explosion on the abundance profiles, as we have seen for other isotopes, due to the steep temperature gradient (Fig.~\ref{fig:t_MASS}).
% , the 20 \msun model therefore has explosive carbon burning below the carbon burning ashes, and no explosive helium burning. 
Comparing the 15 \msun model (Fig.~\ref{fig:long_abu_plot_15}) with the 25 \msun model shown in Fig.~\ref{fig:long_abu_plot_25}, the nucleosynthesis profiles are similar, with a more limited depletion of \Pbslr in the 25 \msun model during explosive He-burning.

%\marco{[MP: the analogous of Fig 12 is not provided for \Pdslr and \Pbslr. Is this a choice made on purpose? If yes, please write a sentence or two highlighting that no big differences has been seen bla bla bla. However, see comment to yields below. Maybe it is needed?]}
%\TL{[TL: I added the equivalent plot, but there is little change from model to model, biggest changes are levels of destruction in explosive He and C burning - Explain why we dont use]}

% \begin{figure}
%  \centering
%  \includegraphics[width=\columnwidth]{Plots/agg_plots/aggregate_plot_M15_Pd107.pdf}
%  \includegraphics[width=\columnwidth]{Plots/agg_plots/aggregate_plot_M15_Pb205.pdf}
%  \caption{Same as Fig.~\ref{fig:agg_plot_al26_fe60_mn53_15}. Isotopes shown are \Pdslr and \Pbslr.}
%  \label{fig:agg_plot_pd_pb_15}
% \end{figure}

As a general trend, we can see from Tables \ref{yield_table:3} and \ref{yield_table:5} that the yields of \Pdslr and \Pbslr %(Tables \ref{yield_table:3} and \ref{yield_table:5}) 
decrease as the explosion energy increases, varying only by as much as a factor of two.
% \marco{[MP: your motivation would work for \Pbslr but not for \Pdslr, since the n-process can make it efficiently.... be careful!] }% due to the impact of this energy on the temperature gradient, which strongly affects explosive He-burning and therefore the neutron burst. 
% There is a small contribution to the final yields of \Pbslr also from explosive C-burning. The exact location of the mass cut determines whether this effect is included in the final yield.
This is due to the fact that by increasing the explosion energy more of the CCSN ejecta are exposed to conditions where both the two SLRs are depleted, and the explosive C-burning peak is moved outward in mass coordinate. 

\subsection{\Snslr} \label{ss:sn}

The nucleosynthesis of \Snslr requires neutron-rich conditions to drive a double neutron capture %to be synthesised 
via the reaction chain $^{124}$Sn(n,$\gamma$)$^{125}$Sn(n,$\gamma$)\Snslr. 
The efficiency of this channel depends on the temperature dependent $\beta$-decay at $^{125}$Sn, whose half-life becomes less than a day in typical He-burning conditions and of the order of an hour or less in C-burning conditions \citep[e.g.,][]{takahashi:87}. Therefore, high neutron densities are required to accumulate a significant amount of \Snslr.
%This reaction chain is strongly dependant on the neutron density, similarly to \Feslr discussed in section \ref{ss:fe60}. 
%Contributions from radiogenic decay can also produce \Snslr, as there is no shielding stable isotope. However, as Sn has a magic proton number of 50 the nearest stable isotope of the two elements with 49 and 48 protons ($^{115}$In and $^{116}$Cd) are ten nucleons lighter than \Snslr, making production by $\beta^{-}$ decay difficult. 
% they are typically not significant here because the neutron burst does not reach that far from the valley of $\beta$-stability to be able to produce isotopes that could contribute to mass 126.
Figure \ref{fig:long_abu_plot_15} shows that in the 15\stmass model \Snslr is marginally produced in the pre-CCSN phase in the C-burning ashes.
During explosive nucleosynthesis, \Snslr is produced in sites of high neutron density, specifically in explosive C burning (around 2.16\stmass) and much more significantly (by three orders of magnitude in this model) in explosive He burning (2.75 -- 3.2\stmass). Thanks to the high neutron densities reached by the n-process, we notice also a small additional radiogenic contribution to \Snslr by $^{126}$In.

% In the fiducial 20\msun model shown in Fig.~\ref{fig:long_abu_plot_20} there is no significant production of \Snslr, since the CCSN production is limited by the lack of a neutron source during a non-active explosive C burning and He-burning.

In the fiducial 20\msun model (shown in Fig.~\ref{fig:long_abu_plot_20}) explosive C- and He-burning are not significantly activated, resulting in a lack of a neutron source and no production of \Snslr.
%since there is no explosive helium burning, and 
Furthermore, the pre-CCSN production of \Snslr is even lower than in the 15\stmass model, by almost two orders of magnitude.
Finally, also in the 25\stmass model in Fig.~\ref{fig:long_abu_plot_25}, %shows that %in the 25\stmass model 
\Snslr production occurs during the explosion, in both the C-burning (with similar abundance as in the 15\stmass model) and He-burning ashes (but an order of magnitude lower than in the 15\stmass model).

%Comparing 
The final yields of the all 15, 20 and 25\stmass models are given in Table \ref{yield_table:4}. In general, the models with high explosion energies produce more \Snslr. This is due to an increase in neutron density in explosive C-burning, and to a broader range of ejecta affected by the n-process during explosive He-burning. %most notably in the 20\stmass models. This
We obtain the largest variation of \Snslr yields for the 20\stmass models, increasing %results in an increase of yields 
from 8.9$\times$10$^{-14}$\stmass in the model with the lowest explosion energy (0.5\FoE) to 5$\times$10$^{-10}$\stmass in the model with the highest explosion energy (8.9\FoE).

\subsection{\Islr and \Csslr}

The neutron-rich SLRs \Islr and \Csslr are also separated from the $\beta^{-}$-valley of stability by one unstable isotope. The radiogenic contribution from their respective neutron-rich unstable isobars can be very significant in explosive conditions, or even dominate the total CCSN ejecta.
%We discuss \Islr and \Csslr together as their nucleosynthesis is similar, both requiring two neutron captures or the $\beta$-decay of unstable isotopes. 
%Regions of high neutron density are therefore required to form both \Islr and \Csslr. 

Using the 15\stmass model shown in Fig.~\ref{fig:long_abu_plot_15} as an example, during the pre-CCSN stages the s-process makes both \Islr and \Csslr by neutron capture through the branching points of $^{128}$I and $^{134}$Cs
%of pre-CCSN nucleosynthesis, we see production 
in the He-burning shell and the C-burning shell, while they are depleted in more advanced stages. %and drop below that during Ne photodisintegration.
% \Islr has a lower abundance pre-CCSN above 3.2\stmass, due to electron capture rate changes.
In the CCSN explosion, the pre-CCSN abundances are only partially modified in the explosive C-burning peak at about 2.1\msun. While, they are depleted by photodisintegration in the deeper parts of the former convective C shell. The n-process triggered by explosive He burning generates a complex pattern of production peaks. 
\Islr and \Csslr are produced directly only in he mildest n-process ejecta, the highest peaks of production are given by radiogenic contribution from higher  neutron-rich  isotopes. For the models considered in this work, 
the ejected isotopes $^{129}$Te, $^{129}$Sb, $^{129}$Sn and $^{129}$In will eventually decay into \Islr.
% \Islr can be ejected eventually as $^{129}$Te, $^{129}$Sb, $^{129}$Sn and $^{129}$In. 
The radioactive isotopes $^{135}$Xe, $^{135}$I and $^{135}$Te may instead contribute to the ejecta of \Csslr.
%During massive star nucleosynthesis \Islr and \Csslr are destroyed during explosive silicon and oxygen burning, with a small peak of production in explosive carbon burning.
%During explosive helium burning \Islr and \Csslr are mostly destroyed, excluding the uppermost region of the explosive ejecta. This is due to the neutron density being high, thus leading to consecutive neutron capture forming \Islr and \Csslr, but then continuing to form $^{130}$I and $^{136}$Cs, respectively. The large production peak is due to contributions from $\beta^{-}$ decay.
% However, while \Islr and \Csslr are not produced directly, we do see a large peak of production from $\beta$-decay from more  neutron-rich  isotopes. only
% This decay provides the large peak of abundance.

In the 20\stmass model shown in Fig.~\ref{fig:long_abu_plot_20}, the explosive contribution to \Islr and \Csslr is more limited compared to the 15\msun model, while a significant fraction of the pre-CCSN yields are ejected mostly unchanged from pre-CCSN abundances for material above 2.2\msun mass coordinate.
%The steep peak temperature profile of the 20\stmass model (Fig.~\ref{fig:long_abu_plot_20}) results in the destruction of \Islr and \Csslr occurring below carbon burning ashes. 
A relevant production peak is still obtained with explosive He-burning, in particular for \Islr. As we discussed for other SLRs before, this is due to the final n-process burst of %The peak in production in the helium burning ash is caused by an increase in 
neutrons from \NeAlphaN. The 25\stmass model in Fig.~\ref{fig:long_abu_plot_25} shows %the same production pre-CCSN 
a similar pre-CCSN production as in the 15\stmass model. However, during explosive nucleosynthesis we see a more complex production in explosive C burning at mass coordinates between 3.3\msun and 4.6\msun, due to the diverse radiogenic contribution to both \Islr and \Csslr. 
%but in the centre of this peak neutron densities reach too high and it is instead destroyed by continued neutron captures. During explosive helium burning the neutron density reaches values that cause destruction.
Comparing the final yields within each 15, 20 and 25\stmass sets of models, there are no significant changes (Table \ref{yield_table:4}). 
The largest variation is obtained between the 15\msun, 20\stmass and 25\stmass models, within the yields of \Islr or \Csslr increasing by factors of two, three and two, respectively. 
% It should be noted that a single 25\stmass model produces an ejected yield of 3.9$\times$10$^{-6}$\stmass, which is four orders of magnitude higher than any other model. However, this large increase seen only in \Snslr and \Islr, as such this is assumed to be an anomalous result.
%\TL{[TL:Is it worth expanding on this odd result, it seems like the issue might be the lowest mass coordinate producing seemingly 1e-3 of both isotopes. - Clean up this model, clean up the final boyo]}

% [MP: Or something like that. Are you sure about this? If this is the case, you are correctly considering the radiogenic contribution etc, this should be justified. My best guess by looking at the models is that CCSN production is compensating the depletion of what was there from the pre-CCSN, but if for instance I am looking at the 25Msun in the figure, I can see a strong production of I129 during the explosion, well above the pre-CCSN reservoir. In the 15msun, I can see the CCSN explosion to increase the Cs135 by a factor of 2 or so. ]

\subsection{\Hfslr} \label{ss:hf}

\Hfslr is produce by neutron capture, similar to \Islr and \Csslr. It may be directly produced from two neutron captures from
%Both for pre-CCSN and explosive nucleosynthesis the production of \Hfslr is defined by neutron capture, either by double neutron capture on 
$^{180}$Hf through the unstable $^{181}$Hf, or as a result of radiogenic contribution via $\beta^{-}$ decay from its more neutron-rich unstable isobars. However, for the stellar calculations presented in this work we have used the $^{181}$Hf $\beta^{-}$ decay rate by \cite{goriely:99}, which does not take into account of experimental data by \cite{bondarenko:02}. 
Therefore, we underestimated the half-life of $^{181}$Hf.
\cite{lugaro:14} showed that the different half-life leads to an increase of \Hfslr in CCSN by 7\% in their 15\stmass model and a factor of 2.6 for their 25\stmass model.

% This results to a strong underestimation of the half-life of $^{181}$Hf, and to a reduced direct production of \Hfslr in s-process conditions \citep[][]{lugaro:14}. 
% If the half-life of $^{181}$Hf were lower, it would decay faster than it ought to. 
% It would therefore be harder to produce \Hfslr via neutron capture on $^{181}$Hf.
% If the \Hfslr were higher there would be an increased chance of neutron capture on $^{181}$Hf, producing more \Hfslr.
% While this does not affect the explosive nucleosynthesis of \Hfslr in our models, we expect a significantly higher \Hfslr amount made in the pre-CCSN stage compared to what we obtain once the updated $^{181}$Hf $\beta^{-}$ decay rate would be used instead. %$\beta^{-}$ decay from unstable isotopes. 

In the 15\msun model shown in Fig.~\ref{fig:long_abu_plot_15}, during the pre-CCSN phase \Hfslr is directly produced via neutron capture in C-burning, and some late production is also visible in the deepest He-burning regions between mass coordinates 2.6\msun and 2.8\msun. %, see Fig.~\ref{fig:long_abu_plot_15} for an example, where neutron densities allow captures from $^{180}$Hf through the unstable $^{181}$Hf. 
As for other neutron-rich SLRs seen in previous sections, the C-burning ashes are the deepest CCSN ejecta carrying some \Hfslr. However, unlike these other SLRs, in this case there is no direct or radiogenic contribution from explosive C-burning feeding \Hfslr. 
% \marco{[MP: this is interesting. Why there is none? I do not see it either. This would be something to find out, maybe on a specific work focused on Hf182. If I would be a referee I would ask about this, but let's keep it simple for now. ]}\TL{[TL: I thought it was due to n-captures on Te and Xe for CS and I, however there is no seed isotope within a single neutron capture for Hf, I will check the flux plots to confirm]}
%During CCSN nucleosynthesis \Hfslr is destroyed in NSE and explosive C-burning via the \gammaproc. 
In explosive He-burning the production of \Hfslr is instead more complex, where the direct production is 
% marginal and the n-process ejecta are dominated 
supplemented
by radiogenic contributions to \Hfslr. Within the models considered in this work, we can identify the contribution of $^{182}$Lu, $^{182}$Yb, $^{182}$Tm and $^{182}$Er. %not due to neutron captures on $^{180}$Hf and $^{181}$Hf, but to the contribution from the $\beta^{-}$decay of $^{182}$Lu, which is in turn fed by the decay of $^{182}$Yb. 

% of Fig.~\ref{fig:long_abu_plot_15}
Compared to the 15\msun model, the 20\msun model shown in Fig.~\ref{fig:long_abu_plot_20} has a lower contribution from explosive nucleosynthesis, and therefore its yields are expected to be more affected by the $^{181}$Hf $\beta^{-}$ decay rate used in the simulations. 
% \TL{[TL: Marco, is this in relation to the Bondarenko rate?]}
%the steep temperature gradient in the 20\stmass model (see Fig.~\ref{fig:t_MASS}) reduces the production of \Hfslr during explosive He-burning. 
% Comparing the above 15\msun model to a 25\msun model (Fig.\ref{fig:long_abu_plot_25}) we see that du .
Within this model, \Hfslr is produced during the CCSN explosion %explosive nucleosynthesis
only in the He-burning regions. On the other hand, the total ejecta are dominated by the pre-CCSN production in the C shell ashes, between about 2.2\msun and 4\msun. Also the \Hfslr ejecta from the 25\stmass model are dominated by the pre-CCSN production. %Comparison the 15\stmass model to the 25\stmass model of Fig.~\ref{fig:long_abu_plot_25} shows destruction in NSE below explosive carbon burning. The inner explosive helium burning produces some \Hfslr. 
Comparing the final yields of the 15\stmass set of models, there are variations of an order of magnitude (Table \ref{yield_table:5}). This is due to the distribution of the CCSN peak temperatures reached within the stellar progenitor structure, %temperature gradients of the models, 
where the more extreme explosion energy models reach the He burning ashes with higher temperatures (see Fig.~\ref{fig:t_MASS}), thus producing more neutrons and more radiogenic contributions to \Hfslr. 
Variations in the yields of \Hfslr for the 20\stmass models set is similar to that of the 15\stmass models, varying by roughly one order of magnitude.
Variations in the yields of \Hfslr for the 25\stmass models are more limited, of a factor of two (Table \ref{yield_table:5}).
% This is due to a reduced 
% \marco{[MP: interesting that the 25Msun does not have the same radiogenic contribution as the 15Msun. They seem to be quite similar from previous comparison. Did you think about it? Is there a specific reason? Food for thoughts. ]}

\begin{figure}
 \centering
 \includegraphics[width=\columnwidth]{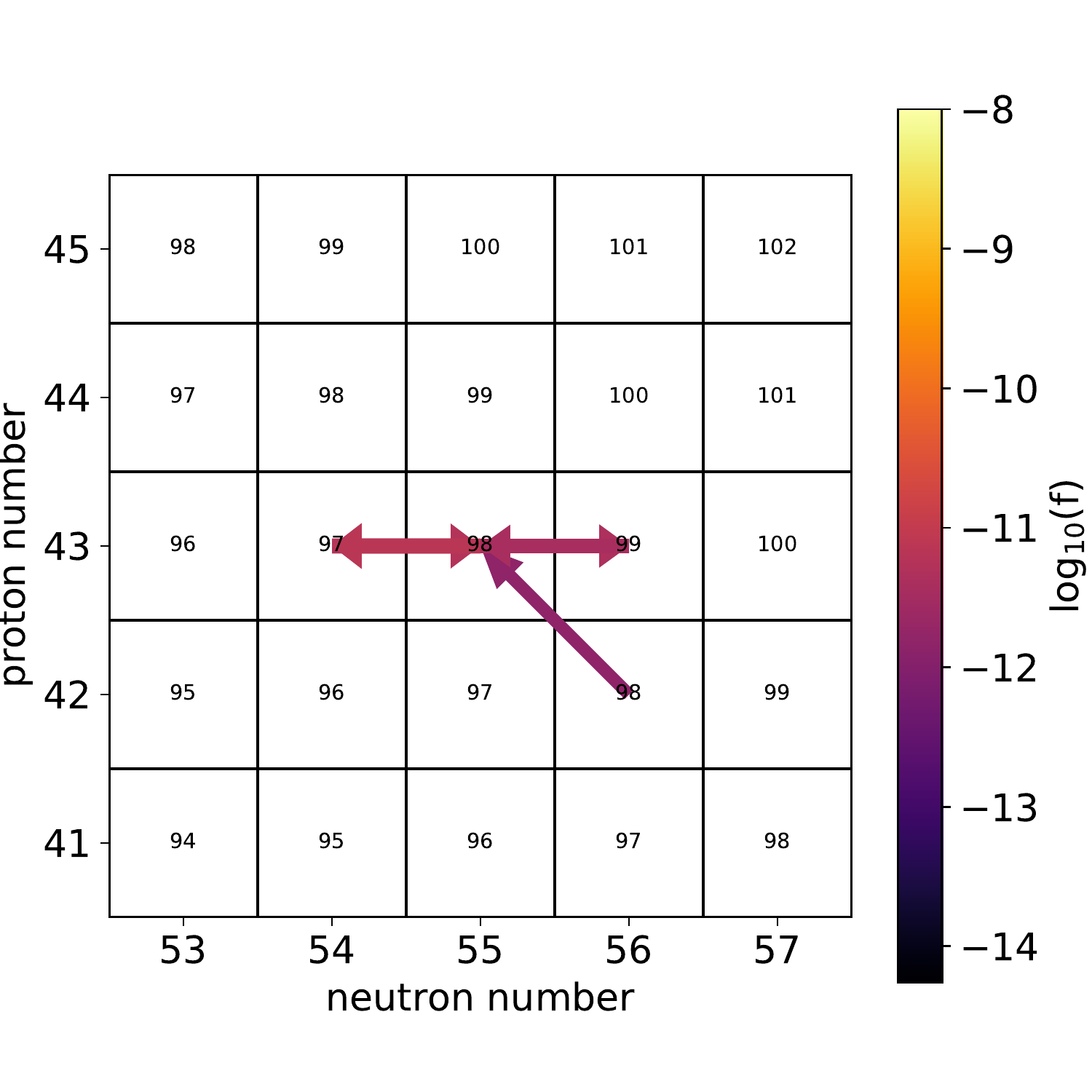}
 \includegraphics[width=\columnwidth]{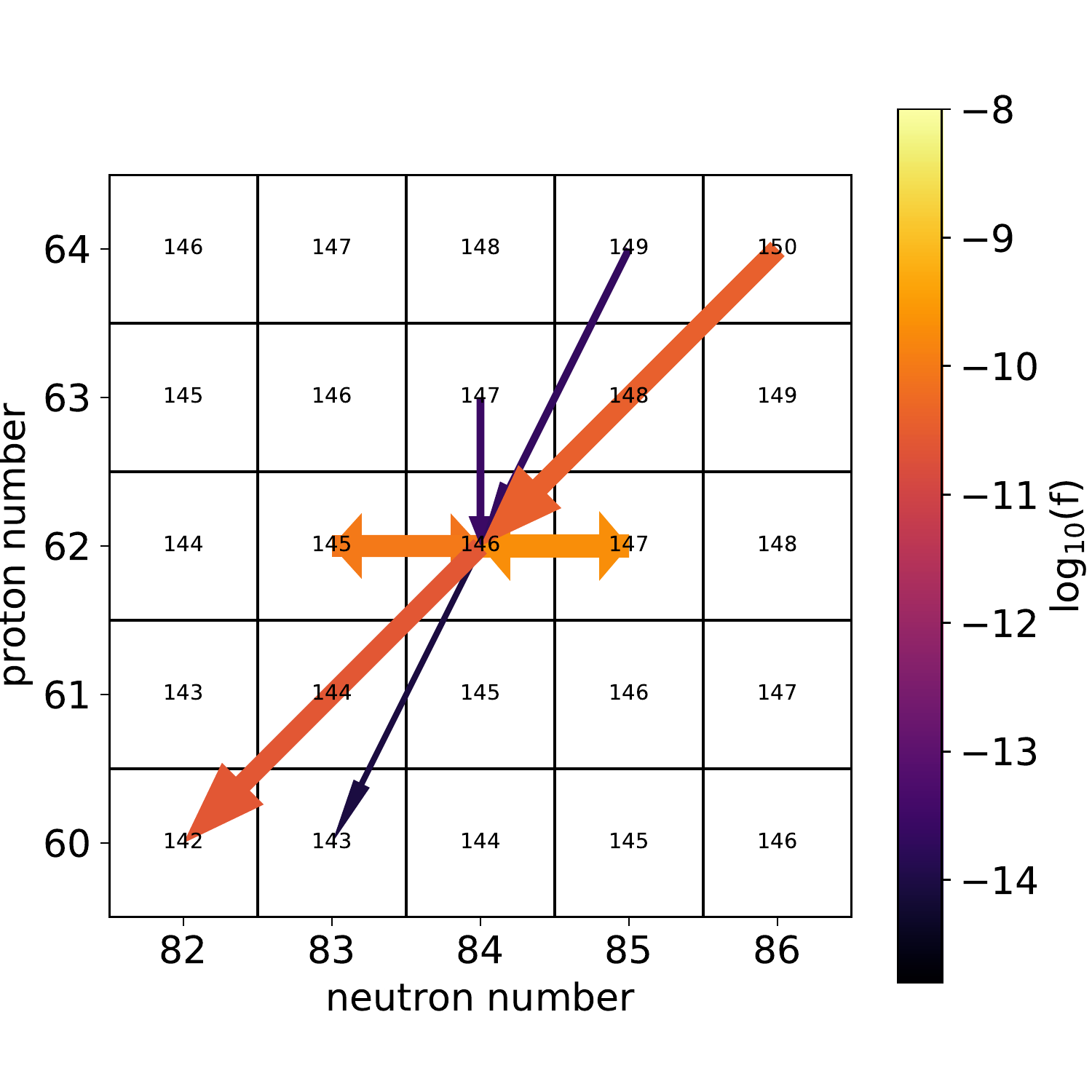}
 \caption{Same as in Fig.~\ref{fig:Flux_matrix_1_al26}, but for 
 \Tcslrseven~(Z=43, top panel) and 
 \Smslr~(Z=62, bottom panel) in explosive Ne burning (mass coordinate 2.1\stmass from the 15\msun model, Fig.~\ref{fig:long_abu_plot_15}).}
 \label{fig:Flux_matrix_2_tc97sm146}
\end{figure}

\begin{figure*}
 \centering
 \includegraphics[width=\columnwidth]{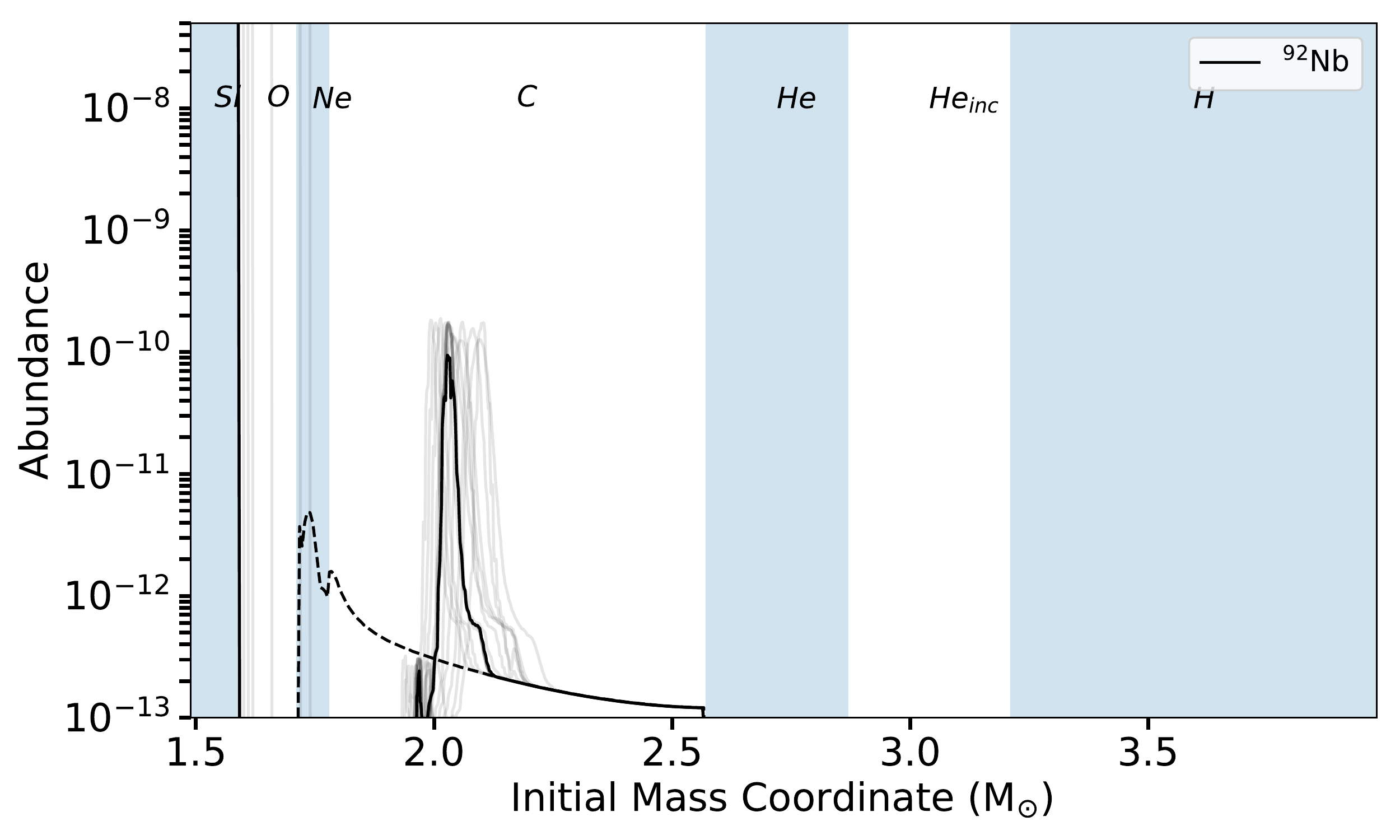}
 \includegraphics[width=\columnwidth]{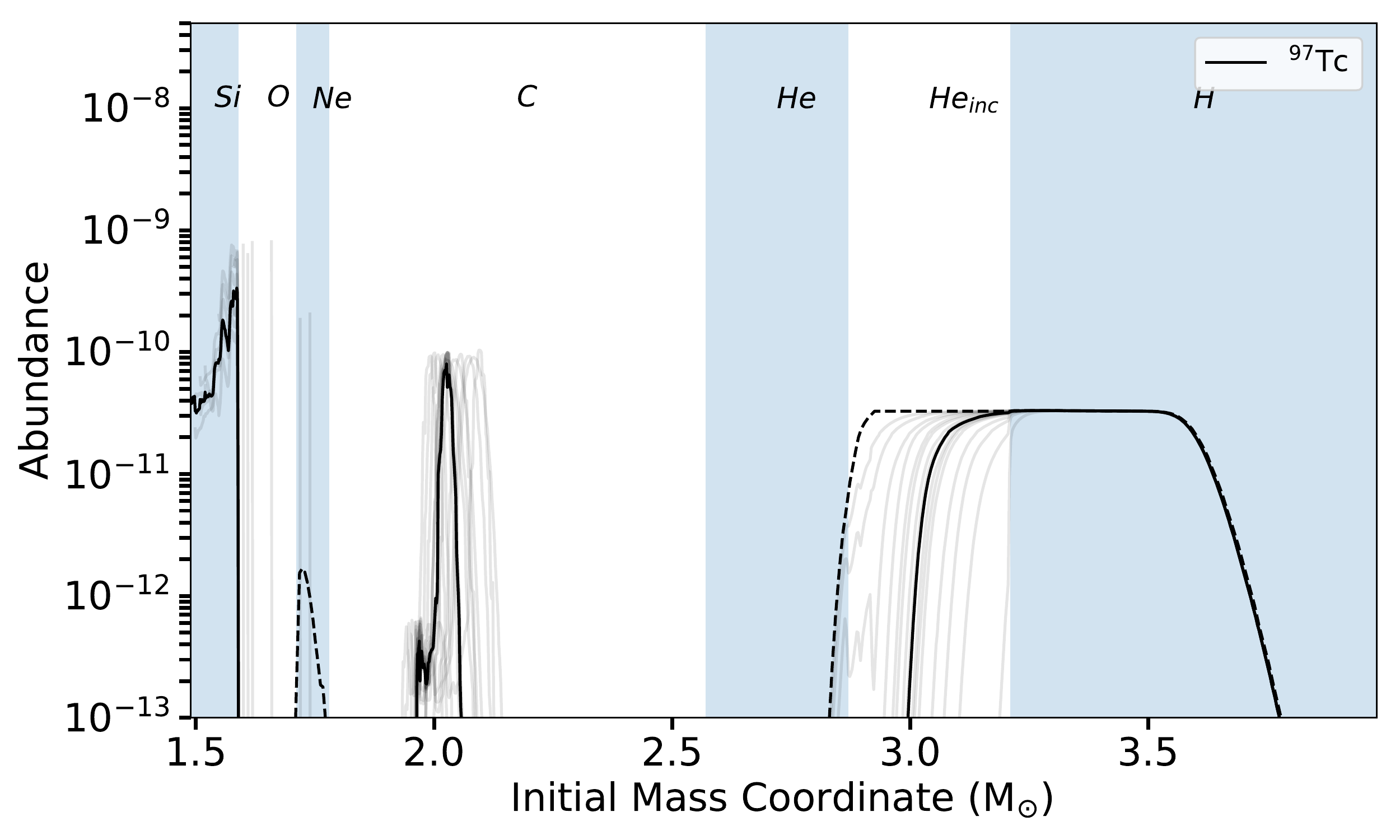}
 \includegraphics[width=\columnwidth]{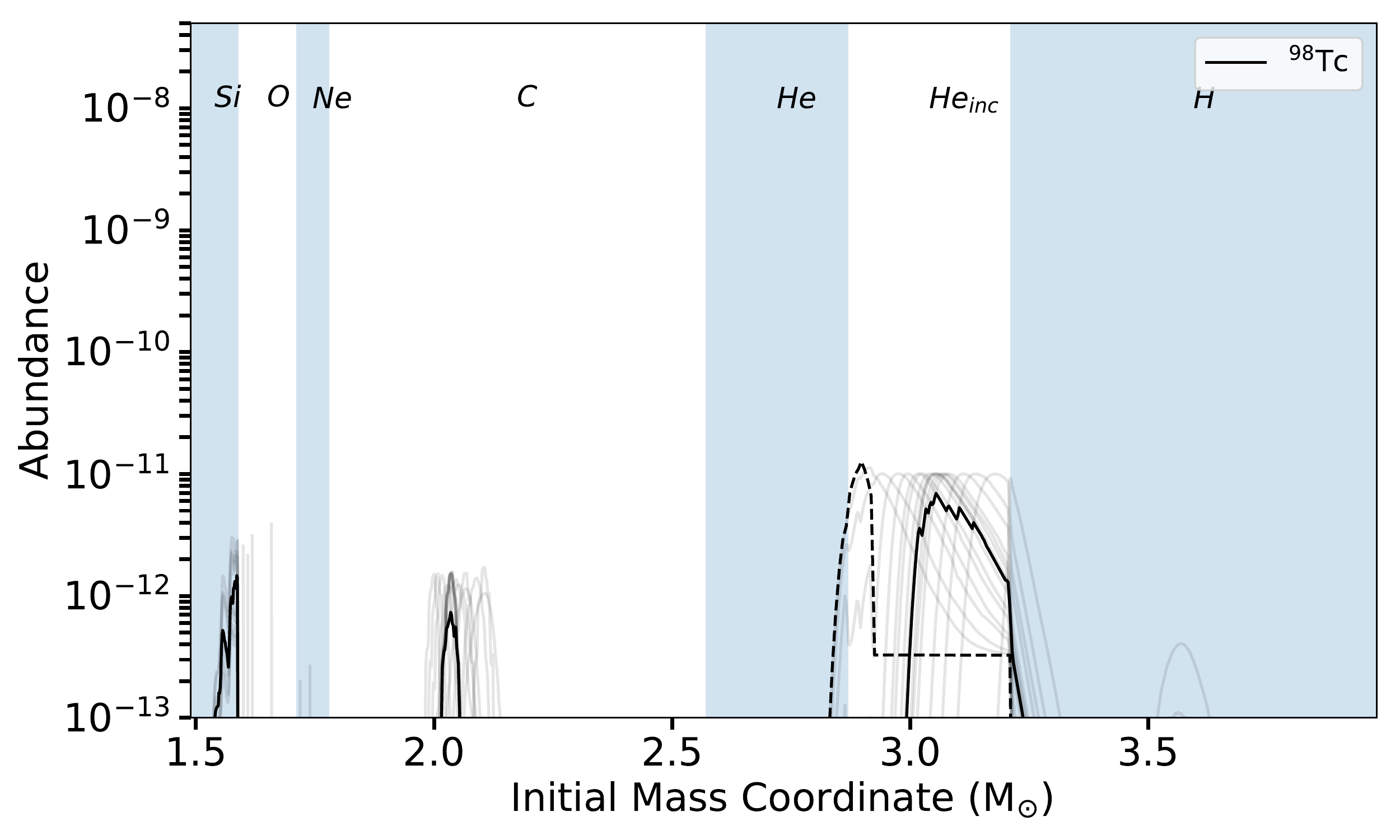}
 \includegraphics[width=\columnwidth]{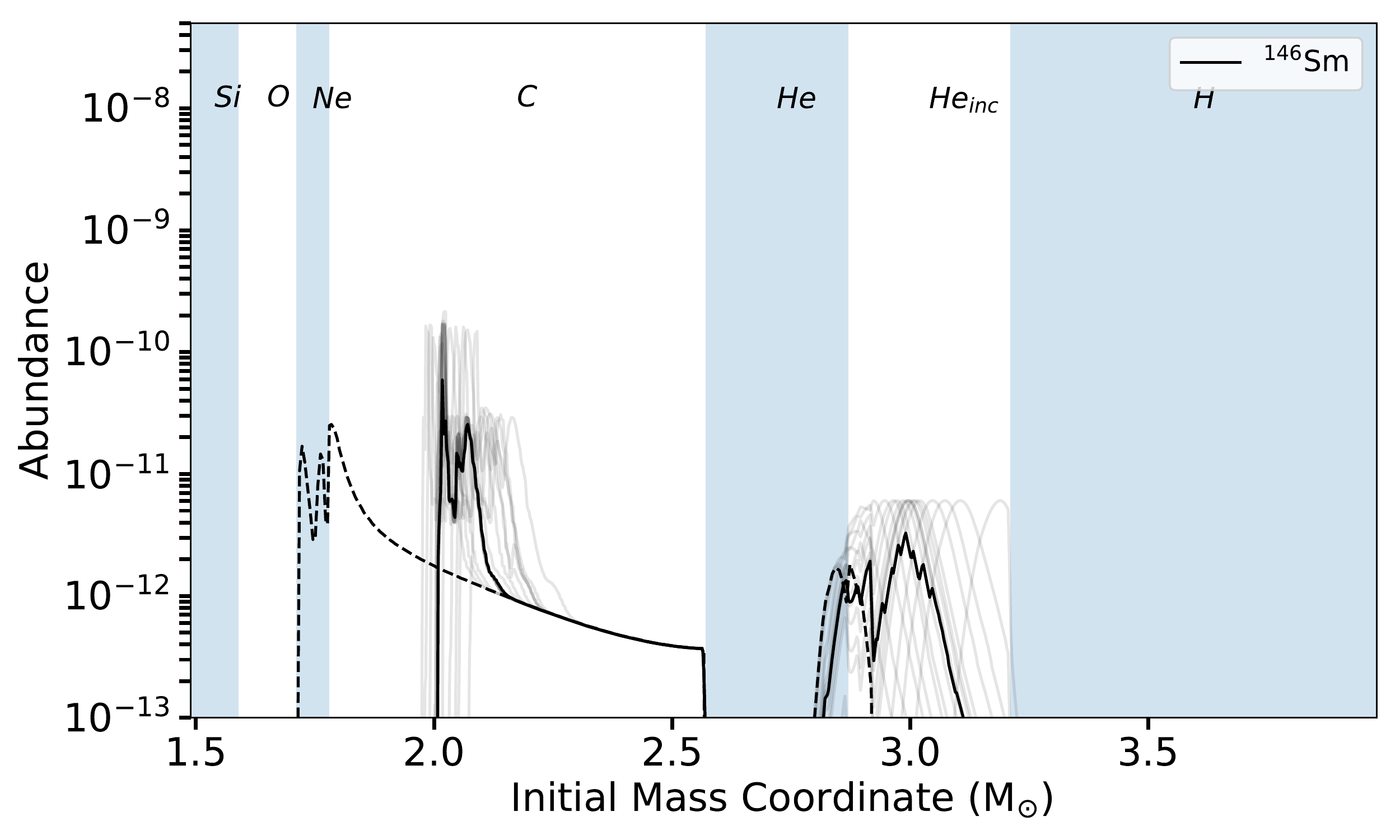}
 \caption{Abundance plots of \Nbslr, \Tcslrseven, \Tcslreight and \Smslr for all 15\stmass models, as in Fig.~\ref{fig:agg_plot_al26_fe60_mn53_15}. It should be noted at \Nbslr abundances %within the Si-burning ashes 
 reach peaks of about 1$\times$10$^{-5}$ mass fraction in $\alpha$-rich freeze-out conditions, for mass coordinates M$<$1.6\stmass.}
 \label{fig:agg_p_isotopes_15}
\end{figure*}

\subsection{Heavy SLR isotopes accessible via the $\gamma$-process: \Nbslr, \Tcslrseven, \Tcslreight, and \Smslr} \label{ss:p_iso}

The four SLRs isotopes \Nbslr, \Tcslrseven, \Tcslreight and \Smslr are all potential products of the $\gamma$ process in CCSNe, where the nucleosynthesis is mostly driven by the photodisintegration of heavier isotopes \citep[e.g.,][]{Woosley1978,Arnould2003,rauscher:13,Pignatari2016a}. 
% However, specific nucleosynthesis properties need to be considered to take into account all the conditions met during the explosion. 
In some of the CCSN ejecta exposed to the highest temperatures and densities, also charged particle reactions and in particular direct proton captures 
% and charged particle reactions 
can be expected to be relevant for the production of \Nbslr, \Tcslrseven and \Tcslreight
% in some of the CCSN ejecta exposed to the highest temperatures and densities 
\citep[e.g.,][]{Hoffman1996,Lugaro2016,Travaglio2018}.
%The important proton capture reactions are $^{91}$Zr(p,$\gamma$)$^{92}$Nb, $^{96}$Mo(p,$\gamma$)$^{97}$Tc and $^{97}$Mo(p,$\gamma$)$^{98}$Tc, respectively.
Moreover, both Tc SLRs may receive a small neutron capture contribution fed from the available abundance of the stable isotope $^{96}$Ru, via the $^{96}$Ru(n,$\gamma$)$^{97}$Ru($\beta^{+}$)$^{97}$Tc(n,$\gamma$)$^{98}$Tc chain. A small amount of \Smslr can also be made by neutron capture, starting from the stable isotope $^{144}$Sm. 
\Nbslr is shielded from any radiogenic contribution from the proton-rich side of the valley of $\beta$ stability by its stable isobar $^{92}$Mo. The same applies for \Tcslreight and its stable isobar $^{98}$Ru. 
A radiogenic contribution to \Smslr can be made by the respective proton-rich unstable isobars, but also by the $\alpha$-decay channel $^{154}$Dy($\alpha$)$^{150}$Gd($\alpha$)$^{146}$Sm after the $\beta$-decay radiogenic contributions to the single radioactive $^{154}$Dy and $^{150}$Gd.
%For \Smslr, a small amount can also be made by neutron capture, starting from the stable isotope $^{144}$Sm. Finally, radiogenic contributions can included for \Tcslrseven and \Smslr. For both \Tcslrseven and \Smslr, the total radiogenic contribution can be impacted by the respective proton-rich unstable isobars. For \Smslr, there is also a contribution from the $\alpha$-decay channel $^{154}$Dy($\gamma$,$\alpha$)$^{150}$Gd($\gamma$,$\alpha$)$^{146}$Sm and the respective $\beta$-decay radiogenic contributions to the single SLRs $^{154}$Dy and $^{150}$Gd. Both \Nbslr and \Tcslreight are shielded from any radiogenic contribution from the proton-rich side of the valley of $\beta$ stability by their stable isobars $^{92}$Mo and $^{98}$Ru, respectively. However, both Tc SLRs may receive a small neutron capture contribution fed from the available abundance of the stable isotope $^{96}$Ru, via the nucleosynthesis channel $^{96}$Ru(n,$\gamma$)$^{97}$Ru($\beta^{+}$)$^{97}$Tc(n,$\gamma$)$^{98}$Tc. 

%\tl{Discussion of the 15Msun Pre-CCSN.}
For the 15\stmass model in Figure \ref{fig:long_abu_plot_15}, all the proton-rich SLRs except for \Tcslreight are present in the pre-CCSN ashes of Ne-burning and partial O-burning. In these regions, between about 1.65\msun and 1.8\msun, the $\gamma$-process is activated \citep[e.g.,][]{arnould:76}. Some \Nbslr and \Smslr can also be found in the pre-CCSN C-burning ashes up to a mass coordinate 2.55\msun, due to the interaction between the former convective O-burning shell and C-burning shell \citep[e.g.,][]{meakin:06,Ritter2018,andrassy:20}.
\Tcslrseven, \Tcslreight, and \Smslr are also produced in the top layers of the He-burning ashes (above 2.8\msun in the figure), due to their neutron capture channels activated by the s-process in these regions.
%\TL{Discussion of the 15Msun Post-CCSN.}

During explosive nucleosynthesis, proton captures are active in the deepest region of the ejecta by $\alpha$-rich freeze-out \citep[e.g.,][]{Hoffman1996,Pignatari2016a}, producing \Nbslr, \Tcslrseven and small traces of \Tcslreight (see in Figure \ref{fig:long_abu_plot_15} the region between the mass cut and 1.6\msun). Note that, the abundance of \Nbslr is so high in this deep region that this isotope is plotted both in the second and the forth panels from the top in Figure \ref{fig:long_abu_plot_15}.
The $\gamma$-process driven by the CCSN explosion is %also 
active %in explosive conditions, with 
at a typical CCSN temperature peak between about 2.5\,GK and 3.2\,GK \citep[e.g.,][]{rapp:06}. This temperature range is reached in the explosive Ne-burning and partial O-burning regions, where the isotopes discussed in this section are efficiently produced (bottom panel of Fig.~\ref{fig:long_abu_plot_15}). The production flux of %three of the four isotopes 
\Tcslrseven and \Smslr~during explosive Ne-burning is shown in Fig.~\ref{fig:Flux_matrix_2_tc97sm146}. 
In the figure we can see that together with the nucleosynthesis flow triggered by photodisintegration reactions, for \Tcslrseven also some proton capture reactions happen: namely, $^{96}$Mo(p,$\gamma$)$^{97}$Tc. 
Not included in the figure, $^{91}$Zr(p,$\gamma$)$^{92}$Nb and $^{98}$Mo(p,n)$^{98}$Tc drives the relevant production flow for \Nbslr and \Tcslreight, respectively. Instead, in this same mass region \Smslr is mostly driven by ($\gamma$,n) and ($\gamma$,$\alpha$) photodisintegration.
%which illustrate that the reactions that significantly producing the four isotopes are: (p,n) for \Nbslr (not shown in plot), (p,$\gamma$) and (p,n) for \Tcslrseven, (p,n) for \Tcslreight, and ($\gamma$,n) for \Smslr. 
During the explosion most of the pre-CCSN \Tcslrseven, \Tcslreight and \Smslr made in the He ashes are destroyed by neutron captures. However, some of these isotopes are made again in the milder n-process components between 3.1\msun and 3.3\msun, %Within the H ashes, 
during the neutron burst. %, \Tcslreight and \Smslr are produced via neutron captures (see $\approx$3.2\stmass in the lower panel of Fig.~\ref{fig:long_abu_plot_15}). 

%\TL{(Reference to figure with different CCSN explosion for the 15Msun.)}
Figure \ref{fig:agg_p_isotopes_15} shows the abundance profiles of \Nbslr, \Tcslrseven, \Tcslreight and \Smslr for all the 15\stmass models. Qualitatively, the behaviour is consistent between the models. Except for \Nbslr, significant variations in the abundances are seen in explosive He-burning, due to the n-process efficiency varying between the models. %differences in the temperature profiles and position of the mass cut. 
%Due to production via $\alpha$-rich freeze-out in the inner most region, 
The yield of \Nbslr %and \Tcslrseven 
is mostly sensitive to the amount of CCSN fallback of the model, due to production via $\alpha$-rich freeze-out in the innermost regions. For instance, by comparing the yields of \Nbslr from the 15\stmass models with a low and high amount of CCSN fallback, the \Nbslr abundance changes by five orders of magnitude (see Table \ref{yield_table:3}). 
The yield of \Tcslrseven is less affected by the position of the mass cut compared to \Nbslr, as \Tcslrseven is produced in both the $\alpha$-rich freeze-out and by neutron capture in He-burning conditions.
According to Table \ref{yield_table:3}, the \Tcslrseven ejected yield varies by an order of magnitude, depending on the position of the mass cut.

% \marco{[MP: Tom? Are you sure that fallback does not affect Tc97?? Is this still confusing between Tc97 and Tc98? I wrote here the sentence following what you wrote, but from the figure it seems that Tc97 ejecta is dominated by the alpha-rich freeze-out. By how much is changing the ejecta of Tc97? By a factor of... ] }
%is significantly produced during the explosion in the $\alpha$-rich freeze-out, however, its final yield is barely impacted by the position of the mass cut, as the bulk of yield comes from the hydrogen ashes (see Table \ref{yield_table:3}). 

From comparing the 20\stmass (Fig.~\ref{fig:long_abu_plot_20}) and 25\stmass (Fig.~\ref{fig:long_abu_plot_25}) models with the 15\stmass model (Fig.~\ref{fig:long_abu_plot_15}), %we see the same behaviours, excluding the production in innermost core, again due to mass cut position.
the first relevant difference is that the higher mass progenitors do not eject any material exposed to $\alpha$-rich freeze-out conditions. In the 20\msun model, the $\gamma$-process production is limited to between mass coordinates 2\msun and 2.2\msun. The milder explosive He-burning experienced in this model with respect to the 15\msun model, causes a weaker efficiency of the n-process. These conditions favour the production of \Tcslreight and \Smslr in the ejecta between 4.8\msun and 5.2\msun. The s-process component of \Tcslrseven made during the pre-CCSN phase appears to be more relevant than the $\gamma$-process in the ejecta of this model. In the 25\msun star shown in Fig.~\ref{fig:long_abu_plot_25}, the $\gamma$-process production becomes extremely important for all the SLR isotopes considered here. For instance, in this model \Tcslrseven is mostly made in these conditions (around 3.4\stmass). For \Smslr, there is also a clear contribution from explosive C-burning made by neutron captures, above mass coordinate 3.6\msun. \Tcslreight shows a comparable production due to photodisintegration and by neutron captures.

%\TL{Final yield discussion with numbers}
The yields of all 15, 20 and 25\stmass models are listed in Tables \ref{yield_table:2} (\Nbslr), \ref{yield_table:3} (\Tcslrseven and \Tcslreight) and \ref{yield_table:5}(\Smslr). As mentioned, \Nbslr shows the largest %a significant 
difference in yields, depending on the location of the mass cut, varying from 2.8$\times$10$^{-12}$\stmass to 5.2$\times$10$^{-7}$\stmass in the 15\stmass models. 
The 20\stmass models have no inclusion of the $\alpha$-rich freeze-out in ejecta, and some of these models do not eject the results of explosive carbon burning either. This causes the ejected yield to also vary considerably, from 2.8$\times$10$^{-15}$\stmass to 1.8$\times$10$^{-7}$\stmass.
The final yields of \Tcslrseven and \Tcslreight also change with the mass cut by about an order of magnitude, as there is production in the \alp-rich freeze-out that is not included in models with more CCSN fallback. 
As \Tcslrseven is more impacted by $\alpha$-rich freeze-out its ejected yield varies by an order of magnitude in the 15\stmass models and by a factor of two in the 20\stmass. As \Tcslreight is less impacted it has a similar variation between models across the 15\stmass, 20\stmass and 25\stmass models, of around a factor of two.
% Due to the impact ofexplosive C-burning on \Smslr it is most impacted by the position of the mass cut. As the mass cut location varies by a small amount in the 15\stmass
% The variations in the 20\msun and 25\msun models are more limited, up to a factor of \marco{xxx, xxx and xxx for \Nbslr, \Tcslrseven and \Tcslreight, respectively [MP: Tom, see here.]. } %the high mass cut. 
\Smslr shows more limited variation in the yield across the 15\stmass and 25\stmass models, up to a factor of 20\%. The 20\stmass and 25\stmass models with the mass cut above the location of explosive C-burning have reduced yields, from 8.3$\times$10$^{-11}$\stmass in the 20\stmass model with the lowest mass cut 
% (1.74\stmass in mass coordinates) 
to 1.0$\times$10$^{-13}$\stmass in the 20\stmass model with the highest mass cut.
% (3.40\stmass).

%%%%%%%%%%%%%%%%%%%%%%%%%%%%%%
\section{Discussion} \label{sec:disc}

Starting from the analysis provided in the previous section, we discuss here the possibility of using SLRs abundances to track the contribution of the different components in the CCSN ejecta. %In this section we summarize the specific nucleosynthesis sites in the CCSN ejecta that provide the largest contributions to the final ejected yield of SLRs. 
%, and identify the SLRs that are produced in this site. After formulating these diagnostic techniques, we could use them to locate which sites have been activated in observed abundance patterns, e.g. the ESS. We also use the diagnostic techniques to compare our yields to other yield sets available in the literature. [MP: I do not think it makes much sense what you write here. What technique are you using to compare? Are you not just comparing? ]
We then compare our results with other stellar yield sets available in the literature. 

\subsection{Abundances of SLR isotopes: a diagnostic for CCSN nucleosynthesis?} \label{ss:diag}

In Section \ref{sec:results}, we have described in detail the nucleosynthesis of SLRs in the different components of CCSN ejecta
and have shown
% . While at least a simplified picture for the nucleosynthesis of those isotopes is well known from the literature, it becomes clear from our analysis 
that the same SLR can be found in different parts of the ejecta. In some cases (e.g., \Clslr and \Caslr) the nucleosynthesis production paths change from one region to the other, with different nuclear reactions relevant for the production in the different regions. In other cases, the important nuclear reactions are the same (e.g., \Feslr), but they are triggered at different temperature and density conditions. Furthermore, the stellar half-life of most of SLRs is long enough to have a significant amount of their pre-CCSN production still present in the final ejecta (e.g., \Alslr and \Hfslr). Overall six main nucleosynthesis components in CCSN ejecta can be identified from our analysis for the SLRs: (1) $\alpha$-rich freeze-out; (2) explosive Si-burning; (3) explosive O-burning; (4) explosive Ne-burning with partial O-burning; (5) explosive C-burning together with pre-CCSN C-burning ashes; and (6) the most external ejecta with He-burning and H-burning products, from both pre-CCSN and explosive nucleosynthesis. The production of isotopes in H-burning and He-burning conditions differ from each other, although, depending on the evolution history of the progenitor star, it is possible to find significant overlap between the ejecta of these two components. Additionally, a precise discussion of the SLR production in the He-burning region requires the He-burning ejecta to be split into sub-components. Therefore, for the sake of simplicity we consider these most external components together. 

\begin{figure*}
 \centering
 \includegraphics[width=\linewidth]{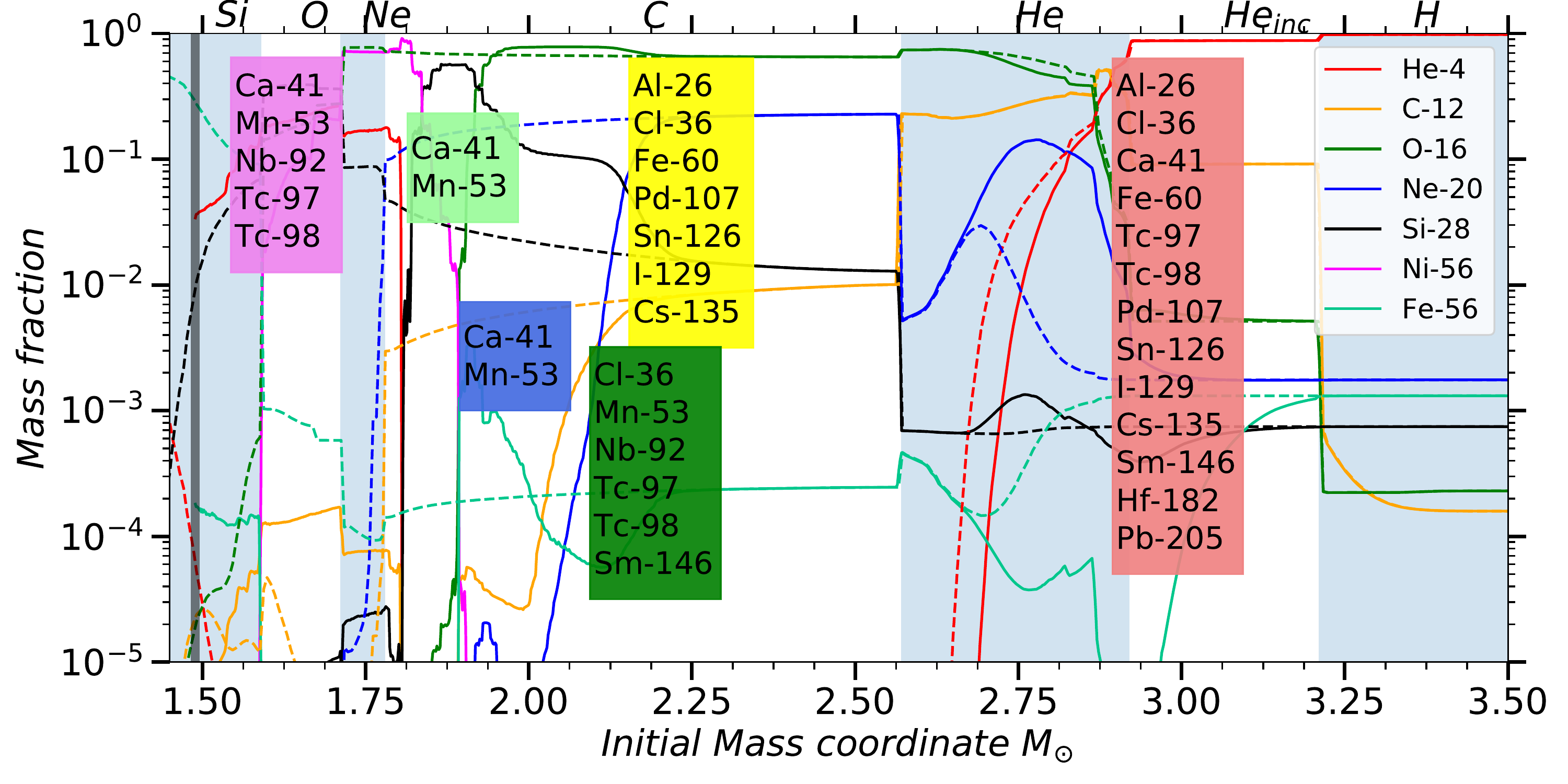}
 \caption{As a summary, we highlight the main production layers in the CCSN ejecta for the each SLR. We use as a background structure the 15\msun model (shown in the top panel of  Fig.~\ref{fig:long_abu_plot_15}). Each colour defines a specific explosive nucleosynthesis regime: pink is $\alpha$-rich freeze-out (1.5\msun (location of the mass cut) $\lesssim$ \msun $\lesssim$ 1.67\msun), light green explosive Si-burning (1.67\msun $\lesssim$ \msun $\lesssim$ 1.82\msun), blue explosive O-burning (1.82\msun $\lesssim$ \msun $\lesssim$ 1.95\msun), dark green explosive Ne-burning (1.95\msun $\lesssim$ \msun $\lesssim$ 2.1\msun), yellow explosive C-burning (2.1\msun $\lesssim$ \msun $\lesssim$ 2.55\msun), and red explosive He-burning (\msun $\gtrsim$ 2.55\msun).
 The position of each box does not depict the exact mass range of nucleosynthesis, rather it is indicative of the rough position.
% (2) 1.67\msun $\lesssim$ \msun $\lesssim$ 1.82\msun; (3) 1.82\msun $\lesssim$ \msun $\lesssim$ 1.95\msun; (4) 1.95\msun $\lesssim$ \msun $\lesssim$ 2.1\msun; (5) 2.1\msun $\lesssim$ \msun $\lesssim$ 2.55\msun; (6) \msun $\gtrsim$ 2.55\msun
 %\marco{[MP: This scheme needs to be reviewed. There are also typos. For instance, Mn53 in alpha-rich freeze-out is missing. Nb92 is not explosive Si-burning product. The magenta box should not overlap with the explosive Si-burning region (the bluish bar at 1.75\msun coordinate, etc. What about H-burning ejecta, making on top of all Al26? Etc. Also, would be good to make this plot really nice, for presentations or else. We can leave it for now and customize later.]\TL{[TL: My plan was to show the impact of the 6 sites described in 4.1, otherwise it seems confusing to me to include these and not discuss them in the section the plot is placed. I can make another plot that shows ALL SLR impact sites, but I'm not sure where that would be placed.]}}
 }
 \label{fig:summary_15_SLRs}
\end{figure*}

In Fig.~\ref{fig:summary_15_SLRs}, as a summary, we show the 15\msun CCSN ejecta, highlighting the main nucleosynthesis components for the different SLRs. Using the simplified scheme of the six components mentioned above, the components can be roughly divided as 
described in the caption of Fig.~\ref{fig:summary_15_SLRs}.
% follows: (1) 1.5\msun (location of the mass cut) $\lesssim$ \msun $\lesssim$ 1.67\msun; (2) 1.67\msun $\lesssim$ \msun $\lesssim$ 1.82\msun; (3) 1.82\msun $\lesssim$ \msun $\lesssim$ 1.95\msun; (4) 1.95\msun $\lesssim$ \msun $\lesssim$ 2.1\msun; (5) 2.1\msun $\lesssim$ \msun $\lesssim$ 2.55\msun; (6) \msun $\gtrsim$ 2.55\msun.

The $\alpha$-rich freeze-out component (1) is present in most of the 15\msun CCSN models of our set, and in one model of the 20\msun star, as the inner most regions of most models are swallowed by the mass cut. 
% [MP: is it present in some outlying models of the 20\msun? check. There is at least one maybe?]
Although our models tend to have the $\alpha$-rich freeze-out component in smaller progenitor masses, there is no clear correlation with the progenitor mass or linear dependence with explosion energy. In the following section, we will see that in other stellar sets this component is not ejected from any stellar masses. 
We have seen that the SLR isotopes \Nbslr and \Tcslrseven are efficiently made by component (1), together with a contribution from the explosive Ne-burning, component (4), among others. 
% From the analysis in the following section, we see that 
Our \Tcslrseven abundances are not particularly enhanced compared to other stellar sets without $\alpha$-rich freeze-out in 15\msun stars (see following section). This is because the \Tcslrseven production by component (1) is less than an order of magnitude larger than by produced component (4), with additional contributions from explosive C-burning (5) and He-burning (6), respectively. 
Therefore, only the \Nbslr abundance can be used as a diagnostic of component (1) contribution, as it dominates its production by component (4) by up to several orders of magnitudes. An important caveat to keep in mind is that one-dimensional CCSN models tend to under-produce the $\gamma$-process production in the \Nbslr region by about an order of magnitude, when compared to the abundances of stable Mo and Ru proton-rich isotopes in the Solar System. This puzzle is still a matter of debate \citep[e.g.,][]{Travaglio2018}. However, the production of \Nbslr is so high in zone (1) of our models, that the diagnostic power of the SLR for this component can be safely derived.

The SLRs \Alslr, \Clslr, \Caslr, \Feslr, \Pdslr and \Pbslr are made by multiple components (see Figure \ref{fig:summary_15_SLRs}), both from pre-CCSN and explosive nucleosynthesis. This makes them unsuited as a diagnostic of the CCSN properties, as the impact of real physics properties, stellar uncertainties, and nuclear physics uncertainties are difficult to disentangle. \Csslr may also be included in this list of SLRs as the explosive contribution during explosive He-burning causes the contribution from component (6) to match that of component (5). This impact is dependant on the explosive energy of the model in question. 

We also include \Hfslr in this category of SLRs with limited diagnostic capability. In our models, we have seen that the CCSN production of this isotope dominates the pre-CCSN production due to neutron captures in (5) and (6). 
However, as we mentioned in Section \ref{ss:hf}, our models are calculated using an older stellar half-life of the branching point $^{181}$Hf, which reduces the s-process production of \Hfslr in pre-CCSN conditions \citep{lugaro:14}. In practice, by using the updated nuclear reaction rates the pre-CCSN production will increase, while the CCSN production will not be affected by such a modification and its effective contribution to the ejected yields will be reduced. New simulations are needed using the updated nuclear reaction to establish whether \Hfslr can be used a diagnostic of CCSN properties. 

\Mnslr is produced by the deepest components (1) - (4). Therefore, an efficient production is obtained for different progenitor masses and over a wide range of explosion energies. Faint supernova models with high fall-back mass will not eject these internal zones \citep[e.g.,][]{Takigawa2008,Nomoto2013}, and therefore will be \Mnslr-poor. In our stellar sample this is the case for three 25\msun models.
% [MP: I can see in the lower errorbar tail the outlying models with high fallback. Is this just 1 model, or you got more there?]. 

As we discussed in Section \ref{ss:p_iso}, the typical $\gamma$-process SLRs \Tcslrseven, \Tcslreight and \Smslr (by component 4) can also be made by neutron captures within component (6). Additionally, \Smslr is also made by component (5). In our models, for all progenitor masses and explosion setups, the $\gamma$-process production by component (4) is by far the most efficient for \Smslr. Therefore, we can consider \Smslr as the only safe diagnostic to track the contribution from the $\gamma$-process and this region of the ejecta.

Finally, according to our models, \Snslr is a convincing diagnostic of the high neutron densities reached during the CCSN explosion by the $^{22}$Ne($\alpha$,n)$^{25}$Mg activation within component (5) and (6). The production associated to the final neutron density is orders of magnitude larger than the pre-CCSN production, and the \Snslr abundance is extremely low in models with the weakest CCSN explosion energies. This is the case for all the progenitor masses. It is less clear that the same is true for \Islr. While for instance we can derive similar conclusions for the 25\msun models (Figure \ref{fig:long_abu_plot_25}), this is less clear for the 15\msun and 20\msun models (Figures \ref{fig:long_abu_plot_15} and \ref{fig:long_abu_plot_20}, respectively), where a significant pre-CCSN production is obtained in zone (5). A more detailed analysis of the \Islr production
and a careful handling of present nuclear uncertainties may be useful to clarify the diagnostic potential of this SLR isotope.

\subsection{Comparison with other studies} \label{ss:Comparison}

We compare our yields to other sets of CCSN models found in literature: \citet[][R02]{Rauscher2002}, \citet[][L18]{Limongi2018}, \citet[][S18]{Sieverding2018}, and \citet[][C19]{Curtis2019}. The models described in R02 are calculated using Kepler for both progenitor and explosion calculations, with an initial composition based on \cite{Anders1989} at Z=0.02. The models described in L18 are calculated using the FRANEC code with initial composition also based on \cite{Anders1989} at Z=0.02. The models described in S18 are calculated using the Kepler code \citep{Weaver1978} with an initial composition based on \cite{Lodders2003}. The explosion is simulated with a piston as described in \cite{Woosley1995}. The models described in C19 are calculated using the Kepler code for the progenitor \citep{Woosley2007} and the PUSH model for the explosion \citep[][]{Perego2015}. These models have an initial composition as described in \cite{Lodders2003}, at Z=0.013. We note that the C19 study considers only the innermost regions of the star when calculating the explosive nucleosynthesis. S18 includes all SLRs found in Table \ref{tab:ess_ref_table}, while R02, L18 and C19 do not, in particular: R02 does not include \Tcslreight; L18 does not include \Tcslrseven, \Tcslreight, \Pdslr, \Snslr, \Islr, \Smslr and \Hfslr; and C19 does not include \Clslr, \Nbslr, \Pdslr, \Snslr, \Islr, \Csslr, \Smslr, \Hfslr, and \Pbslr.

\begin{figure}
 \centering
 \includegraphics[width=\columnwidth]{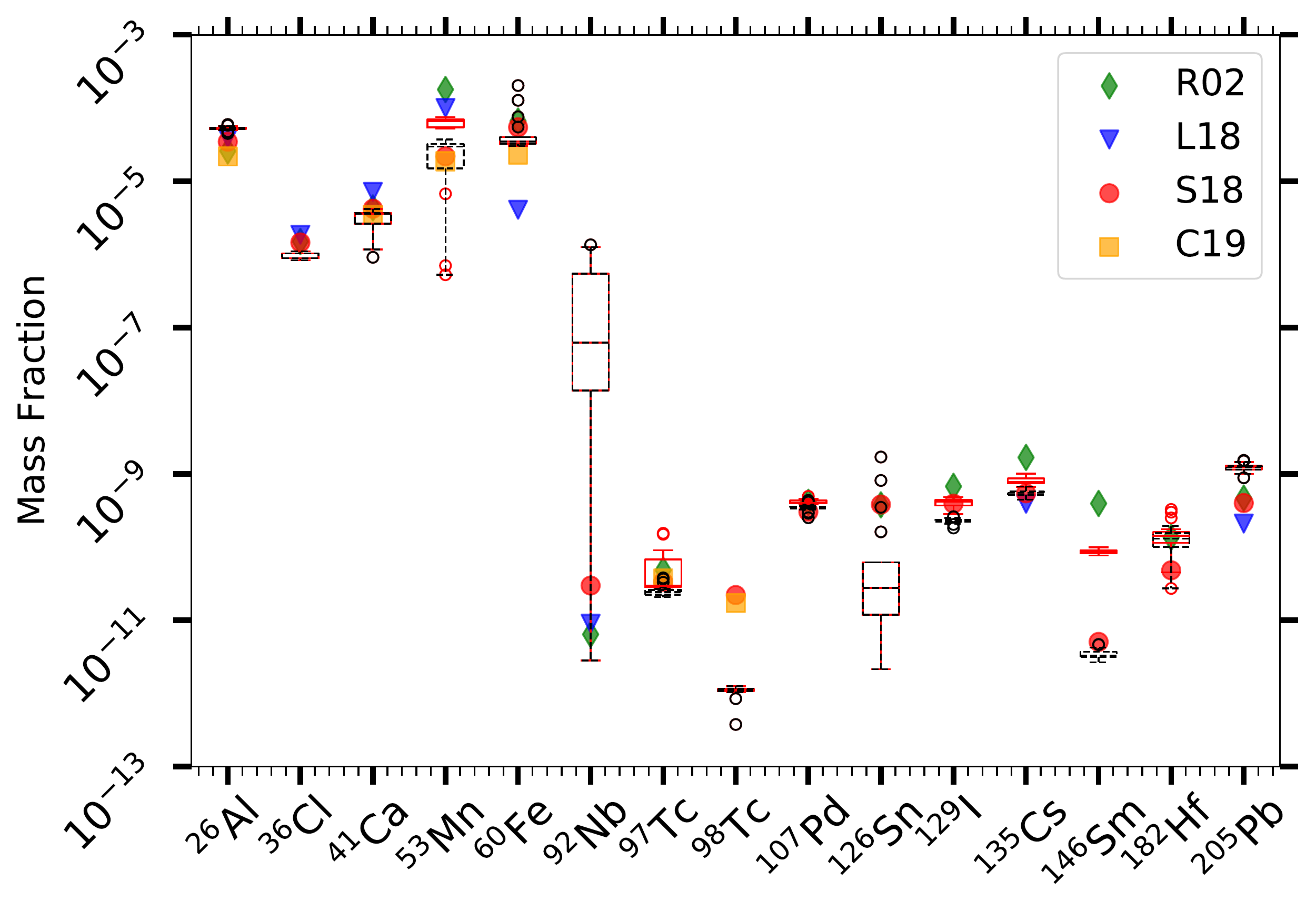}
 \includegraphics[width=\columnwidth]{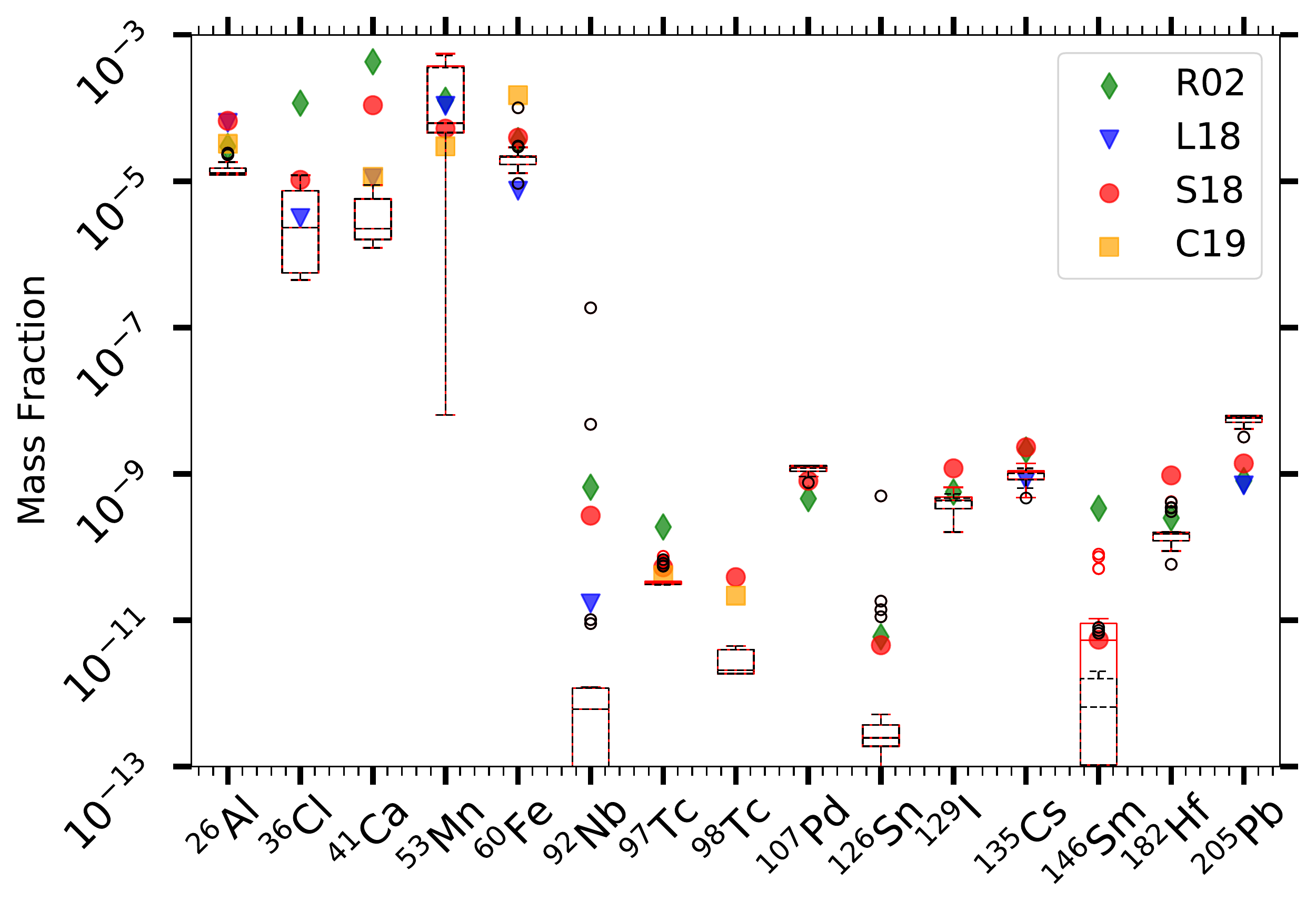}
 \includegraphics[width=\columnwidth]{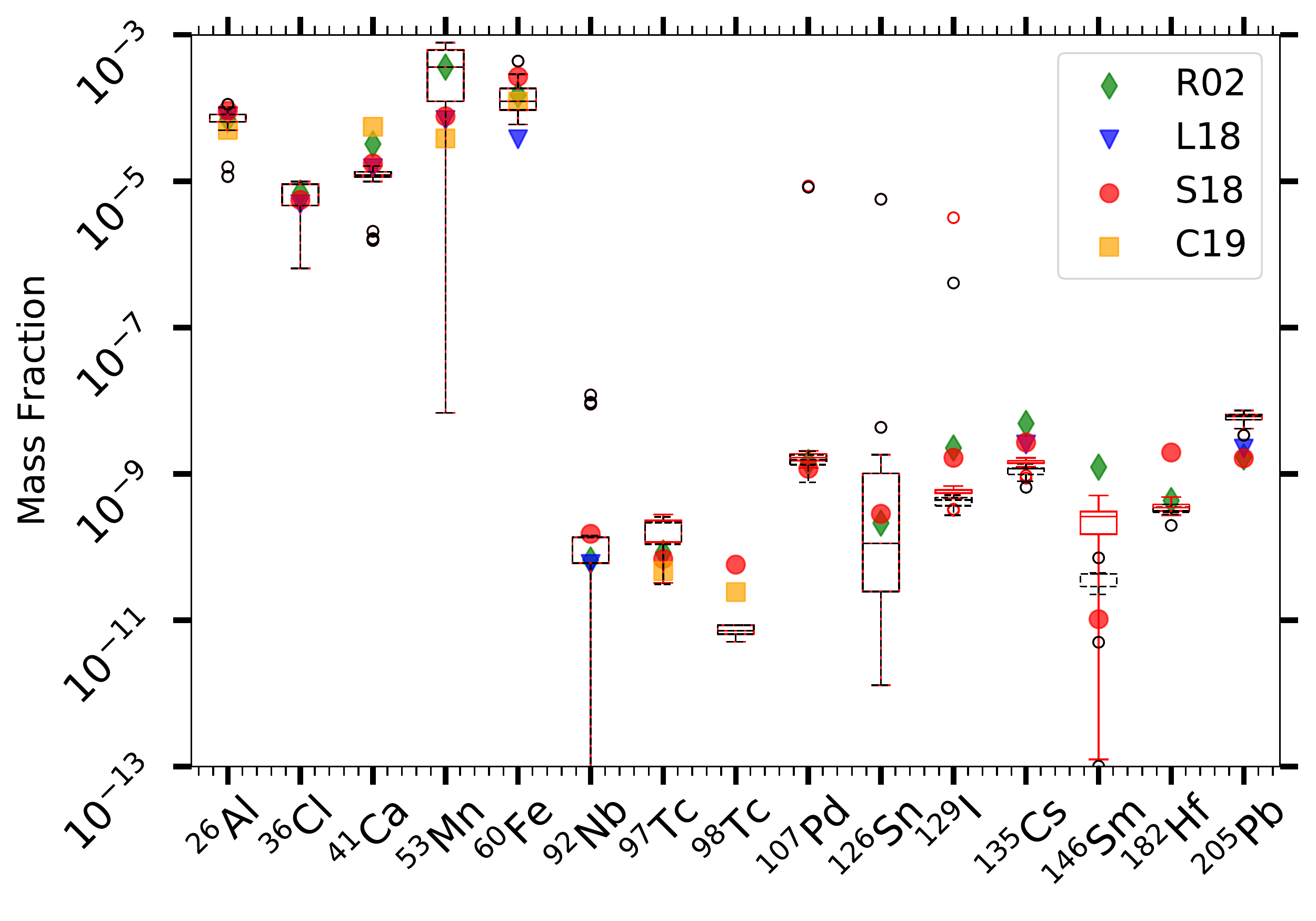}
 \caption{Comparison of all yield sets for each SLR isotope.
 Literature models plotted here are: \citet[][R02]{Rauscher2002}, \citet[][L18]{Limongi2018}, \citet[][S18]{Sieverding2018}, and \citet[][C19]{Curtis2019}.
 The upper panel shows the 15\stmass models, the middle panel the 20\stmass models and the lower panel the 25\stmass models. Yields presented in this work are shown as a two \textit{boxplots}, in red (yields including their respective radiogenic contributions) and in dashed black (yields not including radiogenic contributions). 
 As by the definition of a \textit{boxplot} the box size shows the range of values between the 25$^{\rm{th}}$ and the 75$^{\rm{th}}$ percentile of all the models in the dataset.
 The error bar on the boxplot denote the lower (upper) quartiles, and are given by subtracting (adding) to the position of the 25$^{\rm{th}}$ (75$^{\rm{th}}$) percentile 1.5 times the interquartile range (i.e., the box size).
%  , unless nonphysical values are obtained (e.g., negative abundances). In this case, the most extreme model is used as reference. 
 All values above (below) these thresholds are plotted as outliers.
%  Each quartile represents a quarter of the dataset.
 Outlying models outside the outer quartiles are shown as small circles for both decayed and undecayed results, using the same color scheme as the boxes.}
 \label{fig:comparison}
\end{figure}

In Figure \ref{fig:comparison} we compare the yields of the SLRs listed in Table \ref{tab:ess_ref_table} to the literature yields mentioned above. For each yield data set (R02, L18, S18, and C19) we compare the yields from the same initial mass. 
For completion we include the yields that include radiogenic contributions. %We show two versions of our yield sets. The first version, shown as a red box-plot, includes radiogenic contributions to the SLRs like in the rest of the paper. The second version, shown as a black box-plot, does not include radiogenic contributions, which corresponds to the yields of the other data sets.

We first examine the 15\stmass models (upper panel of Figure~\ref{fig:comparison}).
%The range within most of the SLR yields of all models is compact, resulting in a small(short?\marco{what?}) box plot for our models, as the explosion conditions of the innermost star are similar in all models. 
%\marco{[I do not know what compact means in this context.]}
In our calculations, \Mnslr, \Tcslrseven, \Islr, \Csslr, \Smslr and \Hfslr show a relevant or dominant radiogenic production, as already described in Section \ref{sec:results}. While, the other SLRs are mostly present already in the CCSN ejecta. The interquartile range (i.e., the range between the 25$^{th}$ and 75$^{th}$ percentiles) is limited within a factor of two for most of the isotopes. Exceptions are \Nbslr, \Tcslrseven and \Snslr. In particular, the largest interquartile range is obtained for \Nbslr, varying by of a factor of around two orders of magnitude.
The lower quartile range can be very large towards lower abundances, where the full range of outlying models are considered within this range (see figure caption for details). \Nbslr is again the most remarkable case in the figure, with an effective lower quartile range of about six orders of magnitude. \Nbslr is effectively produced only in the innermost region of the explosion ejecta. Therefore, the models that do not include this region will have drastically lower final yields.
In the figure, \Alslr, \Clslr, \Caslr, \Tcslrseven, \Pdslr, \Islr and \Hfslr show remarkable similarities by comparing our results with different yield sets from the literature. For other isotopes large discrepancies can be obtained.

All literature models that include \Nbslr show yields that lie within the lower quartile of our models, indicating a higher mass cut than many of our 15\stmass models. 
Our \Mnslr yields lie in between the high production obtained by R02 and L18, and the lower abundances by S18 and C19. However, S18 and C19 would still be well within the production range of some of our outlying models. The \Feslr yields are consistent for all models, except for L18 whose abundances are roughly one order of magnitude lower than all others. The lower \Feslr abundance in L18 compared to those found in our models is likely due to a reduced neutron burst. Our \Tcslreight yields are about factor of 20 smaller than S18 and C19. Our models also tend to produce less \Snslr compared to R02 and S18, although outlying models show similar (or even higher) production. The \Smslr yields vary by two orders of magnitude between R02 and S18. Our decayed yields are in between these values, but with low variances between models. Similarly, we obtain small variations between our models for \Pbslr, while considering all the models plotted the range of predictions shown is about one order of magnitude. Nuclear and stellar physics uncertainties of different origin can be the source of these discrepancies, in particular for cases where we see little variations in our set of CCSN models. 
% More detailed analysis is required in the future for these isotopes.
%\Snslr on the other hand is not produced in the inner regions, but further out during explosive He-burning. Compared to other explosive He-burning isotopes, \Snslr has a smaller inclusion form explosive C-burning. Therefore, a variance in neutron density during explosive He-burning due to different temperature gradients influences the final yield of \Snslr more than the other n-capture isotopes (e.g., \Islr). Explaining the range in yields for \Tcslreight and \Smslr is difficult as we do not have access to the details concerning the calculations of the other yield data sets. Explanations for the differences could be found in the initial abundances (for \Tcslreight) used or in partially decayed contributions (\Smslr).

The 20\stmass models are shown in the central panel of Figure~\ref{fig:comparison}. Only for \Smslr do we obtain a relevant radiogenic contribution, while this was the case for six SLRs in the 15\stmass models above. Boxes larger than a factor of two are obtained for \Clslr, \Caslr, \Mnslr, \Nbslr, \Tcslreight, and \Smslr. A large lower quartile range is also obtained for \Mnslr, due to the contribution of outlying models with low concentrations of this isotope.

When we compare all the model sets we find that \Alslr, \Caslr, \Feslr, \Tcslrseven, \Tcslreight, \Pdslr, \Islr, \Csslr, \Hfslr, and \Pbslr show a total range smaller than an order of magnitude. Instead, for \Clslr our results are consistent with L18 and S18, but they are much smaller than R02. R02 also provided much higher yields for \Caslr and \Nbslr, although for \Nbslr there are outlying models in our sample with much larger values and covering the full abundance range. Our upper interquartile limit for \Nbslr is one order of magnitude smaller than L18 and up to three order of magnitude smaller than R02. 
% [MP: It is interesting that most of our models has such a lot Nb92. Is it because of the weaker explosive nucleosynthesis triggered by the gamma-process?] 
As for the 15\msun models in the upper panel of the figure, most of our models show a smaller production of \Tcslreight and \Snslr by about an order of magnitude compared to S18 and C19. Finally, once the radiogenic contribution is taken into account, our \Smslr median is consistent with S18, but it is almost two orders of magnitude smaller than R02. 

The 25\stmass models are shown in the lower panel of Figure~\ref{fig:comparison}. We obtain relevant radiogenic contribution for \Islr, \Csslr and \Smslr, and we derived interquartile ranges larger than a factor of two only for \Mnslr, \Nbslr, \Snslr and \Smslr. 
Large lower quartile ranges are also obtained for the same SLRs and for \Nbslr. This is due to the efficient fallback of the 25\msun models in our stellar sample, where the products of e.g., explosive O-burning is often not ejected.
This explanation, however, would not apply to \Snslr. Indeed, this SLR is produced primarily in explosive He-burning (see Section \ref{ss:sn}). In this case, the large variance in the ejected yield can be explained by the position of the ignition point of this burning phase.
Comparing our 25\msun models to those in literature, we find overall a good consistency up to \Tcslrseven, with some significant variation for \Mnslr. We tend to produce less \Tcslreight, \Islr and \Csslr compared to other stellar sets, while for \Smslr most of our data are in between R02 and S18 (with about two order of magnitude of variation between the two models). 

% We compared our results with different sets of CCSN models available in the literature. 
In general, for SLRs lighter than \Feslr there is agreement within an order of magnitude between different 15\msun and 25\msun models. The 20\msun models show a much larger variation, with up to three orders of magnitude difference for \Caslr. For SLRs heavier than \Feslr, \Smslr shows the largest variation of about two orders of magnitude, varying from S18 up to R02 yields. Other isotope yields are consistent within an order of magnitude.
There are three exceptions: \Nbslr, \Tcslreight, and \Snslr. For all the progenitor masses considered, in our calculations we produce about an order of magnitude less \Tcslreight compared to other stellar sets. The same applies for \Snslr if we only consider the 15\msun and 20\msun stars, although in this case we have outlying models consistent or even exceeding other published results. \Nbslr abundances show the most variance in ejected yields, due to the extreme sensitivity to mass cut. If we use our median yield as a reference, the 15\msun models overproduce the SLR isotope by at least three orders of magnitude (even if we have outlying models with low \Nbslr abundance). The median of the 20\msun models is instead about an order of magnitude lower than the closest set (L18), while the 25 \msun models median and other sets are in good agreement.

% In summary, we find that the models presented in R02 generally have a larger yield of SLRs compared to our models, excluding \Pbslr and \Pdslr in our 15 and 20\stmass models, respectively.
% The models presented in L18 produce less \Feslr and \Pbslr across all masses, when compared to our models.
% The models of S18 generally produce a larger yield of SLRs, however \Pbslr is under-produced across all masses.
% C19 generally have higher yields of SLRs compared to our models, excluding \Mnslr across all masses. 
% Some stand out isotopes are \Nbslr and \Tcslreight. 
% Comparing our 15 and 20\stmass models to literature models we find that
% \Nbslr has higher (four orders of magnitude) and lower (between 1--3 orders of magnitude) yields, respectively. 
% \Tcslreight has much lower yields (roughly one order of magnitude) in our models compared to those found in literature.
% Comparing all masses of our models to literature models

%%%%%%%%%%%%%%%%%%%%%%%%%%%%%%
\section{Conclusions} 
\label{sec:Conc}

We presented the CCSN yields of 62 one-dimensional models for three progenitors with initial masses of 15, 20 and 25\msun and of solar metallicity \citep[Z=0.02,][]{Grevasse1993}. The impact on the nucleosynthesis of different explosion parameters is explored for each progenitor mass. This work is part of a set of studies focused on the production of radioactive isotopes in the same set of CCSNe models: \cite{Jones2019} analyzed the nucleosynthesis of \Feslr and \cite{Andrews2020} focused on the production of radioisotopes that are relevant for the next generation of $\gamma$-ray astronomical observations. 
% These are: $^{43}$K, $^{47}$Ca, $^{44}$Sc and $^{47}$Sc, $^{48}$V, $^{48}$Cr and $^{51}$Cr, $^{52}$Mn, $^{59}$Fe, $^{56}$Co and $^{57}$Co, and $^{57}$Ni. 
In this work, we study the production of short-lived radioactive isotopes (SLRs) that have been detected in the early Solar System (ESS), with half lives between 0.1 and 100~Myr.
% : \Alslr, \Clslr, \Caslr, \Mnslr, \Feslr, \Nbslr, \Tcslrseven, \Tcslreight, \Pdslr, \Snslr, \Islr, \Csslr, \Smslr, \Hfslr and \Pbslr. 
Note that compared to \cite{Jones2019} and \cite{Andrews2020}, calculations have been redone to correctly take into account the %presence updated handling of the 
\Alslr isomer during the explosion, resulting in an increase of the \Alslr yields compared to these previous data sets.

Using the CCSN models presented in this work, we explore for the first time in detail the nucleosynthesis for the fifteen SLRs both in the progenitor and during the CCSN explosion. In particular, we examine the main production and destruction sites for each isotope individually in the CCSN ejecta.
We have identified the most favourable stellar conditions to produce each SLR for the CCSN models considered in this work. In particular, several SLRs are co-produced under the same explosive nucleosynthesis conditions in the CCSN ejecta, as summarised in Fig.~\ref{fig:summary_15_SLRs}.
% For instance, ...[MP: Here, are there some SLRs that are produced in exact same conditions?]. 
% On the other hand, other SLRs share only part of their relevant nucleosynthesis conditions, and therefore more variations between theoretical models and in nature can be expected. For instance,... [MP: a couple of example here of SLR that have one site in common, but others not.] 

% We identify the SLRs that are potential diagnostic of the nature of the parent CCSN explosion when compared to early Solar System data: \marco{[MP: list here]} 
% %These sites are used to create a collection of diagnostic tools that can be used to aid in the identification of explosive yields when compared to observational data. 
% For example, the inclusion of large \Nbslr abundance in either a theoretical yield or observational data set may indicate a CCSN that has ejected its innermost material, and as such has a high explosion energy. \marco{[MP: I do not understand this sentence above. What does it mean?]}
%list some results.. certain SLRs are very sensitive to mass cut, others to expl energy, others seem very stable?

We compared our results with different sets of CCSN models available in the literature and find the following:
\begin{itemize}
    \item SLRs lighter than and including \Feslr are in agreement within an order of magnitude across all masses.
    % \item the 20\stmass moel shows the largest variance
    \item SLRs heavier than \Feslr are generally consistent within an order of magnitude, excluding \Nbslr, \Tcslreight, \Snslr, and \Smslr.
    \item Regarding \Nbslr, the 15\msun models overproduce the SLR isotope by three orders of magnitude in the 15\stmass models and by an order of magnitude in the 20\stmass models.
    \item We produce an order of magnitude less \Tcslreight than in other stellar sets.
    \item We produce an order of magnitude less \Snslr relative to other 15 and 20\stmass models.
    \item \cite{Rauscher2002} produces more \Smslr than our models (up to two orders of magnitude) and \cite{Sieverding2018} produces less than our models (up to an order of magnitude).
\end{itemize}
% In general, for SLRs lighter than \Feslr (included) there is agreement within an order of magnitude between different 15\msun and 25\msun models. The 20\msun models show a much larger variation, with up to three orders of magnitude difference for \Caslr. For SLRs heavier than \Feslr, \Smslr shows the largest variation of about two orders of magnitude, varying from S18 up to R02 yields. Other isotope yields are consistent within an order of magnitude.

% However there are three exceptions: \Nbslr, \Tcslreight and \Snslr. For all the progenitor masses considered, in our calculations we produce about an order of magnitude less \Tcslreight compared to other stellar sets. The same applies for \Snslr if we only consider the 15\msun and 20\msun stars, although in this case we have outlying models consistent or even exceeding other published results. \Nbslr abundances show the most variance in ejected yields, due to the extreme sensitivity to mass cut. If we use our median yield as a reference, the 15\msun models overproduce the SLR isotope by at least three orders of magnitude (even if we have outlying models with low \Nbslr abundance). The median of the 20\msun models is instead about an order of magnitude lower than the closest set (L18), while the 25 \msun models median and other sets are in good agreement.

Such variations in the production of SLRs in one-dimensional CCSN models can be understood for nuclei like \Nbslr and \Mnslr due to their creation in the inner regions of the CCSN ejecta, and the impact of assumptions made in the models and stellar uncertainties. Also, typical \gammaproc products like \Smslr and neutron-capture products like \Snslr show variations well above an order of magnitude.  Because our set of CCSN explosions cover a large parameter space, we were able to provide a first assessment of the impact of these differences.
% , once also outlying models are considered in the analysis. 
More detailed works are required in the future in order to understand and disentangle the source of these changes. In particular, for a number of SLRs (\Caslr, \Nbslr, \Tcslreight, \Snslr, and \Smslr) other stellar sets show yields higher than our range of ejected abundances. Other sources of variations need to be carefully taken into account in future works, like e.g., nuclear reaction rate uncertainties and the use of different stellar progenitors.
Future work will apply this broad nucleosynthesis analysis to observed ESS abundances, in order to determine if a single CCSN could populate the SLRs found in meteorites.

% In the next paper we will compare our results to the SLR abundances of the ESS in order to determine the CCSN parameters that best replicate the observed ESS abundances.

%These isotopes are all created in the inner regions of the CCSN ejecta, and therefore they are more affected from the assumptions and parametrization made in the models. %explosive shock, 
%and as such are sensitive to high mass cut values as is seen in most of the literature models examined.
%Future work includes matching the ESS values with our large set of models to determine what type of CCSN would best fit the observed ratios of our investigated SLRs vs their daughter isotope.

%%%%%%%%%%%%%%%%%%%%%%%%%%%%%%%%%%%%%%%%%%%%%%%%%%

\section*{Acknowledgements}

% We thank A. Sieverding for providing yields for 
This work is supported by the ERC Consolidator Grant (Hungary) programme (RADIOSTAR, G.A. n. 724560). TVL, MP, and BKG acknowledge the financial support of NuGrid/JINA-CEE (NSF Grant PHY-1430152) and STFC (through the University of Hull’s Consolidated Grant ST/R000840/1), and ongoing access to {\tt viper}, the University of Hull High Performance Computing Facility. MP acknowledges the support from the "Lend{\"u}let-2014" Programme of the Hungarian Academy of Sciences (Hungary). We thank the ChETEC COST Action (CA16117), supported by the European Cooperation in Science and Technology. This work was supported by the European Union’s Horizon 2020 research and innovation programme (ChETEC-INFRA -- Project no. 101008324), and the IReNA network supported by US NSF AccelNet.

\section*{Data availability}

The data underlying this article will be shared on reasonable request to the corresponding author.
% The data presented in \cite{Fryer2018} and \cite{,Andrews2020} can be found here: \hyperlink{https://ccsweb.lanl.gov/astro/nucleosynthesis/nucleosynthesis_astro.html}{https://ccsweb.lanl.gov/astro/nucleosynthesis/nucleosynthesis\_astro.html}. This does not include changes to the \Alslr yields, as outlined in Section \ref{ss:Methods}.


\bibliographystyle{mnras}
\bibliography{library} 

%%%%%%%%%%%%%%%%%%%%%%%%%%%%%%%%%%%%%%%%%%%%%%%%%%

%%%%%%%%%%%%%%%%% APPENDICES %%%%%%%%%%%%%%%%%%%%%
\newpage
\appendix

% \section{SLR production ratio Table for each model}

% \input{tables/Summary_data_table1.tex}

% \input{tables/Summary_data_table2.tex}

% \section{SLR yield Table for each model}
\section{Table of ejected yields}

\begin{table*}
\centering
\begin{tabular}{rrr|rrrrrr}
M$_{prog}$& E$_{exp}$& M$_{rem}$ & $^{26}$Al & $^{27}$Al & $^{36}$Cl & $^{35}$Cl & $^{41}$Ca & $^{40}$Ca \\ 
(M$_{\odot})$ & (\FoE) & (M$_{\odot})$  &  &  &  &  &  &  \\ 
\hline
15 & 1.90 & 1.62  & 5.391$\times$10$^{-5}$ & 5.533$\times$10$^{-3}$ & 8.916$\times$10$^{-7}$ & 1.169$\times$10$^{-4}$ & 3.649$\times$10$^{-6}$ & 4.658$\times$10$^{-3}$ \\ 
15 & 1.86 & 1.63  & 5.396$\times$10$^{-5}$ & 5.534$\times$10$^{-3}$ & 8.917$\times$10$^{-7}$ & 1.167$\times$10$^{-4}$ & 3.631$\times$10$^{-6}$ & 4.658$\times$10$^{-3}$ \\ 
15 & 1.94 & 1.61  & 5.387$\times$10$^{-5}$ & 5.533$\times$10$^{-3}$ & 8.920$\times$10$^{-7}$ & 1.171$\times$10$^{-4}$ & 3.687$\times$10$^{-6}$ & 4.659$\times$10$^{-3}$ \\ 
15 & 0.54 & 1.91  & 5.813$\times$10$^{-5}$ & 5.733$\times$10$^{-3}$ & 8.689$\times$10$^{-7}$ & 9.238$\times$10$^{-5}$ & 1.170$\times$10$^{-6}$ & 6.427$\times$10$^{-4}$ \\ 
15 & 0.92 & 1.75  & 5.334$\times$10$^{-5}$ & 5.657$\times$10$^{-3}$ & 1.059$\times$10$^{-6}$ & 1.273$\times$10$^{-4}$ & 2.636$\times$10$^{-6}$ & 4.484$\times$10$^{-3}$ \\ 
15 & 1.69 & 1.52  & 5.525$\times$10$^{-5}$ & 5.632$\times$10$^{-3}$ & 9.117$\times$10$^{-7}$ & 1.184$\times$10$^{-4}$ & 3.612$\times$10$^{-6}$ & 4.627$\times$10$^{-3}$ \\ 
15 & 0.34 & 1.94  & 5.958$\times$10$^{-5}$ & 5.772$\times$10$^{-3}$ & 8.348$\times$10$^{-7}$ & 8.395$\times$10$^{-5}$ & 9.121$\times$10$^{-7}$ & 5.997$\times$10$^{-4}$ \\ 
15 & 3.43 & 1.51  & 5.147$\times$10$^{-5}$ & 5.605$\times$10$^{-3}$ & 1.046$\times$10$^{-6}$ & 1.277$\times$10$^{-4}$ & 3.293$\times$10$^{-6}$ & 4.680$\times$10$^{-3}$ \\ 
15 & 2.06 & 1.59  & 5.382$\times$10$^{-5}$ & 5.531$\times$10$^{-3}$ & 8.918$\times$10$^{-7}$ & 1.170$\times$10$^{-4}$ & 3.724$\times$10$^{-6}$ & 4.661$\times$10$^{-3}$ \\ 
15 & 2.24 & 1.56  & 5.372$\times$10$^{-5}$ & 5.529$\times$10$^{-3}$ & 8.910$\times$10$^{-7}$ & 1.166$\times$10$^{-4}$ & 3.727$\times$10$^{-6}$ & 4.660$\times$10$^{-3}$ \\ 
15 & 2.60 & 1.52  & 5.388$\times$10$^{-5}$ & 5.527$\times$10$^{-3}$ & 8.837$\times$10$^{-7}$ & 1.152$\times$10$^{-4}$ & 3.666$\times$10$^{-6}$ & 4.677$\times$10$^{-3}$ \\ 
15 & 0.74 & 1.73  & 5.268$\times$10$^{-5}$ & 5.672$\times$10$^{-3}$ & 1.093$\times$10$^{-6}$ & 1.328$\times$10$^{-4}$ & 2.626$\times$10$^{-6}$ & 4.521$\times$10$^{-3}$ \\ 
15 & 0.82 & 1.88  & 5.666$\times$10$^{-5}$ & 5.665$\times$10$^{-3}$ & 8.818$\times$10$^{-7}$ & 1.090$\times$10$^{-4}$ & 2.544$\times$10$^{-6}$ & 1.599$\times$10$^{-3}$ \\ 
15 & 10.7 & 1.53  & 4.487$\times$10$^{-5}$ & 5.106$\times$10$^{-3}$ & 9.929$\times$10$^{-7}$ & 1.251$\times$10$^{-4}$ & 4.193$\times$10$^{-6}$ & 4.668$\times$10$^{-3}$ \\ 
15 & 2.47 & 1.52  & 4.999$\times$10$^{-5}$ & 5.354$\times$10$^{-3}$ & 9.239$\times$10$^{-7}$ & 1.205$\times$10$^{-4}$ & 3.523$\times$10$^{-6}$ & 4.904$\times$10$^{-3}$ \\ 
15 & 4.79 & 1.50  & 4.766$\times$10$^{-5}$ & 5.477$\times$10$^{-3}$ & 1.098$\times$10$^{-6}$ & 1.329$\times$10$^{-4}$ & 3.231$\times$10$^{-6}$ & 4.863$\times$10$^{-3}$ \\ 
15 & 2.63 & 1.53  & 4.740$\times$10$^{-5}$ & 5.335$\times$10$^{-3}$ & 1.032$\times$10$^{-6}$ & 1.279$\times$10$^{-4}$ & 3.714$\times$10$^{-6}$ & 4.913$\times$10$^{-3}$ \\ 
20 & 2.85 & 1.74  & 1.426$\times$10$^{-5}$ & 6.682$\times$10$^{-3}$ & 1.106$\times$10$^{-5}$ & 1.029$\times$10$^{-3}$ & 5.912$\times$10$^{-6}$ & 2.834$\times$10$^{-3}$ \\ 
20 & 1.47 & 2.23  & 1.303$\times$10$^{-5}$ & 6.081$\times$10$^{-3}$ & 6.083$\times$10$^{-7}$ & 7.007$\times$10$^{-5}$ & 1.842$\times$10$^{-6}$ & 7.157$\times$10$^{-4}$ \\ 
20 & 2.50 & 1.93  & 1.831$\times$10$^{-5}$ & 6.970$\times$10$^{-3}$ & 1.161$\times$10$^{-5}$ & 1.148$\times$10$^{-3}$ & 2.442$\times$10$^{-6}$ & 7.286$\times$10$^{-4}$ \\ 
20 & 5.03 & 1.74  & 2.305$\times$10$^{-5}$ & 7.163$\times$10$^{-3}$ & 3.460$\times$10$^{-6}$ & 3.112$\times$10$^{-4}$ & 7.032$\times$10$^{-6}$ & 3.253$\times$10$^{-3}$ \\ 
20 & 2.76 & 1.76  & 1.473$\times$10$^{-5}$ & 6.740$\times$10$^{-3}$ & 1.190$\times$10$^{-5}$ & 1.073$\times$10$^{-3}$ & 6.048$\times$10$^{-6}$ & 2.783$\times$10$^{-3}$ \\ 
20 & 2.60 & 1.90  & 1.533$\times$10$^{-5}$ & 6.788$\times$10$^{-3}$ & 1.211$\times$10$^{-5}$ & 1.024$\times$10$^{-3}$ & 5.260$\times$10$^{-6}$ & 1.004$\times$10$^{-3}$ \\ 
20 & 2.43 & 1.86  & 1.438$\times$10$^{-5}$ & 6.684$\times$10$^{-3}$ & 1.102$\times$10$^{-5}$ & 1.017$\times$10$^{-3}$ & 5.502$\times$10$^{-6}$ & 1.085$\times$10$^{-3}$ \\ 
20 & 1.04 & 2.47  & 1.225$\times$10$^{-5}$ & 5.361$\times$10$^{-3}$ & 5.697$\times$10$^{-7}$ & 6.916$\times$10$^{-5}$ & 1.680$\times$10$^{-6}$ & 7.102$\times$10$^{-4}$ \\ 
20 & 4.15 & 1.85  & 2.301$\times$10$^{-5}$ & 7.134$\times$10$^{-3}$ & 3.448$\times$10$^{-6}$ & 3.091$\times$10$^{-4}$ & 6.626$\times$10$^{-6}$ & 3.069$\times$10$^{-3}$ \\ 
20 & 0.65 & 3.03  & 1.221$\times$10$^{-5}$ & 3.707$\times$10$^{-3}$ & 5.808$\times$10$^{-7}$ & 6.833$\times$10$^{-5}$ & 1.402$\times$10$^{-6}$ & 6.977$\times$10$^{-4}$ \\ 
20 & 0.78 & 2.85  & 1.222$\times$10$^{-5}$ & 4.238$\times$10$^{-3}$ & 5.111$\times$10$^{-7}$ & 6.773$\times$10$^{-5}$ & 1.490$\times$10$^{-6}$ & 7.017$\times$10$^{-4}$ \\ 
20 & 0.84 & 2.62  & 1.224$\times$10$^{-5}$ & 4.916$\times$10$^{-3}$ & 5.457$\times$10$^{-7}$ & 6.860$\times$10$^{-5}$ & 1.603$\times$10$^{-6}$ & 7.068$\times$10$^{-4}$ \\ 
20 & 1.00 & 2.35  & 1.228$\times$10$^{-5}$ & 5.716$\times$10$^{-3}$ & 5.893$\times$10$^{-7}$ & 6.965$\times$10$^{-5}$ & 1.745$\times$10$^{-6}$ & 7.129$\times$10$^{-4}$ \\ 
20 & 1.65 & 1.78  & 1.285$\times$10$^{-5}$ & 6.629$\times$10$^{-3}$ & 7.647$\times$10$^{-6}$ & 8.548$\times$10$^{-4}$ & 4.864$\times$10$^{-6}$ & 1.018$\times$10$^{-3}$ \\ 
20 & 0.75 & 2.76  & 1.223$\times$10$^{-5}$ & 4.504$\times$10$^{-3}$ & 5.257$\times$10$^{-7}$ & 6.815$\times$10$^{-5}$ & 1.534$\times$10$^{-6}$ & 7.037$\times$10$^{-4}$ \\ 
20 & 0.53 & 3.40  & 1.219$\times$10$^{-5}$ & 2.617$\times$10$^{-3}$ & 4.463$\times$10$^{-7}$ & 6.843$\times$10$^{-5}$ & 1.224$\times$10$^{-6}$ & 6.896$\times$10$^{-4}$ \\ 
20 & 4.33 & 1.87  & 2.344$\times$10$^{-5}$ & 7.186$\times$10$^{-3}$ & 3.148$\times$10$^{-6}$ & 3.176$\times$10$^{-4}$ & 6.677$\times$10$^{-6}$ & 2.318$\times$10$^{-3}$ \\ 
20 & 0.81 & 2.70  & 1.223$\times$10$^{-5}$ & 4.685$\times$10$^{-3}$ & 5.343$\times$10$^{-7}$ & 6.837$\times$10$^{-5}$ & 1.565$\times$10$^{-6}$ & 7.051$\times$10$^{-4}$ \\ 
20 & 1.19 & 2.28  & 1.232$\times$10$^{-5}$ & 5.923$\times$10$^{-3}$ & 6.016$\times$10$^{-7}$ & 6.996$\times$10$^{-5}$ & 1.785$\times$10$^{-6}$ & 7.145$\times$10$^{-4}$ \\ 
20 & 1.52 & 1.97  & 1.294$\times$10$^{-5}$ & 6.605$\times$10$^{-3}$ & 7.173$\times$10$^{-6}$ & 7.646$\times$10$^{-4}$ & 2.253$\times$10$^{-6}$ & 7.268$\times$10$^{-4}$ \\ 
20 & 1.39 & 1.93  & 1.287$\times$10$^{-5}$ & 6.613$\times$10$^{-3}$ & 6.694$\times$10$^{-6}$ & 7.309$\times$10$^{-4}$ & 2.342$\times$10$^{-6}$ & 7.296$\times$10$^{-4}$ \\ 
20 & 8.86 & 1.74  & 2.426$\times$10$^{-5}$ & 7.100$\times$10$^{-3}$ & 2.324$\times$10$^{-6}$ & 3.280$\times$10$^{-4}$ & 8.908$\times$10$^{-6}$ & 3.730$\times$10$^{-3}$ \\ 
20 & 0.85 & 2.62  & 1.224$\times$10$^{-5}$ & 4.918$\times$10$^{-3}$ & 5.459$\times$10$^{-7}$ & 6.860$\times$10$^{-5}$ & 1.604$\times$10$^{-6}$ & 7.068$\times$10$^{-4}$ \\ 
25 & 1.92 & 3.13  & 8.203$\times$10$^{-5}$ & 1.588$\times$10$^{-2}$ & 9.113$\times$10$^{-6}$ & 6.555$\times$10$^{-4}$ & 1.278$\times$10$^{-5}$ & 5.735$\times$10$^{-3}$ \\ 
25 & 3.07 & 1.83  & 8.203$\times$10$^{-5}$ & 1.588$\times$10$^{-2}$ & 9.113$\times$10$^{-6}$ & 6.556$\times$10$^{-4}$ & 1.279$\times$10$^{-5}$ & 5.736$\times$10$^{-3}$ \\ 
25 & 3.30 & 2.35  & 8.144$\times$10$^{-5}$ & 1.582$\times$10$^{-2}$ & 9.157$\times$10$^{-6}$ & 6.399$\times$10$^{-4}$ & 1.201$\times$10$^{-5}$ & 1.937$\times$10$^{-3}$ \\ 
25 & 1.04 & 1.84  & 6.468$\times$10$^{-5}$ & 1.433$\times$10$^{-2}$ & 9.909$\times$10$^{-6}$ & 6.003$\times$10$^{-4}$ & 1.342$\times$10$^{-5}$ & 4.121$\times$10$^{-3}$ \\ 
25 & 9.73 & 2.35  & 8.052$\times$10$^{-5}$ & 1.566$\times$10$^{-2}$ & 8.701$\times$10$^{-6}$ & 6.538$\times$10$^{-4}$ & 1.157$\times$10$^{-5}$ & 1.615$\times$10$^{-3}$ \\ 
25 & 1.20 & 1.84  & 6.468$\times$10$^{-5}$ & 1.435$\times$10$^{-2}$ & 9.909$\times$10$^{-6}$ & 6.003$\times$10$^{-4}$ & 1.342$\times$10$^{-5}$ & 4.121$\times$10$^{-3}$ \\ 
25 & 7.42 & 2.37  & 1.025$\times$10$^{-4}$ & 1.550$\times$10$^{-2}$ & 3.331$\times$10$^{-6}$ & 4.503$\times$10$^{-4}$ & 1.377$\times$10$^{-5}$ & 9.066$\times$10$^{-3}$ \\ 
25 & 18.4 & 2.35  & 8.176$\times$10$^{-5}$ & 1.580$\times$10$^{-2}$ & 8.754$\times$10$^{-6}$ & 6.537$\times$10$^{-4}$ & 1.145$\times$10$^{-5}$ & 1.596$\times$10$^{-3}$ \\ 
25 & 14.8 & 2.35  & 1.124$\times$10$^{-4}$ & 1.753$\times$10$^{-2}$ & 8.768$\times$10$^{-6}$ & 8.551$\times$10$^{-4}$ & 1.610$\times$10$^{-5}$ & 9.571$\times$10$^{-3}$ \\ 
25 & 2.78 & 2.35  & 8.145$\times$10$^{-5}$ & 1.583$\times$10$^{-2}$ & 9.157$\times$10$^{-6}$ & 6.399$\times$10$^{-4}$ & 1.201$\times$10$^{-5}$ & 1.937$\times$10$^{-3}$ \\ 
25 & 0.89 & 4.66  & 1.161$\times$10$^{-5}$ & 4.534$\times$10$^{-3}$ & 6.913$\times$10$^{-7}$ & 6.005$\times$10$^{-5}$ & 1.641$\times$10$^{-6}$ & 3.969$\times$10$^{-4}$ \\ 
25 & 7.08 & 2.35  & 8.141$\times$10$^{-5}$ & 1.581$\times$10$^{-2}$ & 9.157$\times$10$^{-6}$ & 6.399$\times$10$^{-4}$ & 1.201$\times$10$^{-5}$ & 1.937$\times$10$^{-3}$ \\ 
25 & 0.92 & 1.84  & 6.468$\times$10$^{-5}$ & 1.435$\times$10$^{-2}$ & 9.909$\times$10$^{-6}$ & 6.003$\times$10$^{-4}$ & 1.342$\times$10$^{-5}$ & 4.121$\times$10$^{-3}$ \\ 
25 & 2.64 & 2.35  & 8.145$\times$10$^{-5}$ & 1.582$\times$10$^{-2}$ & 9.157$\times$10$^{-6}$ & 6.399$\times$10$^{-4}$ & 1.201$\times$10$^{-5}$ & 1.937$\times$10$^{-3}$ \\ 
25 & 6.17 & 2.38  & 9.464$\times$10$^{-5}$ & 1.520$\times$10$^{-2}$ & 3.243$\times$10$^{-6}$ & 4.443$\times$10$^{-4}$ & 1.371$\times$10$^{-5}$ & 8.928$\times$10$^{-3}$ \\ 
25 & 4.73 & 2.38  & 4.979$\times$10$^{-5}$ & 1.211$\times$10$^{-2}$ & 2.616$\times$10$^{-6}$ & 3.302$\times$10$^{-4}$ & 1.138$\times$10$^{-5}$ & 7.712$\times$10$^{-3}$ \\ 
25 & 1.52 & 1.83  & 6.468$\times$10$^{-5}$ & 1.435$\times$10$^{-2}$ & 9.909$\times$10$^{-6}$ & 6.003$\times$10$^{-4}$ & 1.342$\times$10$^{-5}$ & 4.121$\times$10$^{-3}$ \\ 
25 & 1.57 & 3.73  & 1.563$\times$10$^{-5}$ & 7.618$\times$10$^{-3}$ & 6.514$\times$10$^{-7}$ & 4.398$\times$10$^{-5}$ & 2.066$\times$10$^{-6}$ & 4.145$\times$10$^{-4}$ \\ 
25 & 0.99 & 4.89  & 1.158$\times$10$^{-5}$ & 3.790$\times$10$^{-3}$ & 6.436$\times$10$^{-7}$ & 5.823$\times$10$^{-5}$ & 1.549$\times$10$^{-6}$ & 3.922$\times$10$^{-4}$ \\ 
25 & 8.40 & 2.38  & 8.034$\times$10$^{-5}$ & 1.567$\times$10$^{-2}$ & 8.630$\times$10$^{-6}$ & 6.139$\times$10$^{-4}$ & 9.901$\times$10$^{-6}$ & 1.312$\times$10$^{-3}$ \\ 
25 & 2.53 & 2.35  & 8.145$\times$10$^{-5}$ & 1.583$\times$10$^{-2}$ & 9.157$\times$10$^{-6}$ & 6.399$\times$10$^{-4}$ & 1.201$\times$10$^{-5}$ & 1.937$\times$10$^{-3}$ \\ 
25 & 4.72 & 2.35  & 8.143$\times$10$^{-5}$ & 1.581$\times$10$^{-2}$ & 9.157$\times$10$^{-6}$ & 6.399$\times$10$^{-4}$ & 1.201$\times$10$^{-5}$ & 1.937$\times$10$^{-3}$ \\ 
\end{tabular}
\caption{Complete table of decayed isotopic yields for all models in mass fraction. M$_{prog}$ is the progenitor mass, E$_{exp}$ is the explosion energy, M$_{rem}$ is the remnant mass. 
% \textcolor{red}{[Can change the sig figs in these to be smaller, I assume the online version will have a .CSV table to read easier, for now I leave as is]}
}
\label{yield_table:1}
\end{table*}

\begin{table*}
\centering
\begin{tabular}{rrr|rrrrrr}
M$_{prog}$& E$_{exp}$& M$_{rem}$ & $^{53}$Mn & $^{55}$Mn & $^{60}$Fe & $^{56}$Fe & $^{92}$Nb & $^{92}$Mo \\ 
(M$_{\odot})$ & (\FoE) & (M$_{\odot})$  &  &  &  &  &  &  \\ 
\hline
15 & 1.90 & 1.62  & 6.915$\times$10$^{-5}$ & 3.118$\times$10$^{-4}$ & 3.426$\times$10$^{-5}$ & 5.139$\times$10$^{-2}$ & 5.466$\times$10$^{-8}$ & 3.856$\times$10$^{-6}$ \\ 
15 & 1.86 & 1.63  & 6.908$\times$10$^{-5}$ & 3.144$\times$10$^{-4}$ & 3.339$\times$10$^{-5}$ & 5.140$\times$10$^{-2}$ & 6.232$\times$10$^{-8}$ & 3.293$\times$10$^{-6}$ \\ 
15 & 1.94 & 1.61  & 6.937$\times$10$^{-5}$ & 3.118$\times$10$^{-4}$ & 3.488$\times$10$^{-5}$ & 5.133$\times$10$^{-2}$ & 5.133$\times$10$^{-8}$ & 4.230$\times$10$^{-6}$ \\ 
15 & 0.54 & 1.91  & 7.023$\times$10$^{-7}$ & 1.183$\times$10$^{-4}$ & 3.021$\times$10$^{-5}$ & 1.188$\times$10$^{-2}$ & 2.821$\times$10$^{-12}$ & 8.406$\times$10$^{-9}$ \\ 
15 & 0.92 & 1.75  & 5.459$\times$10$^{-5}$ & 3.309$\times$10$^{-4}$ & 3.318$\times$10$^{-5}$ & 4.090$\times$10$^{-2}$ & 1.383$\times$10$^{-8}$ & 2.217$\times$10$^{-6}$ \\ 
15 & 1.69 & 1.52  & 7.376$\times$10$^{-5}$ & 3.163$\times$10$^{-4}$ & 3.548$\times$10$^{-5}$ & 5.069$\times$10$^{-2}$ & 8.207$\times$10$^{-7}$ & 8.583$\times$10$^{-5}$ \\ 
15 & 0.34 & 1.94  & 5.264$\times$10$^{-7}$ & 1.209$\times$10$^{-4}$ & 3.032$\times$10$^{-5}$ & 1.175$\times$10$^{-2}$ & 2.819$\times$10$^{-12}$ & 8.429$\times$10$^{-9}$ \\ 
15 & 3.43 & 1.51  & 6.329$\times$10$^{-5}$ & 3.158$\times$10$^{-4}$ & 1.275$\times$10$^{-4}$ & 5.146$\times$10$^{-2}$ & 5.548$\times$10$^{-7}$ & 7.983$\times$10$^{-5}$ \\ 
15 & 2.06 & 1.59  & 7.267$\times$10$^{-5}$ & 3.118$\times$10$^{-4}$ & 3.625$\times$10$^{-5}$ & 5.110$\times$10$^{-2}$ & 3.600$\times$10$^{-7}$ & 4.755$\times$10$^{-5}$ \\ 
15 & 2.24 & 1.56  & 7.536$\times$10$^{-5}$ & 3.060$\times$10$^{-4}$ & 4.002$\times$10$^{-5}$ & 5.087$\times$10$^{-2}$ & 1.255$\times$10$^{-6}$ & 1.820$\times$10$^{-4}$ \\ 
15 & 2.60 & 1.52  & 7.369$\times$10$^{-5}$ & 3.058$\times$10$^{-4}$ & 5.501$\times$10$^{-5}$ & 5.064$\times$10$^{-2}$ & 1.351$\times$10$^{-6}$ & 2.302$\times$10$^{-4}$ \\ 
15 & 0.74 & 1.73  & 5.220$\times$10$^{-5}$ & 3.358$\times$10$^{-4}$ & 3.232$\times$10$^{-5}$ & 4.741$\times$10$^{-2}$ & 1.178$\times$10$^{-8}$ & 2.420$\times$10$^{-6}$ \\ 
15 & 0.82 & 1.88  & 6.729$\times$10$^{-6}$ & 1.325$\times$10$^{-4}$ & 3.043$\times$10$^{-5}$ & 1.197$\times$10$^{-2}$ & 2.933$\times$10$^{-12}$ & 8.360$\times$10$^{-9}$ \\ 
15 & 10.7 & 1.53  & 6.301$\times$10$^{-5}$ & 3.088$\times$10$^{-4}$ & 3.087$\times$10$^{-5}$ & 5.236$\times$10$^{-2}$ & 3.969$\times$10$^{-7}$ & 4.268$\times$10$^{-5}$ \\ 
15 & 2.47 & 1.52  & 6.653$\times$10$^{-5}$ & 3.345$\times$10$^{-4}$ & 3.685$\times$10$^{-5}$ & 5.327$\times$10$^{-2}$ & 6.125$\times$10$^{-8}$ & 2.264$\times$10$^{-6}$ \\ 
15 & 4.79 & 1.50  & 5.896$\times$10$^{-5}$ & 3.184$\times$10$^{-4}$ & 2.030$\times$10$^{-4}$ & 5.325$\times$10$^{-2}$ & 5.236$\times$10$^{-7}$ & 1.051$\times$10$^{-4}$ \\ 
15 & 2.63 & 1.53  & 6.964$\times$10$^{-5}$ & 3.268$\times$10$^{-4}$ & 7.601$\times$10$^{-5}$ & 5.279$\times$10$^{-2}$ & 5.454$\times$10$^{-7}$ & 6.486$\times$10$^{-5}$ \\ 
20 & 2.85 & 1.74  & 4.955$\times$10$^{-4}$ & 6.749$\times$10$^{-4}$ & 2.139$\times$10$^{-5}$ & 2.600$\times$10$^{-2}$ & 1.145$\times$10$^{-12}$ & 9.238$\times$10$^{-9}$ \\ 
20 & 1.47 & 2.23  & 6.496$\times$10$^{-5}$ & 3.043$\times$10$^{-4}$ & 2.093$\times$10$^{-5}$ & 1.318$\times$10$^{-2}$ & 8.751$\times$10$^{-15}$ & 8.972$\times$10$^{-9}$ \\ 
20 & 2.50 & 1.93  & 5.005$\times$10$^{-6}$ & 2.319$\times$10$^{-4}$ & 2.201$\times$10$^{-5}$ & 1.373$\times$10$^{-2}$ & 1.019$\times$10$^{-12}$ & 9.226$\times$10$^{-9}$ \\ 
20 & 5.03 & 1.74  & 4.616$\times$10$^{-4}$ & 6.219$\times$10$^{-4}$ & 3.025$\times$10$^{-5}$ & 2.914$\times$10$^{-2}$ & 4.756$\times$10$^{-9}$ & 2.087$\times$10$^{-7}$ \\ 
20 & 2.76 & 1.76  & 5.022$\times$10$^{-4}$ & 6.626$\times$10$^{-4}$ & 2.163$\times$10$^{-5}$ & 2.443$\times$10$^{-2}$ & 1.225$\times$10$^{-12}$ & 9.242$\times$10$^{-9}$ \\ 
20 & 2.60 & 1.90  & 5.537$\times$10$^{-5}$ & 1.792$\times$10$^{-4}$ & 2.217$\times$10$^{-5}$ & 1.590$\times$10$^{-2}$ & 1.222$\times$10$^{-12}$ & 9.235$\times$10$^{-9}$ \\ 
20 & 2.43 & 1.86  & 2.732$\times$10$^{-4}$ & 2.909$\times$10$^{-4}$ & 2.139$\times$10$^{-5}$ & 1.931$\times$10$^{-2}$ & 1.140$\times$10$^{-12}$ & 9.245$\times$10$^{-9}$ \\ 
20 & 1.04 & 2.47  & 6.232$\times$10$^{-5}$ & 2.961$\times$10$^{-4}$ & 1.840$\times$10$^{-5}$ & 1.340$\times$10$^{-2}$ & 6.253$\times$10$^{-15}$ & 8.972$\times$10$^{-9}$ \\ 
20 & 4.15 & 1.85  & 4.640$\times$10$^{-4}$ & 6.686$\times$10$^{-4}$ & 2.895$\times$10$^{-5}$ & 2.362$\times$10$^{-2}$ & 8.999$\times$10$^{-12}$ & 9.225$\times$10$^{-9}$ \\ 
20 & 0.65 & 3.03  & 3.575$\times$10$^{-7}$ & 1.130$\times$10$^{-4}$ & 1.287$\times$10$^{-5}$ & 1.248$\times$10$^{-2}$ & 4.025$\times$10$^{-15}$ & 8.972$\times$10$^{-9}$ \\ 
20 & 0.78 & 2.85  & 4.169$\times$10$^{-5}$ & 2.234$\times$10$^{-4}$ & 1.461$\times$10$^{-5}$ & 1.376$\times$10$^{-2}$ & 4.730$\times$10$^{-15}$ & 8.972$\times$10$^{-9}$ \\ 
20 & 0.84 & 2.62  & 5.270$\times$10$^{-5}$ & 2.706$\times$10$^{-4}$ & 1.687$\times$10$^{-5}$ & 1.349$\times$10$^{-2}$ & 5.631$\times$10$^{-15}$ & 8.972$\times$10$^{-9}$ \\ 
20 & 1.00 & 2.35  & 6.696$\times$10$^{-5}$ & 3.052$\times$10$^{-4}$ & 1.962$\times$10$^{-5}$ & 1.328$\times$10$^{-2}$ & 6.741$\times$10$^{-15}$ & 8.972$\times$10$^{-9}$ \\ 
20 & 1.65 & 1.78  & 2.798$\times$10$^{-4}$ & 2.877$\times$10$^{-4}$ & 2.132$\times$10$^{-5}$ & 1.964$\times$10$^{-2}$ & 6.240$\times$10$^{-13}$ & 9.326$\times$10$^{-9}$ \\ 
20 & 0.75 & 2.76  & 5.051$\times$10$^{-5}$ & 2.495$\times$10$^{-4}$ & 1.548$\times$10$^{-5}$ & 1.366$\times$10$^{-2}$ & 5.083$\times$10$^{-15}$ & 8.972$\times$10$^{-9}$ \\ 
20 & 0.53 & 3.40  & 6.396$\times$10$^{-9}$ & 1.100$\times$10$^{-4}$ & 9.311$\times$10$^{-6}$ & 1.238$\times$10$^{-2}$ & 2.838$\times$10$^{-15}$ & 8.974$\times$10$^{-9}$ \\ 
20 & 4.33 & 1.87  & 5.017$\times$10$^{-4}$ & 5.269$\times$10$^{-4}$ & 2.941$\times$10$^{-5}$ & 2.205$\times$10$^{-2}$ & 1.019$\times$10$^{-11}$ & 9.393$\times$10$^{-9}$ \\ 
20 & 0.81 & 2.70  & 5.512$\times$10$^{-5}$ & 2.710$\times$10$^{-4}$ & 1.609$\times$10$^{-5}$ & 1.366$\times$10$^{-2}$ & 5.324$\times$10$^{-15}$ & 8.972$\times$10$^{-9}$ \\ 
20 & 1.19 & 2.28  & 6.470$\times$10$^{-5}$ & 3.022$\times$10$^{-4}$ & 2.035$\times$10$^{-5}$ & 1.321$\times$10$^{-2}$ & 7.082$\times$10$^{-15}$ & 8.972$\times$10$^{-9}$ \\ 
20 & 1.52 & 1.97  & 1.011$\times$10$^{-5}$ & 2.421$\times$10$^{-4}$ & 2.132$\times$10$^{-5}$ & 1.367$\times$10$^{-2}$ & 6.932$\times$10$^{-13}$ & 9.209$\times$10$^{-9}$ \\ 
20 & 1.39 & 1.93  & 1.559$\times$10$^{-6}$ & 1.763$\times$10$^{-4}$ & 2.132$\times$10$^{-5}$ & 1.393$\times$10$^{-2}$ & 6.069$\times$10$^{-13}$ & 9.299$\times$10$^{-9}$ \\ 
20 & 8.86 & 1.74  & 5.526$\times$10$^{-4}$ & 6.588$\times$10$^{-4}$ & 1.008$\times$10$^{-4}$ & 3.112$\times$10$^{-2}$ & 1.858$\times$10$^{-7}$ & 1.409$\times$10$^{-6}$ \\ 
20 & 0.85 & 2.62  & 5.501$\times$10$^{-5}$ & 2.776$\times$10$^{-4}$ & 1.688$\times$10$^{-5}$ & 1.353$\times$10$^{-2}$ & 5.634$\times$10$^{-15}$ & 8.972$\times$10$^{-9}$ \\ 
25 & 1.92 & 3.13  & 7.808$\times$10$^{-4}$ & 5.953$\times$10$^{-4}$ & 1.274$\times$10$^{-4}$ & 4.792$\times$10$^{-2}$ & 6.085$\times$10$^{-11}$ & 1.289$\times$10$^{-8}$ \\ 
25 & 3.07 & 1.83  & 7.810$\times$10$^{-4}$ & 5.935$\times$10$^{-4}$ & 1.277$\times$10$^{-4}$ & 4.748$\times$10$^{-2}$ & 1.198$\times$10$^{-8}$ & 1.308$\times$10$^{-8}$ \\ 
25 & 3.30 & 2.35  & 3.633$\times$10$^{-4}$ & 2.342$\times$10$^{-4}$ & 1.234$\times$10$^{-4}$ & 1.637$\times$10$^{-2}$ & 5.987$\times$10$^{-11}$ & 1.265$\times$10$^{-8}$ \\ 
25 & 1.04 & 1.84  & 5.222$\times$10$^{-4}$ & 4.567$\times$10$^{-4}$ & 9.402$\times$10$^{-5}$ & 4.215$\times$10$^{-2}$ & 9.485$\times$10$^{-9}$ & 1.539$\times$10$^{-8}$ \\ 
25 & 9.73 & 2.35  & 4.420$\times$10$^{-5}$ & 9.458$\times$10$^{-5}$ & 2.069$\times$10$^{-4}$ & 1.075$\times$10$^{-2}$ & 6.123$\times$10$^{-11}$ & 1.487$\times$10$^{-8}$ \\ 
25 & 1.20 & 1.84  & 6.260$\times$10$^{-4}$ & 4.565$\times$10$^{-4}$ & 9.402$\times$10$^{-5}$ & 4.215$\times$10$^{-2}$ & 9.015$\times$10$^{-9}$ & 1.543$\times$10$^{-8}$ \\ 
25 & 7.42 & 2.37  & 6.758$\times$10$^{-4}$ & 8.669$\times$10$^{-4}$ & 2.380$\times$10$^{-4}$ & 2.868$\times$10$^{-2}$ & 2.667$\times$10$^{-11}$ & 9.166$\times$10$^{-9}$ \\ 
25 & 18.4 & 2.35  & 4.348$\times$10$^{-5}$ & 9.310$\times$10$^{-5}$ & 2.898$\times$10$^{-4}$ & 1.058$\times$10$^{-2}$ & 6.182$\times$10$^{-11}$ & 1.492$\times$10$^{-8}$ \\ 
25 & 14.8 & 2.35  & 6.762$\times$10$^{-4}$ & 8.919$\times$10$^{-4}$ & 4.381$\times$10$^{-4}$ & 3.042$\times$10$^{-2}$ & 1.179$\times$10$^{-10}$ & 1.750$\times$10$^{-8}$ \\ 
25 & 2.78 & 2.35  & 3.634$\times$10$^{-4}$ & 2.340$\times$10$^{-4}$ & 1.234$\times$10$^{-4}$ & 1.637$\times$10$^{-2}$ & 5.973$\times$10$^{-11}$ & 1.265$\times$10$^{-8}$ \\ 
25 & 0.89 & 4.66  & 7.694$\times$10$^{-9}$ & 6.139$\times$10$^{-5}$ & 5.999$\times$10$^{-5}$ & 6.803$\times$10$^{-3}$ & 2.556$\times$10$^{-14}$ & 4.650$\times$10$^{-9}$ \\ 
25 & 7.08 & 2.35  & 3.633$\times$10$^{-4}$ & 2.341$\times$10$^{-4}$ & 1.234$\times$10$^{-4}$ & 1.637$\times$10$^{-2}$ & 5.989$\times$10$^{-11}$ & 1.265$\times$10$^{-8}$ \\ 
25 & 0.92 & 1.84  & 6.260$\times$10$^{-4}$ & 4.565$\times$10$^{-4}$ & 9.402$\times$10$^{-5}$ & 4.215$\times$10$^{-2}$ & 9.015$\times$10$^{-9}$ & 1.543$\times$10$^{-8}$ \\ 
25 & 2.64 & 2.35  & 3.634$\times$10$^{-4}$ & 2.340$\times$10$^{-4}$ & 1.234$\times$10$^{-4}$ & 1.637$\times$10$^{-2}$ & 5.971$\times$10$^{-11}$ & 1.265$\times$10$^{-8}$ \\ 
25 & 6.17 & 2.38  & 6.681$\times$10$^{-4}$ & 8.691$\times$10$^{-4}$ & 2.039$\times$10$^{-4}$ & 2.741$\times$10$^{-2}$ & 2.670$\times$10$^{-11}$ & 9.156$\times$10$^{-9}$ \\ 
25 & 4.73 & 2.38  & 4.989$\times$10$^{-4}$ & 6.807$\times$10$^{-4}$ & 1.265$\times$10$^{-4}$ & 3.242$\times$10$^{-2}$ & 1.423$\times$10$^{-10}$ & 2.099$\times$10$^{-8}$ \\ 
25 & 1.52 & 1.83  & 6.260$\times$10$^{-4}$ & 4.565$\times$10$^{-4}$ & 9.402$\times$10$^{-5}$ & 4.215$\times$10$^{-2}$ & 9.015$\times$10$^{-9}$ & 1.543$\times$10$^{-8}$ \\ 
25 & 1.57 & 3.73  & 1.169$\times$10$^{-8}$ & 6.298$\times$10$^{-5}$ & 6.926$\times$10$^{-5}$ & 6.956$\times$10$^{-3}$ & 3.245$\times$10$^{-14}$ & 4.638$\times$10$^{-9}$ \\ 
25 & 0.99 & 4.89  & 6.809$\times$10$^{-9}$ & 6.097$\times$10$^{-5}$ & 5.979$\times$10$^{-5}$ & 6.759$\times$10$^{-3}$ & 2.412$\times$10$^{-14}$ & 4.649$\times$10$^{-9}$ \\ 
25 & 8.40 & 2.38  & 1.776$\times$10$^{-5}$ & 8.317$\times$10$^{-5}$ & 2.131$\times$10$^{-4}$ & 9.851$\times$10$^{-3}$ & 6.121$\times$10$^{-11}$ & 1.485$\times$10$^{-8}$ \\ 
25 & 2.53 & 2.35  & 3.634$\times$10$^{-4}$ & 2.340$\times$10$^{-4}$ & 1.234$\times$10$^{-4}$ & 1.637$\times$10$^{-2}$ & 5.968$\times$10$^{-11}$ & 1.265$\times$10$^{-8}$ \\ 
25 & 4.72 & 2.35  & 3.633$\times$10$^{-4}$ & 2.343$\times$10$^{-4}$ & 1.234$\times$10$^{-4}$ & 1.637$\times$10$^{-2}$ & 5.989$\times$10$^{-11}$ & 1.265$\times$10$^{-8}$ \\ 
\end{tabular}
\caption{Complete table of decayed isotopic yields for all models in mass fraction. M$_{prog}$ is the progenitor mass, E$_{exp}$ is the explosion energy, M$_{rem}$ is the remnant mass.}
\label{yield_table:2}
\end{table*}

\begin{table*}
\centering
\begin{tabular}{rrr|rrrrrr}
M$_{prog}$& E$_{exp}$& M$_{rem}$ & $^{97}$Tc & $^{98}$Ru & $^{98}$Tc & $^{98}$Ru & $^{107}$Pd & $^{108}$Pd \\ 
(M$_{\odot})$ & (\FoE) & (M$_{\odot})$  &  &  &  &  &  &  \\ 
\hline
15 & 1.90 & 1.62  & 2.839$\times$10$^{-11}$ & 9.794$\times$10$^{-10}$ & 1.129$\times$10$^{-12}$ & 9.794$\times$10$^{-10}$ & 4.017$\times$10$^{-10}$ & 1.116$\times$10$^{-8}$ \\ 
15 & 1.86 & 1.63  & 2.876$\times$10$^{-11}$ & 9.795$\times$10$^{-10}$ & 1.123$\times$10$^{-12}$ & 9.795$\times$10$^{-10}$ & 4.022$\times$10$^{-10}$ & 1.119$\times$10$^{-8}$ \\ 
15 & 1.94 & 1.61  & 2.861$\times$10$^{-11}$ & 9.808$\times$10$^{-10}$ & 1.133$\times$10$^{-12}$ & 9.808$\times$10$^{-10}$ & 4.015$\times$10$^{-10}$ & 1.116$\times$10$^{-8}$ \\ 
15 & 0.54 & 1.91  & 2.881$\times$10$^{-11}$ & 9.627$\times$10$^{-10}$ & 1.255$\times$10$^{-12}$ & 9.627$\times$10$^{-10}$ & 4.516$\times$10$^{-10}$ & 1.150$\times$10$^{-8}$ \\ 
15 & 0.92 & 1.75  & 2.864$\times$10$^{-11}$ & 9.842$\times$10$^{-10}$ & 1.067$\times$10$^{-12}$ & 9.842$\times$10$^{-10}$ & 3.981$\times$10$^{-10}$ & 1.120$\times$10$^{-8}$ \\ 
15 & 1.69 & 1.52  & 9.054$\times$10$^{-11}$ & 1.201$\times$10$^{-9}$ & 1.158$\times$10$^{-12}$ & 1.201$\times$10$^{-9}$ & 3.983$\times$10$^{-10}$ & 1.117$\times$10$^{-8}$ \\ 
15 & 0.34 & 1.94  & 2.925$\times$10$^{-11}$ & 9.576$\times$10$^{-10}$ & 1.192$\times$10$^{-12}$ & 9.576$\times$10$^{-10}$ & 4.534$\times$10$^{-10}$ & 1.152$\times$10$^{-8}$ \\ 
15 & 3.43 & 1.51  & 6.779$\times$10$^{-11}$ & 1.164$\times$10$^{-9}$ & 8.463$\times$10$^{-13}$ & 1.164$\times$10$^{-9}$ & 4.342$\times$10$^{-10}$ & 1.093$\times$10$^{-8}$ \\ 
15 & 2.06 & 1.59  & 6.222$\times$10$^{-11}$ & 1.060$\times$10$^{-9}$ & 1.150$\times$10$^{-12}$ & 1.060$\times$10$^{-9}$ & 4.003$\times$10$^{-10}$ & 1.115$\times$10$^{-8}$ \\ 
15 & 2.24 & 1.56  & 1.495$\times$10$^{-10}$ & 1.294$\times$10$^{-9}$ & 1.183$\times$10$^{-12}$ & 1.294$\times$10$^{-9}$ & 3.960$\times$10$^{-10}$ & 1.115$\times$10$^{-8}$ \\ 
15 & 2.60 & 1.52  & 1.545$\times$10$^{-10}$ & 1.345$\times$10$^{-9}$ & 1.123$\times$10$^{-12}$ & 1.345$\times$10$^{-9}$ & 3.834$\times$10$^{-10}$ & 1.113$\times$10$^{-8}$ \\ 
15 & 0.74 & 1.73  & 2.959$\times$10$^{-11}$ & 9.900$\times$10$^{-10}$ & 1.038$\times$10$^{-12}$ & 9.900$\times$10$^{-10}$ & 4.031$\times$10$^{-10}$ & 1.124$\times$10$^{-8}$ \\ 
15 & 0.82 & 1.88  & 2.789$\times$10$^{-11}$ & 9.652$\times$10$^{-10}$ & 1.140$\times$10$^{-12}$ & 9.652$\times$10$^{-10}$ & 4.414$\times$10$^{-10}$ & 1.146$\times$10$^{-8}$ \\ 
15 & 10.7 & 1.53  & 5.590$\times$10$^{-11}$ & 1.176$\times$10$^{-9}$ & 1.093$\times$10$^{-12}$ & 1.176$\times$10$^{-9}$ & 3.856$\times$10$^{-10}$ & 1.096$\times$10$^{-8}$ \\ 
15 & 2.47 & 1.52  & 2.837$\times$10$^{-11}$ & 1.006$\times$10$^{-9}$ & 1.122$\times$10$^{-12}$ & 1.006$\times$10$^{-9}$ & 3.944$\times$10$^{-10}$ & 1.106$\times$10$^{-8}$ \\ 
15 & 4.79 & 1.50  & 7.164$\times$10$^{-11}$ & 1.239$\times$10$^{-9}$ & 3.760$\times$10$^{-13}$ & 1.239$\times$10$^{-9}$ & 4.935$\times$10$^{-10}$ & 1.081$\times$10$^{-8}$ \\ 
15 & 2.63 & 1.53  & 6.674$\times$10$^{-11}$ & 1.192$\times$10$^{-9}$ & 1.032$\times$10$^{-12}$ & 1.192$\times$10$^{-9}$ & 3.814$\times$10$^{-10}$ & 1.092$\times$10$^{-8}$ \\ 
20 & 2.85 & 1.74  & 3.259$\times$10$^{-11}$ & 9.824$\times$10$^{-10}$ & 3.968$\times$10$^{-12}$ & 9.824$\times$10$^{-10}$ & 1.300$\times$10$^{-9}$ & 1.659$\times$10$^{-8}$ \\ 
20 & 1.47 & 2.23  & 3.088$\times$10$^{-11}$ & 9.184$\times$10$^{-10}$ & 1.848$\times$10$^{-12}$ & 9.184$\times$10$^{-10}$ & 1.278$\times$10$^{-9}$ & 1.636$\times$10$^{-8}$ \\ 
20 & 2.50 & 1.93  & 3.336$\times$10$^{-11}$ & 9.793$\times$10$^{-10}$ & 4.012$\times$10$^{-12}$ & 9.793$\times$10$^{-10}$ & 1.287$\times$10$^{-9}$ & 1.674$\times$10$^{-8}$ \\ 
20 & 5.03 & 1.74  & 5.724$\times$10$^{-11}$ & 1.072$\times$10$^{-9}$ & 4.115$\times$10$^{-12}$ & 1.072$\times$10$^{-9}$ & 1.188$\times$10$^{-9}$ & 1.629$\times$10$^{-8}$ \\ 
20 & 2.76 & 1.76  & 3.286$\times$10$^{-11}$ & 9.836$\times$10$^{-10}$ & 3.965$\times$10$^{-12}$ & 9.836$\times$10$^{-10}$ & 1.295$\times$10$^{-9}$ & 1.658$\times$10$^{-8}$ \\ 
20 & 2.60 & 1.90  & 3.282$\times$10$^{-11}$ & 9.862$\times$10$^{-10}$ & 3.956$\times$10$^{-12}$ & 9.862$\times$10$^{-10}$ & 1.290$\times$10$^{-9}$ & 1.656$\times$10$^{-8}$ \\ 
20 & 2.43 & 1.86  & 3.322$\times$10$^{-11}$ & 9.808$\times$10$^{-10}$ & 3.692$\times$10$^{-12}$ & 9.808$\times$10$^{-10}$ & 1.300$\times$10$^{-9}$ & 1.658$\times$10$^{-8}$ \\ 
20 & 1.04 & 2.47  & 3.085$\times$10$^{-11}$ & 9.184$\times$10$^{-10}$ & 1.886$\times$10$^{-12}$ & 9.184$\times$10$^{-10}$ & 1.171$\times$10$^{-9}$ & 1.575$\times$10$^{-8}$ \\ 
20 & 4.15 & 1.85  & 5.807$\times$10$^{-11}$ & 1.071$\times$10$^{-9}$ & 4.459$\times$10$^{-12}$ & 1.071$\times$10$^{-9}$ & 1.219$\times$10$^{-9}$ & 1.639$\times$10$^{-8}$ \\ 
20 & 0.65 & 3.03  & 3.088$\times$10$^{-11}$ & 9.186$\times$10$^{-10}$ & 1.850$\times$10$^{-12}$ & 9.186$\times$10$^{-10}$ & 9.223$\times$10$^{-10}$ & 1.434$\times$10$^{-8}$ \\ 
20 & 0.78 & 2.85  & 3.088$\times$10$^{-11}$ & 9.185$\times$10$^{-10}$ & 1.854$\times$10$^{-12}$ & 9.185$\times$10$^{-10}$ & 1.002$\times$10$^{-9}$ & 1.479$\times$10$^{-8}$ \\ 
20 & 0.84 & 2.62  & 3.087$\times$10$^{-11}$ & 9.185$\times$10$^{-10}$ & 1.855$\times$10$^{-12}$ & 9.185$\times$10$^{-10}$ & 1.104$\times$10$^{-9}$ & 1.537$\times$10$^{-8}$ \\ 
20 & 1.00 & 2.35  & 3.086$\times$10$^{-11}$ & 9.184$\times$10$^{-10}$ & 1.869$\times$10$^{-12}$ & 9.184$\times$10$^{-10}$ & 1.225$\times$10$^{-9}$ & 1.605$\times$10$^{-8}$ \\ 
20 & 1.65 & 1.78  & 3.373$\times$10$^{-11}$ & 9.790$\times$10$^{-10}$ & 2.174$\times$10$^{-12}$ & 9.790$\times$10$^{-10}$ & 1.306$\times$10$^{-9}$ & 1.654$\times$10$^{-8}$ \\ 
20 & 0.75 & 2.76  & 3.088$\times$10$^{-11}$ & 9.185$\times$10$^{-10}$ & 1.852$\times$10$^{-12}$ & 9.185$\times$10$^{-10}$ & 1.042$\times$10$^{-9}$ & 1.502$\times$10$^{-8}$ \\ 
20 & 0.53 & 3.40  & 3.091$\times$10$^{-11}$ & 9.190$\times$10$^{-10}$ & 1.849$\times$10$^{-12}$ & 9.190$\times$10$^{-10}$ & 7.578$\times$10$^{-10}$ & 1.342$\times$10$^{-8}$ \\ 
20 & 4.33 & 1.87  & 6.347$\times$10$^{-11}$ & 1.192$\times$10$^{-9}$ & 4.429$\times$10$^{-12}$ & 1.192$\times$10$^{-9}$ & 1.198$\times$10$^{-9}$ & 1.630$\times$10$^{-8}$ \\ 
20 & 0.81 & 2.70  & 3.087$\times$10$^{-11}$ & 9.185$\times$10$^{-10}$ & 1.857$\times$10$^{-12}$ & 9.185$\times$10$^{-10}$ & 1.069$\times$10$^{-9}$ & 1.517$\times$10$^{-8}$ \\ 
20 & 1.19 & 2.28  & 3.082$\times$10$^{-11}$ & 9.184$\times$10$^{-10}$ & 1.928$\times$10$^{-12}$ & 9.184$\times$10$^{-10}$ & 1.256$\times$10$^{-9}$ & 1.622$\times$10$^{-8}$ \\ 
20 & 1.52 & 1.97  & 3.385$\times$10$^{-11}$ & 9.654$\times$10$^{-10}$ & 2.187$\times$10$^{-12}$ & 9.654$\times$10$^{-10}$ & 1.305$\times$10$^{-9}$ & 1.654$\times$10$^{-8}$ \\ 
20 & 1.39 & 1.93  & 3.370$\times$10$^{-11}$ & 9.759$\times$10$^{-10}$ & 2.072$\times$10$^{-12}$ & 9.759$\times$10$^{-10}$ & 1.306$\times$10$^{-9}$ & 1.654$\times$10$^{-8}$ \\ 
20 & 8.86 & 1.74  & 7.480$\times$10$^{-11}$ & 1.467$\times$10$^{-9}$ & 4.075$\times$10$^{-12}$ & 1.467$\times$10$^{-9}$ & 1.022$\times$10$^{-9}$ & 1.519$\times$10$^{-8}$ \\ 
20 & 0.85 & 2.62  & 3.087$\times$10$^{-11}$ & 9.185$\times$10$^{-10}$ & 1.859$\times$10$^{-12}$ & 9.185$\times$10$^{-10}$ & 1.105$\times$10$^{-9}$ & 1.537$\times$10$^{-8}$ \\ 
25 & 1.92 & 3.13  & 2.352$\times$10$^{-10}$ & 3.175$\times$10$^{-9}$ & 8.409$\times$10$^{-12}$ & 3.175$\times$10$^{-9}$ & 1.666$\times$10$^{-9}$ & 1.608$\times$10$^{-8}$ \\ 
25 & 3.07 & 1.83  & 2.362$\times$10$^{-10}$ & 3.175$\times$10$^{-9}$ & 8.548$\times$10$^{-12}$ & 3.175$\times$10$^{-9}$ & 8.621$\times$10$^{-6}$ & 3.193$\times$10$^{-5}$ \\ 
25 & 3.30 & 2.35  & 2.325$\times$10$^{-10}$ & 3.157$\times$10$^{-9}$ & 8.548$\times$10$^{-12}$ & 3.157$\times$10$^{-9}$ & 1.676$\times$10$^{-9}$ & 1.615$\times$10$^{-8}$ \\ 
25 & 1.04 & 1.84  & 1.609$\times$10$^{-10}$ & 1.925$\times$10$^{-9}$ & 7.192$\times$10$^{-12}$ & 1.925$\times$10$^{-9}$ & 1.995$\times$10$^{-9}$ & 1.757$\times$10$^{-8}$ \\ 
25 & 9.73 & 2.35  & 2.316$\times$10$^{-10}$ & 3.007$\times$10$^{-9}$ & 6.995$\times$10$^{-12}$ & 3.007$\times$10$^{-9}$ & 1.525$\times$10$^{-9}$ & 1.517$\times$10$^{-8}$ \\ 
25 & 1.20 & 1.84  & 1.609$\times$10$^{-10}$ & 1.925$\times$10$^{-9}$ & 7.192$\times$10$^{-12}$ & 1.925$\times$10$^{-9}$ & 2.058$\times$10$^{-9}$ & 1.758$\times$10$^{-8}$ \\ 
25 & 7.42 & 2.37  & 1.018$\times$10$^{-10}$ & 1.569$\times$10$^{-9}$ & 5.071$\times$10$^{-12}$ & 1.569$\times$10$^{-9}$ & 1.499$\times$10$^{-9}$ & 1.470$\times$10$^{-8}$ \\ 
25 & 18.4 & 2.35  & 2.316$\times$10$^{-10}$ & 3.046$\times$10$^{-9}$ & 6.502$\times$10$^{-12}$ & 3.046$\times$10$^{-9}$ & 1.530$\times$10$^{-9}$ & 1.496$\times$10$^{-8}$ \\ 
25 & 14.8 & 2.35  & 2.787$\times$10$^{-10}$ & 4.974$\times$10$^{-9}$ & 6.623$\times$10$^{-12}$ & 4.974$\times$10$^{-9}$ & 1.210$\times$10$^{-9}$ & 1.221$\times$10$^{-8}$ \\ 
25 & 2.78 & 2.35  & 2.325$\times$10$^{-10}$ & 3.157$\times$10$^{-9}$ & 8.551$\times$10$^{-12}$ & 3.157$\times$10$^{-9}$ & 1.697$\times$10$^{-9}$ & 1.617$\times$10$^{-8}$ \\ 
25 & 0.89 & 4.66  & 3.514$\times$10$^{-11}$ & 4.980$\times$10$^{-10}$ & 5.453$\times$10$^{-12}$ & 4.980$\times$10$^{-10}$ & 1.413$\times$10$^{-9}$ & 1.267$\times$10$^{-8}$ \\ 
25 & 7.08 & 2.35  & 2.325$\times$10$^{-10}$ & 3.157$\times$10$^{-9}$ & 8.541$\times$10$^{-12}$ & 3.157$\times$10$^{-9}$ & 1.669$\times$10$^{-9}$ & 1.613$\times$10$^{-8}$ \\ 
25 & 0.92 & 1.84  & 1.609$\times$10$^{-10}$ & 1.925$\times$10$^{-9}$ & 7.192$\times$10$^{-12}$ & 1.925$\times$10$^{-9}$ & 2.068$\times$10$^{-9}$ & 1.758$\times$10$^{-8}$ \\ 
25 & 2.64 & 2.35  & 2.325$\times$10$^{-10}$ & 3.157$\times$10$^{-9}$ & 8.551$\times$10$^{-12}$ & 3.157$\times$10$^{-9}$ & 1.695$\times$10$^{-9}$ & 1.617$\times$10$^{-8}$ \\ 
25 & 6.17 & 2.38  & 1.018$\times$10$^{-10}$ & 1.555$\times$10$^{-9}$ & 5.235$\times$10$^{-12}$ & 1.555$\times$10$^{-9}$ & 1.519$\times$10$^{-9}$ & 1.495$\times$10$^{-8}$ \\ 
25 & 4.73 & 2.38  & 9.483$\times$10$^{-11}$ & 1.496$\times$10$^{-9}$ & 5.849$\times$10$^{-12}$ & 1.496$\times$10$^{-9}$ & 1.859$\times$10$^{-9}$ & 1.675$\times$10$^{-8}$ \\ 
25 & 1.52 & 1.83  & 1.609$\times$10$^{-10}$ & 1.925$\times$10$^{-9}$ & 7.192$\times$10$^{-12}$ & 1.925$\times$10$^{-9}$ & 2.058$\times$10$^{-9}$ & 1.758$\times$10$^{-8}$ \\ 
25 & 1.57 & 3.73  & 3.235$\times$10$^{-11}$ & 4.964$\times$10$^{-10}$ & 6.448$\times$10$^{-12}$ & 4.964$\times$10$^{-10}$ & 1.889$\times$10$^{-9}$ & 1.560$\times$10$^{-8}$ \\ 
25 & 0.99 & 4.89  & 3.481$\times$10$^{-11}$ & 4.979$\times$10$^{-10}$ & 5.792$\times$10$^{-12}$ & 4.979$\times$10$^{-10}$ & 1.284$\times$10$^{-9}$ & 1.186$\times$10$^{-8}$ \\ 
25 & 8.40 & 2.38  & 2.305$\times$10$^{-10}$ & 3.004$\times$10$^{-9}$ & 6.866$\times$10$^{-12}$ & 3.004$\times$10$^{-9}$ & 1.530$\times$10$^{-9}$ & 1.518$\times$10$^{-8}$ \\ 
25 & 2.53 & 2.35  & 2.325$\times$10$^{-10}$ & 3.157$\times$10$^{-9}$ & 8.551$\times$10$^{-12}$ & 3.157$\times$10$^{-9}$ & 1.694$\times$10$^{-9}$ & 1.618$\times$10$^{-8}$ \\ 
25 & 4.72 & 2.35  & 2.326$\times$10$^{-10}$ & 3.157$\times$10$^{-9}$ & 8.546$\times$10$^{-12}$ & 3.157$\times$10$^{-9}$ & 1.670$\times$10$^{-9}$ & 1.614$\times$10$^{-8}$ \\ 
\end{tabular}
\caption{Complete table of decayed isotopic yields for all models in mass fraction. M$_{prog}$ is the progenitor mass, E$_{exp}$ is the explosion energy, M$_{rem}$ is the remnant mass.}
\label{yield_table:3}
\end{table*}

\begin{table*}
\centering
\begin{tabular}{rrr|rrrrrr}
M$_{prog}$& E$_{exp}$& M$_{rem}$ & $^{126}$Sn & $^{124}$Sn & $^{129}$I & $^{127}$I & $^{135}$Cs & $^{133}$Cs \\ 
(M$_{\odot})$ & (\FoE) & (M$_{\odot})$  &  &  &  &  &  &  \\ 
\hline
15 & 1.90 & 1.62  & 2.494$\times$10$^{-11}$ & 6.873$\times$10$^{-9}$ & 4.344$\times$10$^{-10}$ & 1.138$\times$10$^{-8}$ & 7.688$\times$10$^{-10}$ & 1.172$\times$10$^{-8}$ \\ 
15 & 1.86 & 1.63  & 1.900$\times$10$^{-11}$ & 6.839$\times$10$^{-9}$ & 4.205$\times$10$^{-10}$ & 1.139$\times$10$^{-8}$ & 7.504$\times$10$^{-10}$ & 1.173$\times$10$^{-8}$ \\ 
15 & 1.94 & 1.61  & 2.873$\times$10$^{-11}$ & 6.897$\times$10$^{-9}$ & 4.400$\times$10$^{-10}$ & 1.137$\times$10$^{-8}$ & 7.806$\times$10$^{-10}$ & 1.171$\times$10$^{-8}$ \\ 
15 & 0.54 & 1.91  & 2.260$\times$10$^{-12}$ & 6.734$\times$10$^{-9}$ & 3.054$\times$10$^{-10}$ & 1.146$\times$10$^{-8}$ & 7.522$\times$10$^{-10}$ & 1.167$\times$10$^{-8}$ \\ 
15 & 0.92 & 1.75  & 1.186$\times$10$^{-11}$ & 6.824$\times$10$^{-9}$ & 3.913$\times$10$^{-10}$ & 1.142$\times$10$^{-8}$ & 7.315$\times$10$^{-10}$ & 1.176$\times$10$^{-8}$ \\ 
15 & 1.69 & 1.52  & 2.773$\times$10$^{-11}$ & 6.901$\times$10$^{-9}$ & 4.325$\times$10$^{-10}$ & 1.136$\times$10$^{-8}$ & 7.726$\times$10$^{-10}$ & 1.170$\times$10$^{-8}$ \\ 
15 & 0.34 & 1.94  & 2.139$\times$10$^{-12}$ & 6.736$\times$10$^{-9}$ & 2.824$\times$10$^{-10}$ & 1.147$\times$10$^{-8}$ & 7.342$\times$10$^{-10}$ & 1.168$\times$10$^{-8}$ \\ 
15 & 3.43 & 1.51  & 8.130$\times$10$^{-10}$ & 8.856$\times$10$^{-9}$ & 3.737$\times$10$^{-10}$ & 1.133$\times$10$^{-8}$ & 9.662$\times$10$^{-10}$ & 1.164$\times$10$^{-8}$ \\ 
15 & 2.06 & 1.59  & 3.811$\times$10$^{-11}$ & 6.953$\times$10$^{-9}$ & 4.492$\times$10$^{-10}$ & 1.135$\times$10$^{-8}$ & 8.110$\times$10$^{-10}$ & 1.169$\times$10$^{-8}$ \\ 
15 & 2.24 & 1.56  & 6.181$\times$10$^{-11}$ & 7.106$\times$10$^{-9}$ & 4.529$\times$10$^{-10}$ & 1.132$\times$10$^{-8}$ & 8.787$\times$10$^{-10}$ & 1.163$\times$10$^{-8}$ \\ 
15 & 2.60 & 1.52  & 1.613$\times$10$^{-10}$ & 7.655$\times$10$^{-9}$ & 4.176$\times$10$^{-10}$ & 1.130$\times$10$^{-8}$ & 1.011$\times$10$^{-9}$ & 1.154$\times$10$^{-8}$ \\ 
15 & 0.74 & 1.73  & 8.060$\times$10$^{-12}$ & 6.799$\times$10$^{-9}$ & 3.617$\times$10$^{-10}$ & 1.143$\times$10$^{-8}$ & 7.220$\times$10$^{-10}$ & 1.173$\times$10$^{-8}$ \\ 
15 & 0.82 & 1.88  & 2.873$\times$10$^{-12}$ & 6.737$\times$10$^{-9}$ & 3.352$\times$10$^{-10}$ & 1.145$\times$10$^{-8}$ & 7.661$\times$10$^{-10}$ & 1.168$\times$10$^{-8}$ \\ 
15 & 10.7 & 1.53  & 1.790$\times$10$^{-11}$ & 6.824$\times$10$^{-9}$ & 3.751$\times$10$^{-10}$ & 1.137$\times$10$^{-8}$ & 6.664$\times$10$^{-10}$ & 1.167$\times$10$^{-8}$ \\ 
15 & 2.47 & 1.52  & 4.953$\times$10$^{-11}$ & 7.016$\times$10$^{-9}$ & 4.471$\times$10$^{-10}$ & 1.134$\times$10$^{-8}$ & 8.315$\times$10$^{-10}$ & 1.167$\times$10$^{-8}$ \\ 
15 & 4.79 & 1.50  & 1.700$\times$10$^{-9}$ & 9.111$\times$10$^{-9}$ & 4.826$\times$10$^{-10}$ & 1.150$\times$10$^{-8}$ & 8.658$\times$10$^{-10}$ & 1.173$\times$10$^{-8}$ \\ 
15 & 2.63 & 1.53  & 3.496$\times$10$^{-10}$ & 8.200$\times$10$^{-9}$ & 3.703$\times$10$^{-10}$ & 1.129$\times$10$^{-8}$ & 1.000$\times$10$^{-9}$ & 1.154$\times$10$^{-8}$ \\ 
20 & 2.85 & 1.74  & 3.038$\times$10$^{-13}$ & 7.555$\times$10$^{-9}$ & 4.829$\times$10$^{-10}$ & 1.384$\times$10$^{-8}$ & 1.093$\times$10$^{-9}$ & 1.416$\times$10$^{-8}$ \\ 
20 & 1.47 & 2.23  & 2.412$\times$10$^{-13}$ & 7.549$\times$10$^{-9}$ & 4.221$\times$10$^{-10}$ & 1.392$\times$10$^{-8}$ & 1.028$\times$10$^{-9}$ & 1.427$\times$10$^{-8}$ \\ 
20 & 2.50 & 1.93  & 3.940$\times$10$^{-13}$ & 7.565$\times$10$^{-9}$ & 5.158$\times$10$^{-10}$ & 1.378$\times$10$^{-8}$ & 1.117$\times$10$^{-9}$ & 1.413$\times$10$^{-8}$ \\ 
20 & 5.03 & 1.74  & 1.822$\times$10$^{-11}$ & 7.849$\times$10$^{-9}$ & 6.289$\times$10$^{-10}$ & 1.357$\times$10$^{-8}$ & 1.251$\times$10$^{-9}$ & 1.393$\times$10$^{-8}$ \\ 
20 & 2.76 & 1.76  & 3.481$\times$10$^{-13}$ & 7.561$\times$10$^{-9}$ & 4.801$\times$10$^{-10}$ & 1.384$\times$10$^{-8}$ & 1.088$\times$10$^{-9}$ & 1.416$\times$10$^{-8}$ \\ 
20 & 2.60 & 1.90  & 5.173$\times$10$^{-13}$ & 7.574$\times$10$^{-9}$ & 4.789$\times$10$^{-10}$ & 1.384$\times$10$^{-8}$ & 1.084$\times$10$^{-9}$ & 1.416$\times$10$^{-8}$ \\ 
20 & 2.43 & 1.86  & 2.850$\times$10$^{-13}$ & 7.556$\times$10$^{-9}$ & 4.679$\times$10$^{-10}$ & 1.387$\times$10$^{-8}$ & 1.080$\times$10$^{-9}$ & 1.419$\times$10$^{-8}$ \\ 
20 & 1.04 & 2.47  & 2.090$\times$10$^{-13}$ & 7.524$\times$10$^{-9}$ & 3.680$\times$10$^{-10}$ & 1.380$\times$10$^{-8}$ & 9.136$\times$10$^{-10}$ & 1.413$\times$10$^{-8}$ \\ 
20 & 4.15 & 1.85  & 1.116$\times$10$^{-11}$ & 7.813$\times$10$^{-9}$ & 6.135$\times$10$^{-10}$ & 1.361$\times$10$^{-8}$ & 1.236$\times$10$^{-9}$ & 1.395$\times$10$^{-8}$ \\ 
20 & 0.65 & 3.03  & 1.366$\times$10$^{-13}$ & 7.467$\times$10$^{-9}$ & 2.430$\times$10$^{-10}$ & 1.349$\times$10$^{-8}$ & 6.485$\times$10$^{-10}$ & 1.378$\times$10$^{-8}$ \\ 
20 & 0.78 & 2.85  & 1.598$\times$10$^{-13}$ & 7.486$\times$10$^{-9}$ & 2.831$\times$10$^{-10}$ & 1.359$\times$10$^{-8}$ & 7.337$\times$10$^{-10}$ & 1.390$\times$10$^{-8}$ \\ 
20 & 0.84 & 2.62  & 1.895$\times$10$^{-13}$ & 7.509$\times$10$^{-9}$ & 3.343$\times$10$^{-10}$ & 1.372$\times$10$^{-8}$ & 8.422$\times$10$^{-10}$ & 1.404$\times$10$^{-8}$ \\ 
20 & 1.00 & 2.35  & 2.245$\times$10$^{-13}$ & 7.537$\times$10$^{-9}$ & 3.947$\times$10$^{-10}$ & 1.387$\times$10$^{-8}$ & 9.704$\times$10$^{-10}$ & 1.420$\times$10$^{-8}$ \\ 
20 & 1.65 & 1.78  & 2.480$\times$10$^{-13}$ & 7.556$\times$10$^{-9}$ & 4.387$\times$10$^{-10}$ & 1.395$\times$10$^{-8}$ & 1.061$\times$10$^{-9}$ & 1.427$\times$10$^{-8}$ \\ 
20 & 0.75 & 2.76  & 1.714$\times$10$^{-13}$ & 7.495$\times$10$^{-9}$ & 3.032$\times$10$^{-10}$ & 1.364$\times$10$^{-8}$ & 7.762$\times$10$^{-10}$ & 1.395$\times$10$^{-8}$ \\ 
20 & 0.53 & 3.40  & 8.880$\times$10$^{-14}$ & 7.430$\times$10$^{-9}$ & 1.604$\times$10$^{-10}$ & 1.328$\times$10$^{-8}$ & 4.729$\times$10$^{-10}$ & 1.353$\times$10$^{-8}$ \\ 
20 & 4.33 & 1.87  & 1.394$\times$10$^{-11}$ & 7.834$\times$10$^{-9}$ & 6.182$\times$10$^{-10}$ & 1.358$\times$10$^{-8}$ & 1.238$\times$10$^{-9}$ & 1.393$\times$10$^{-8}$ \\ 
20 & 0.81 & 2.70  & 1.794$\times$10$^{-13}$ & 7.501$\times$10$^{-9}$ & 3.168$\times$10$^{-10}$ & 1.368$\times$10$^{-8}$ & 8.052$\times$10$^{-10}$ & 1.399$\times$10$^{-8}$ \\ 
20 & 1.19 & 2.28  & 2.336$\times$10$^{-13}$ & 7.544$\times$10$^{-9}$ & 4.104$\times$10$^{-10}$ & 1.390$\times$10$^{-8}$ & 1.004$\times$10$^{-9}$ & 1.424$\times$10$^{-8}$ \\ 
20 & 1.52 & 1.97  & 2.479$\times$10$^{-13}$ & 7.555$\times$10$^{-9}$ & 4.384$\times$10$^{-10}$ & 1.396$\times$10$^{-8}$ & 1.061$\times$10$^{-9}$ & 1.429$\times$10$^{-8}$ \\ 
20 & 1.39 & 1.93  & 2.479$\times$10$^{-13}$ & 7.556$\times$10$^{-9}$ & 4.382$\times$10$^{-10}$ & 1.396$\times$10$^{-8}$ & 1.060$\times$10$^{-9}$ & 1.429$\times$10$^{-8}$ \\ 
20 & 8.86 & 1.74  & 4.995$\times$10$^{-10}$ & 9.871$\times$10$^{-9}$ & 6.549$\times$10$^{-10}$ & 1.327$\times$10$^{-8}$ & 1.394$\times$10$^{-9}$ & 1.363$\times$10$^{-8}$ \\ 
20 & 0.85 & 2.62  & 1.896$\times$10$^{-13}$ & 7.509$\times$10$^{-9}$ & 3.344$\times$10$^{-10}$ & 1.372$\times$10$^{-8}$ & 8.426$\times$10$^{-10}$ & 1.404$\times$10$^{-8}$ \\ 
25 & 1.92 & 3.13  & 1.399$\times$10$^{-10}$ & 7.171$\times$10$^{-9}$ & 6.040$\times$10$^{-10}$ & 7.715$\times$10$^{-9}$ & 1.397$\times$10$^{-9}$ & 8.671$\times$10$^{-9}$ \\ 
25 & 3.07 & 1.83  & 5.669$\times$10$^{-6}$ & 3.943$\times$10$^{-6}$ & 3.172$\times$10$^{-6}$ & 5.095$\times$10$^{-7}$ & 1.396$\times$10$^{-9}$ & 5.909$\times$10$^{-8}$ \\ 
25 & 3.30 & 2.35  & 1.119$\times$10$^{-10}$ & 6.919$\times$10$^{-9}$ & 6.014$\times$10$^{-10}$ & 7.779$\times$10$^{-9}$ & 1.408$\times$10$^{-9}$ & 8.760$\times$10$^{-9}$ \\ 
25 & 1.04 & 1.84  & 2.468$\times$10$^{-11}$ & 5.372$\times$10$^{-9}$ & 5.244$\times$10$^{-10}$ & 8.311$\times$10$^{-9}$ & 1.429$\times$10$^{-9}$ & 9.398$\times$10$^{-9}$ \\ 
25 & 9.73 & 2.35  & 1.051$\times$10$^{-9}$ & 1.083$\times$10$^{-8}$ & 5.732$\times$10$^{-10}$ & 7.447$\times$10$^{-9}$ & 1.541$\times$10$^{-9}$ & 8.392$\times$10$^{-9}$ \\ 
25 & 1.20 & 1.84  & 2.468$\times$10$^{-11}$ & 5.372$\times$10$^{-9}$ & 5.434$\times$10$^{-10}$ & 8.317$\times$10$^{-9}$ & 1.433$\times$10$^{-9}$ & 9.399$\times$10$^{-9}$ \\ 
25 & 7.42 & 2.37  & 1.251$\times$10$^{-9}$ & 1.184$\times$10$^{-8}$ & 6.158$\times$10$^{-10}$ & 7.296$\times$10$^{-9}$ & 1.564$\times$10$^{-9}$ & 8.152$\times$10$^{-9}$ \\ 
25 & 18.4 & 2.35  & 1.823$\times$10$^{-9}$ & 1.221$\times$10$^{-8}$ & 5.800$\times$10$^{-10}$ & 7.403$\times$10$^{-9}$ & 1.557$\times$10$^{-9}$ & 8.333$\times$10$^{-9}$ \\ 
25 & 14.8 & 2.35  & 4.324$\times$10$^{-9}$ & 1.626$\times$10$^{-8}$ & 6.155$\times$10$^{-10}$ & 6.891$\times$10$^{-9}$ & 1.288$\times$10$^{-9}$ & 7.429$\times$10$^{-9}$ \\ 
25 & 2.78 & 2.35  & 1.119$\times$10$^{-10}$ & 6.919$\times$10$^{-9}$ & 6.073$\times$10$^{-10}$ & 7.779$\times$10$^{-9}$ & 1.409$\times$10$^{-9}$ & 8.760$\times$10$^{-9}$ \\ 
25 & 0.89 & 4.66  & 1.290$\times$10$^{-12}$ & 4.151$\times$10$^{-9}$ & 3.290$\times$10$^{-10}$ & 7.688$\times$10$^{-9}$ & 9.497$\times$10$^{-10}$ & 8.545$\times$10$^{-9}$ \\ 
25 & 7.08 & 2.35  & 1.119$\times$10$^{-10}$ & 6.919$\times$10$^{-9}$ & 5.942$\times$10$^{-10}$ & 7.778$\times$10$^{-9}$ & 1.406$\times$10$^{-9}$ & 8.760$\times$10$^{-9}$ \\ 
25 & 0.92 & 1.84  & 2.468$\times$10$^{-11}$ & 5.372$\times$10$^{-9}$ & 5.472$\times$10$^{-10}$ & 8.318$\times$10$^{-9}$ & 1.434$\times$10$^{-9}$ & 9.399$\times$10$^{-9}$ \\ 
25 & 2.64 & 2.35  & 1.119$\times$10$^{-10}$ & 6.919$\times$10$^{-9}$ & 6.073$\times$10$^{-10}$ & 7.779$\times$10$^{-9}$ & 1.409$\times$10$^{-9}$ & 8.760$\times$10$^{-9}$ \\ 
25 & 6.17 & 2.38  & 9.255$\times$10$^{-10}$ & 1.086$\times$10$^{-8}$ & 6.304$\times$10$^{-10}$ & 7.342$\times$10$^{-9}$ & 1.574$\times$10$^{-9}$ & 8.219$\times$10$^{-9}$ \\ 
25 & 4.73 & 2.38  & 3.921$\times$10$^{-10}$ & 7.656$\times$10$^{-9}$ & 6.826$\times$10$^{-10}$ & 7.693$\times$10$^{-9}$ & 1.651$\times$10$^{-9}$ & 8.748$\times$10$^{-9}$ \\ 
25 & 1.52 & 1.83  & 2.468$\times$10$^{-11}$ & 5.372$\times$10$^{-9}$ & 5.434$\times$10$^{-10}$ & 8.317$\times$10$^{-9}$ & 1.433$\times$10$^{-9}$ & 9.399$\times$10$^{-9}$ \\ 
25 & 1.57 & 3.73  & 8.161$\times$10$^{-12}$ & 4.428$\times$10$^{-9}$ & 4.637$\times$10$^{-10}$ & 8.125$\times$10$^{-9}$ & 1.250$\times$10$^{-9}$ & 9.234$\times$10$^{-9}$ \\ 
25 & 0.99 & 4.89  & 1.781$\times$10$^{-12}$ & 4.166$\times$10$^{-9}$ & 3.175$\times$10$^{-10}$ & 7.512$\times$10$^{-9}$ & 8.712$\times$10$^{-10}$ & 8.321$\times$10$^{-9}$ \\ 
25 & 8.40 & 2.38  & 1.109$\times$10$^{-9}$ & 1.093$\times$10$^{-8}$ & 5.891$\times$10$^{-10}$ & 7.427$\times$10$^{-9}$ & 1.561$\times$10$^{-9}$ & 8.375$\times$10$^{-9}$ \\ 
25 & 2.53 & 2.35  & 1.119$\times$10$^{-10}$ & 6.919$\times$10$^{-9}$ & 6.079$\times$10$^{-10}$ & 7.779$\times$10$^{-9}$ & 1.409$\times$10$^{-9}$ & 8.760$\times$10$^{-9}$ \\ 
25 & 4.72 & 2.35  & 1.119$\times$10$^{-10}$ & 6.919$\times$10$^{-9}$ & 5.981$\times$10$^{-10}$ & 7.779$\times$10$^{-9}$ & 1.407$\times$10$^{-9}$ & 8.761$\times$10$^{-9}$ \\ 
\end{tabular}
\caption{Complete table of decayed isotopic yields for all models in mass fraction. M$_{prog}$ is the progenitor mass, E$_{exp}$ is the explosion energy, M$_{rem}$ is the remnant mass.}
\label{yield_table:4}
\end{table*}

\begin{table*}
\centering
\begin{tabular}{rrr|rrrrrr}
M$_{prog}$& E$_{exp}$& M$_{rem}$ & $^{146}$Sm & $^{144}$Sm & $^{182}$Hf & $^{180}$Hf & $^{205}$Pb & $^{204}$Pb \\ 
(M$_{\odot})$ & (\FoE) & (M$_{\odot})$  &  &  &  &  &  &  \\ 
\hline
15 & 1.90 & 1.62  & 8.319$\times$10$^{-11}$ & 1.170$\times$10$^{-9}$ & 1.425$\times$10$^{-10}$ & 4.693$\times$10$^{-10}$ & 1.232$\times$10$^{-9}$ & 1.798$\times$10$^{-9}$ \\ 
15 & 1.86 & 1.63  & 8.308$\times$10$^{-11}$ & 1.168$\times$10$^{-9}$ & 1.528$\times$10$^{-10}$ & 4.694$\times$10$^{-10}$ & 1.254$\times$10$^{-9}$ & 1.810$\times$10$^{-9}$ \\ 
15 & 1.94 & 1.61  & 8.388$\times$10$^{-11}$ & 1.173$\times$10$^{-9}$ & 1.365$\times$10$^{-10}$ & 4.716$\times$10$^{-10}$ & 1.220$\times$10$^{-9}$ & 1.790$\times$10$^{-9}$ \\ 
15 & 0.54 & 1.91  & 8.885$\times$10$^{-11}$ & 1.095$\times$10$^{-9}$ & 4.533$\times$10$^{-11}$ & 6.686$\times$10$^{-10}$ & 1.503$\times$10$^{-9}$ & 2.008$\times$10$^{-9}$ \\ 
15 & 0.92 & 1.75  & 8.647$\times$10$^{-11}$ & 1.326$\times$10$^{-9}$ & 1.622$\times$10$^{-10}$ & 4.758$\times$10$^{-10}$ & 1.287$\times$10$^{-9}$ & 1.947$\times$10$^{-9}$ \\ 
15 & 1.69 & 1.52  & 7.681$\times$10$^{-11}$ & 1.159$\times$10$^{-9}$ & 1.337$\times$10$^{-10}$ & 4.673$\times$10$^{-10}$ & 1.227$\times$10$^{-9}$ & 1.829$\times$10$^{-9}$ \\ 
15 & 0.34 & 1.94  & 8.906$\times$10$^{-11}$ & 1.067$\times$10$^{-9}$ & 2.718$\times$10$^{-11}$ & 6.747$\times$10$^{-10}$ & 1.529$\times$10$^{-9}$ & 2.053$\times$10$^{-9}$ \\ 
15 & 3.43 & 1.51  & 8.087$\times$10$^{-11}$ & 1.551$\times$10$^{-9}$ & 3.000$\times$10$^{-10}$ & 5.965$\times$10$^{-10}$ & 1.003$\times$10$^{-9}$ & 1.779$\times$10$^{-9}$ \\ 
15 & 2.06 & 1.59  & 8.255$\times$10$^{-11}$ & 1.175$\times$10$^{-9}$ & 1.263$\times$10$^{-10}$ & 4.812$\times$10$^{-10}$ & 1.198$\times$10$^{-9}$ & 1.780$\times$10$^{-9}$ \\ 
15 & 2.24 & 1.56  & 8.171$\times$10$^{-11}$ & 1.166$\times$10$^{-9}$ & 1.120$\times$10$^{-10}$ & 5.117$\times$10$^{-10}$ & 1.161$\times$10$^{-9}$ & 1.766$\times$10$^{-9}$ \\ 
15 & 2.60 & 1.52  & 7.978$\times$10$^{-11}$ & 1.180$\times$10$^{-9}$ & 1.756$\times$10$^{-10}$ & 5.697$\times$10$^{-10}$ & 1.106$\times$10$^{-9}$ & 1.747$\times$10$^{-9}$ \\ 
15 & 0.74 & 1.73  & 8.497$\times$10$^{-11}$ & 1.401$\times$10$^{-9}$ & 1.573$\times$10$^{-10}$ & 4.986$\times$10$^{-10}$ & 1.312$\times$10$^{-9}$ & 1.960$\times$10$^{-9}$ \\ 
15 & 0.82 & 1.88  & 9.004$\times$10$^{-11}$ & 1.131$\times$10$^{-9}$ & 9.491$\times$10$^{-11}$ & 6.329$\times$10$^{-10}$ & 1.449$\times$10$^{-9}$ & 1.945$\times$10$^{-9}$ \\ 
15 & 10.7 & 1.53  & 9.683$\times$10$^{-11}$ & 1.437$\times$10$^{-9}$ & 1.429$\times$10$^{-10}$ & 4.469$\times$10$^{-10}$ & 1.176$\times$10$^{-9}$ & 1.675$\times$10$^{-9}$ \\ 
15 & 2.47 & 1.52  & 9.512$\times$10$^{-11}$ & 1.354$\times$10$^{-9}$ & 1.145$\times$10$^{-10}$ & 4.935$\times$10$^{-10}$ & 1.145$\times$10$^{-9}$ & 1.733$\times$10$^{-9}$ \\ 
15 & 4.79 & 1.50  & 9.049$\times$10$^{-11}$ & 2.026$\times$10$^{-9}$ & 3.270$\times$10$^{-10}$ & 5.840$\times$10$^{-10}$ & 8.844$\times$10$^{-10}$ & 1.721$\times$10$^{-9}$ \\ 
15 & 2.63 & 1.53  & 9.902$\times$10$^{-11}$ & 1.755$\times$10$^{-9}$ & 2.502$\times$10$^{-10}$ & 5.771$\times$10$^{-10}$ & 1.016$\times$10$^{-9}$ & 1.714$\times$10$^{-9}$ \\ 
20 & 2.85 & 1.74  & 8.829$\times$10$^{-12}$ & 4.731$\times$10$^{-10}$ & 1.596$\times$10$^{-10}$ & 2.043$\times$10$^{-9}$ & 6.253$\times$10$^{-9}$ & 7.217$\times$10$^{-9}$ \\ 
20 & 1.47 & 2.23  & 3.026$\times$10$^{-13}$ & 3.264$\times$10$^{-10}$ & 1.510$\times$10$^{-10}$ & 2.042$\times$10$^{-9}$ & 6.133$\times$10$^{-9}$ & 7.058$\times$10$^{-9}$ \\ 
20 & 2.50 & 1.93  & 6.432$\times$10$^{-12}$ & 4.977$\times$10$^{-10}$ & 1.442$\times$10$^{-10}$ & 2.069$\times$10$^{-9}$ & 6.196$\times$10$^{-9}$ & 7.106$\times$10$^{-9}$ \\ 
20 & 5.03 & 1.74  & 5.086$\times$10$^{-11}$ & 9.943$\times$10$^{-10}$ & 4.121$\times$10$^{-10}$ & 1.749$\times$10$^{-9}$ & 5.808$\times$10$^{-9}$ & 6.879$\times$10$^{-9}$ \\ 
20 & 2.76 & 1.76  & 9.208$\times$10$^{-12}$ & 4.776$\times$10$^{-10}$ & 1.581$\times$10$^{-10}$ & 2.039$\times$10$^{-9}$ & 6.232$\times$10$^{-9}$ & 7.192$\times$10$^{-9}$ \\ 
20 & 2.60 & 1.90  & 1.050$\times$10$^{-11}$ & 4.893$\times$10$^{-10}$ & 1.569$\times$10$^{-10}$ & 2.036$\times$10$^{-9}$ & 6.206$\times$10$^{-9}$ & 7.159$\times$10$^{-9}$ \\ 
20 & 2.43 & 1.86  & 8.884$\times$10$^{-12}$ & 4.709$\times$10$^{-10}$ & 1.569$\times$10$^{-10}$ & 2.047$\times$10$^{-9}$ & 6.254$\times$10$^{-9}$ & 7.229$\times$10$^{-9}$ \\ 
20 & 1.04 & 2.47  & 1.096$\times$10$^{-13}$ & 3.264$\times$10$^{-10}$ & 1.337$\times$10$^{-10}$ & 1.888$\times$10$^{-9}$ & 5.530$\times$10$^{-9}$ & 6.396$\times$10$^{-9}$ \\ 
20 & 4.15 & 1.85  & 5.069$\times$10$^{-11}$ & 9.497$\times$10$^{-10}$ & 3.058$\times$10$^{-10}$ & 1.884$\times$10$^{-9}$ & 5.892$\times$10$^{-9}$ & 6.981$\times$10$^{-9}$ \\ 
20 & 0.65 & 3.03  & 9.568$\times$10$^{-14}$ & 3.264$\times$10$^{-10}$ & 8.813$\times$10$^{-11}$ & 1.576$\times$10$^{-9}$ & 4.121$\times$10$^{-9}$ & 4.824$\times$10$^{-9}$ \\ 
20 & 0.78 & 2.85  & 1.001$\times$10$^{-13}$ & 3.264$\times$10$^{-10}$ & 1.027$\times$10$^{-10}$ & 1.676$\times$10$^{-9}$ & 4.574$\times$10$^{-9}$ & 5.329$\times$10$^{-9}$ \\ 
20 & 0.84 & 2.62  & 1.056$\times$10$^{-13}$ & 3.264$\times$10$^{-10}$ & 1.214$\times$10$^{-10}$ & 1.804$\times$10$^{-9}$ & 5.150$\times$10$^{-9}$ & 5.972$\times$10$^{-9}$ \\ 
20 & 1.00 & 2.35  & 1.123$\times$10$^{-13}$ & 3.264$\times$10$^{-10}$ & 1.434$\times$10$^{-10}$ & 1.955$\times$10$^{-9}$ & 5.832$\times$10$^{-9}$ & 6.733$\times$10$^{-9}$ \\ 
20 & 1.65 & 1.78  & 5.378$\times$10$^{-12}$ & 4.466$\times$10$^{-10}$ & 1.577$\times$10$^{-10}$ & 2.066$\times$10$^{-9}$ & 6.304$\times$10$^{-9}$ & 7.300$\times$10$^{-9}$ \\ 
20 & 0.75 & 2.76  & 1.022$\times$10$^{-13}$ & 3.264$\times$10$^{-10}$ & 1.101$\times$10$^{-10}$ & 1.726$\times$10$^{-9}$ & 4.800$\times$10$^{-9}$ & 5.581$\times$10$^{-9}$ \\ 
20 & 0.53 & 3.40  & 1.054$\times$10$^{-13}$ & 3.273$\times$10$^{-10}$ & 5.806$\times$10$^{-11}$ & 1.369$\times$10$^{-9}$ & 3.191$\times$10$^{-9}$ & 3.784$\times$10$^{-9}$ \\ 
20 & 4.33 & 1.87  & 7.359$\times$10$^{-11}$ & 1.499$\times$10$^{-9}$ & 3.478$\times$10$^{-10}$ & 1.823$\times$10$^{-9}$ & 5.809$\times$10$^{-9}$ & 6.897$\times$10$^{-9}$ \\ 
20 & 0.81 & 2.70  & 1.037$\times$10$^{-13}$ & 3.264$\times$10$^{-10}$ & 1.150$\times$10$^{-10}$ & 1.761$\times$10$^{-9}$ & 4.954$\times$10$^{-9}$ & 5.753$\times$10$^{-9}$ \\ 
20 & 1.19 & 2.28  & 1.148$\times$10$^{-13}$ & 3.264$\times$10$^{-10}$ & 1.492$\times$10$^{-10}$ & 1.995$\times$10$^{-9}$ & 6.008$\times$10$^{-9}$ & 6.929$\times$10$^{-9}$ \\ 
20 & 1.52 & 1.97  & 5.638$\times$10$^{-12}$ & 4.365$\times$10$^{-10}$ & 1.577$\times$10$^{-10}$ & 2.065$\times$10$^{-9}$ & 6.301$\times$10$^{-9}$ & 7.301$\times$10$^{-9}$ \\ 
20 & 1.39 & 1.93  & 5.328$\times$10$^{-12}$ & 4.409$\times$10$^{-10}$ & 1.577$\times$10$^{-10}$ & 2.067$\times$10$^{-9}$ & 6.305$\times$10$^{-9}$ & 7.303$\times$10$^{-9}$ \\ 
20 & 8.86 & 1.74  & 8.004$\times$10$^{-11}$ & 2.496$\times$10$^{-9}$ & 4.176$\times$10$^{-10}$ & 1.397$\times$10$^{-9}$ & 4.789$\times$10$^{-9}$ & 6.023$\times$10$^{-9}$ \\ 
20 & 0.85 & 2.62  & 1.057$\times$10$^{-13}$ & 3.264$\times$10$^{-10}$ & 1.215$\times$10$^{-10}$ & 1.805$\times$10$^{-9}$ & 5.152$\times$10$^{-9}$ & 5.975$\times$10$^{-9}$ \\ 
25 & 1.92 & 3.13  & 2.850$\times$10$^{-10}$ & 1.230$\times$10$^{-8}$ & 3.117$\times$10$^{-10}$ & 1.711$\times$10$^{-9}$ & 6.031$\times$10$^{-9}$ & 9.070$\times$10$^{-9}$ \\ 
25 & 3.07 & 1.83  & 2.844$\times$10$^{-10}$ & 1.230$\times$10$^{-8}$ & 3.117$\times$10$^{-10}$ & 1.711$\times$10$^{-9}$ & 6.031$\times$10$^{-9}$ & 9.070$\times$10$^{-9}$ \\ 
25 & 3.30 & 2.35  & 3.073$\times$10$^{-10}$ & 1.212$\times$10$^{-8}$ & 3.031$\times$10$^{-10}$ & 1.735$\times$10$^{-9}$ & 6.110$\times$10$^{-9}$ & 9.142$\times$10$^{-9}$ \\ 
25 & 1.04 & 1.84  & 1.789$\times$10$^{-10}$ & 5.907$\times$10$^{-9}$ & 3.810$\times$10$^{-10}$ & 2.215$\times$10$^{-9}$ & 7.333$\times$10$^{-9}$ & 9.925$\times$10$^{-9}$ \\ 
25 & 9.73 & 2.35  & 2.605$\times$10$^{-10}$ & 1.109$\times$10$^{-8}$ & 2.730$\times$10$^{-10}$ & 1.597$\times$10$^{-9}$ & 5.577$\times$10$^{-9}$ & 8.421$\times$10$^{-9}$ \\ 
25 & 1.20 & 1.84  & 1.789$\times$10$^{-10}$ & 5.907$\times$10$^{-9}$ & 3.810$\times$10$^{-10}$ & 2.215$\times$10$^{-9}$ & 7.333$\times$10$^{-9}$ & 9.925$\times$10$^{-9}$ \\ 
25 & 7.42 & 2.37  & 1.401$\times$10$^{-10}$ & 4.068$\times$10$^{-9}$ & 3.886$\times$10$^{-10}$ & 1.576$\times$10$^{-9}$ & 5.379$\times$10$^{-9}$ & 7.444$\times$10$^{-9}$ \\ 
25 & 18.4 & 2.35  & 2.645$\times$10$^{-10}$ & 1.142$\times$10$^{-8}$ & 4.609$\times$10$^{-10}$ & 1.625$\times$10$^{-9}$ & 5.474$\times$10$^{-9}$ & 8.339$\times$10$^{-9}$ \\ 
25 & 14.8 & 2.35  & 5.072$\times$10$^{-10}$ & 2.205$\times$10$^{-8}$ & 4.837$\times$10$^{-10}$ & 1.006$\times$10$^{-9}$ & 3.374$\times$10$^{-9}$ & 6.887$\times$10$^{-9}$ \\ 
25 & 2.78 & 2.35  & 3.069$\times$10$^{-10}$ & 1.212$\times$10$^{-8}$ & 3.075$\times$10$^{-10}$ & 1.737$\times$10$^{-9}$ & 6.110$\times$10$^{-9}$ & 9.142$\times$10$^{-9}$ \\ 
25 & 0.89 & 4.66  & 1.248$\times$10$^{-13}$ & 1.743$\times$10$^{-10}$ & 3.497$\times$10$^{-10}$ & 1.585$\times$10$^{-9}$ & 4.685$\times$10$^{-9}$ & 5.718$\times$10$^{-9}$ \\ 
25 & 7.08 & 2.35  & 3.073$\times$10$^{-10}$ & 1.212$\times$10$^{-8}$ & 2.926$\times$10$^{-10}$ & 1.730$\times$10$^{-9}$ & 6.110$\times$10$^{-9}$ & 9.142$\times$10$^{-9}$ \\ 
25 & 0.92 & 1.84  & 1.789$\times$10$^{-10}$ & 5.907$\times$10$^{-9}$ & 3.810$\times$10$^{-10}$ & 2.215$\times$10$^{-9}$ & 7.333$\times$10$^{-9}$ & 9.925$\times$10$^{-9}$ \\ 
25 & 2.64 & 2.35  & 3.069$\times$10$^{-10}$ & 1.212$\times$10$^{-8}$ & 3.094$\times$10$^{-10}$ & 1.738$\times$10$^{-9}$ & 6.110$\times$10$^{-9}$ & 9.142$\times$10$^{-9}$ \\ 
25 & 6.17 & 2.38  & 1.350$\times$10$^{-10}$ & 3.942$\times$10$^{-9}$ & 3.500$\times$10$^{-10}$ & 1.592$\times$10$^{-9}$ & 5.492$\times$10$^{-9}$ & 7.463$\times$10$^{-9}$ \\ 
25 & 4.73 & 2.38  & 1.222$\times$10$^{-10}$ & 3.644$\times$10$^{-9}$ & 3.403$\times$10$^{-10}$ & 1.988$\times$10$^{-9}$ & 6.676$\times$10$^{-9}$ & 8.481$\times$10$^{-9}$ \\ 
25 & 1.52 & 1.83  & 1.789$\times$10$^{-10}$ & 5.907$\times$10$^{-9}$ & 3.810$\times$10$^{-10}$ & 2.215$\times$10$^{-9}$ & 7.333$\times$10$^{-9}$ & 9.925$\times$10$^{-9}$ \\ 
25 & 1.57 & 3.73  & 5.140$\times$10$^{-12}$ & 1.697$\times$10$^{-10}$ & 3.720$\times$10$^{-10}$ & 2.058$\times$10$^{-9}$ & 6.608$\times$10$^{-9}$ & 7.903$\times$10$^{-9}$ \\ 
25 & 0.99 & 4.89  & 1.422$\times$10$^{-13}$ & 1.740$\times$10$^{-10}$ & 3.463$\times$10$^{-10}$ & 1.418$\times$10$^{-9}$ & 4.150$\times$10$^{-9}$ & 5.074$\times$10$^{-9}$ \\ 
25 & 8.40 & 2.38  & 2.614$\times$10$^{-10}$ & 1.109$\times$10$^{-8}$ & 3.464$\times$10$^{-10}$ & 1.625$\times$10$^{-9}$ & 5.574$\times$10$^{-9}$ & 8.420$\times$10$^{-9}$ \\ 
25 & 2.53 & 2.35  & 3.069$\times$10$^{-10}$ & 1.212$\times$10$^{-8}$ & 3.113$\times$10$^{-10}$ & 1.739$\times$10$^{-9}$ & 6.110$\times$10$^{-9}$ & 9.142$\times$10$^{-9}$ \\ 
25 & 4.72 & 2.35  & 3.077$\times$10$^{-10}$ & 1.212$\times$10$^{-8}$ & 2.952$\times$10$^{-10}$ & 1.731$\times$10$^{-9}$ & 6.110$\times$10$^{-9}$ & 9.143$\times$10$^{-9}$ \\ 
\end{tabular}
\caption{Complete table of decayed isotopic yields for all models in mass fraction. M$_{prog}$ is the progenitor mass, E$_{exp}$ is the explosion energy, M$_{rem}$ is the remnant mass.}
\label{yield_table:5}
\end{table*}

% \begin{figure}
% \centering
% \includegraphics[width=\columnwidth]{s/aggregate_plot_M25_Al26.pdf}
% \caption{Same as Fig.~\ref{fig:agg_plot_al26_fe60_mn53_15}.}
% \label{fig:agg_appendix_1}
% \end{figure}

% Don't change these lines
\bsp	% typesetting comment
\label{lastpage}
\end{document}